\PassOptionsToPackage{unicode}{hyperref}
\PassOptionsToPackage{hyphens}{url}
\PassOptionsToPackage{dvipsnames,svgnames,x11names}{xcolor}
\documentclass[
  11pt,
]{article}
\usepackage{amsmath,amssymb}
\usepackage{iftex}
\ifPDFTeX
  \usepackage[T1]{fontenc}
  \usepackage[utf8]{inputenc}
  \usepackage{textcomp} 
\else 
  \usepackage{unicode-math} 
  \defaultfontfeatures{Scale=MatchLowercase}
  \defaultfontfeatures[\rmfamily]{Ligatures=TeX,Scale=1}
\fi
\usepackage[]{mathptmx}
\ifPDFTeX\else
\fi
\IfFileExists{upquote.sty}{\usepackage{upquote}}{}
\IfFileExists{microtype.sty}{
  \usepackage[]{microtype}
  \UseMicrotypeSet[protrusion]{basicmath} 
}{}
\makeatletter
\@ifundefined{KOMAClassName}{
  \IfFileExists{parskip.sty}{%
    \usepackage{parskip}
  }{
    \setlength{\parindent}{0pt}
    \setlength{\parskip}{6pt plus 2pt minus 1pt}}
}{
  \KOMAoptions{parskip=half}}
\makeatother
\usepackage{xcolor}
\usepackage{longtable,booktabs,array}
\usepackage{calc} 
\usepackage{etoolbox}
\makeatletter
\patchcmd\longtable{\par}{\if@noskipsec\mbox{}\fi\par}{}{}
\makeatother
\IfFileExists{footnotehyper.sty}{\usepackage{footnotehyper}}{\usepackage{footnote}}
\makesavenoteenv{longtable}
\usepackage{graphicx}
\makeatletter
\def\maxwidth{\ifdim\Gin@nat@width>\linewidth\linewidth\else\Gin@nat@width\fi}
\def\maxheight{\ifdim\Gin@nat@height>\textheight\textheight\else\Gin@nat@height\fi}
\makeatother
\setkeys{Gin}{width=\maxwidth,height=\maxheight,keepaspectratio}
\makeatletter
\def\fps@figure{htbp}
\makeatother
\providecommand{\tightlist}{%
  \setlength{\itemsep}{0pt}\setlength{\parskip}{0pt}}
\usepackage[margin=1in]{geometry}
\usepackage{newunicodechar}
\newunicodechar{−}{\ensuremath{-}}
\newunicodechar{×}{\ensuremath{\times}}
\newunicodechar{≤}{\ensuremath{\leq}}
\newunicodechar{≥}{\ensuremath{\geq}}
\newunicodechar{→}{\ensuremath{\rightarrow}}
\newunicodechar{≈}{\ensuremath{\approx}}
\newunicodechar{±}{\ensuremath{\pm}}
\newunicodechar{–}{--}
\usepackage{etoolbox}
\AtBeginEnvironment{longtable}{\footnotesize\setlength{\tabcolsep}{4pt}}
\usepackage{caption}
\usepackage{microtype}
\usepackage{xurl}
\usepackage{needspace}

\ifLuaTeX
  \usepackage{selnolig}  
\fi
\IfFileExists{bookmark.sty}{\usepackage{bookmark}}{\usepackage{hyperref}}
\IfFileExists{xurl.sty}{\usepackage{xurl}}{} 
\hypersetup{
  colorlinks=true,
  linkcolor={Maroon},
  filecolor={Maroon},
  citecolor={Blue},
  urlcolor={Blue},
  pdfcreator={LaTeX via pandoc}}

\author{}
\date{}

\begin{document}

\hypersetup{pdftitle={NIFT-SQL: Customer-Supervised Table Scoring for Enterprise Text-to-SQL Retrieval}, pdfauthor={Arjun Rao, Nishant Gopinath}}
\title{NIFT-SQL: Customer-Supervised Table Scoring for Enterprise Text-to-SQL Retrieval}
\author{Arjun Rao\\\texttt{\small arjun.rao01@sap.com} \and Nishant Gopinath\\\texttt{\small nishant.gopinath@sap.com}}
\date{November 2026}
\maketitle
\begin{abstract}
Enterprise text-to-SQL starts with finding the tables a question needs,
which is hard when a schema spans several databases and hundreds of
cryptic tables. We ask whether models trained separately on each
customer's example questions, reading only table names, column names,
and declared keys, improve table retrieval. NIFT-SQL (Necessity-Informed
Fusion for Text-to-SQL) fuses hybrid search over a retriever fine-tuned
on the customer's questions with a fine-tuned 560M cross-encoder that
scores every table for whether the query needs it, through a logistic
regression fitted on held-out questions. On BEAVER, LiveSQLBench, and
LiveSQLBench-Large (244 to 971 tables; four splits, three trainings),
NIFT-SQL places every reference table in the top 15 for 69.2\%, 89.5\%,
and 85.6\% of test questions, against 30.4\%, 41.0\%, and 21.8\% for
hybrid search and 61.1\%, 81.7\%, and 70.4\% for the fine-tuned
retriever with the same labels. With off-the-shelf search, the
cross-encoder without fine-tuning is 21 points lower. A zero-shot 8B
reranker needs no labels and, routed by database, shows no resolved
complete@15 difference on two collections; adding its scores as two
fusion features gives a high-accuracy mode that reaches 72.4\%, 92.5\%,
and 88.6\%, ahead of the validation-selected zero-shot reranker variant
on two collections, with no resolved difference on the third. Rank
fusion of the two trained parts, grouped by database, shows no resolved
difference from the learned fusion. Scored once on a newer BEAVER
release, NIFT-SQL beats search and leads the label-matched retriever at
top 8 on the synthetic test, but trails it at top 15 on real questions.

\end{abstract}

\hypertarget{introduction}{%
\subsection{1
Introduction}\label{introduction}}

A text-to-SQL system
can write a correct
query only if it has
the right tables,
and in enterprise
databases, finding
them is the
bottleneck. On the
BEAVER enterprise
benchmark,
state-of-the-art
agentic frameworks
reach 10.8\%
execution accuracy
with GPT-5.2, and
30.1\% when given
oracle hints for
every annotated
subtask.
Schema-linking hints
alone, which say
which tables, join
keys, and columns to
use, roughly double
execution accuracy
(for ReFoRCE
averaged over seven
LLMs, from 9.5\% to
18.9\%; Chen, Yang,
et al., 2024).
Finding the tables
is the first step.

Enterprise schemas
defeat generic
retrieval for four
reasons:

\begin{itemize}
\tightlist
\item
  \textbf{Names and
  keys only.} Table
  names, column
  names, and
  declared keys are
  often the only
  information
  available.
  Descriptions,
  comments, and
  sample values are
  missing,
  restricted, or
  untrustworthy, so
  we rely only on
  names and keys.
\item
  \textbf{Cryptic,
  local vocabulary.}
  Names such as
  \texttt{fclt\_\allowbreak{}org\_\allowbreak{}dlc\_\allowbreak{}key}
  or
  \texttt{arcregistry}
  encode a company's
  conventions that a
  general-purpose
  embedding model
  has never seen.
\item
  \textbf{Size.} A
  customer's schema
  can span several
  databases and
  hundreds to
  thousands of
  tables;
  LiveSQLBench-Large
  has 971 tables in
  18 databases,
  about 54 per
  database.
\item
  \textbf{Multi-table
  questions.} A
  query needs every
  table it joins. On
  BEAVER, the
  average query uses
  4.4 tables, and
  31\% use five or
  more. Finding most
  of them is not
  enough.
\end{itemize}

The usual answers
are a stronger
general model (a
larger embedding
model, a zero-shot
reranker, or an LLM
reading the schema)
or a hand-built
semantic layer. We
take a different
route. Many
enterprise
deployments have
verified example
questions with their
SQL, and a few dozen
to about ninety of
them already give
large gains over
search. We train a
small scorer per
customer to answer
one question for
every table: is this
table needed to
answer the question?
We combine its
judgment with hybrid
search, whose
retriever is
fine-tuned on the
same questions,
through a small
learned fusion. We
study how
customer-supervised
scoring complements
conventional
retrieval, and when
database grouping or
learned fusion adds
value beyond the
scorer alone. Our
central finding is
that
customer-supervised
scoring and
retrieval complement
each other. Each
trained part
improves over
search, and on the
three main
collections together
they outperform a
retriever trained on
the same labels at
both cutoffs. That
lead is narrower
than it looks: in a
frozen-method
holdout on a newer
BEAVER release, it
holds only at
complete@8 on
synthetic questions,
and when the models
are trained only on
synthetic questions,
the fine-tuned
retriever alone is
ahead at complete@15
on real logged
questions, as the
trained scorer
transfers poorly
from synthetic to
real questions
(Section 6.9). A
fine-tuned retriever
is therefore a
strong and cheaper
baseline, and the
scorer's value is
clearest at a tight
context budget. A
zero-shot 8B
reranker, which
needs no labels, is
complementary rather
than a substitute:
on its own it comes
close to the trained
pipeline on two
collections, and as
two extra fusion
features it improves
complete@8 on all
three collections
and complete@15 on
the two LiveSQLBench
collections. Of the
two scorer
architectures we
tested, neither
clearly beats the
other inside the
fused pipeline at
the full training
size, although the
cross-encoder is
stronger on its own
and with less
training data
(Sections 5 and 7);
training each one
improves it
substantially, and
we use the
cross-encoder
because it is
cheaper.

Three empirical
findings organize
the paper. Trained
scoring and
retrieval are
complementary:
together they
complete more
multi-table sets
than the retriever
trained on the same
labels, most clearly
on questions that
need five or more
tables (Sections 6.2
and 6.7). On
collections of small
databases, finding
the right database
matters more than
ranking the tables
inside it (Section
6.5). Per-customer
labels reach
accuracy that a
zero-shot reranker
reaches only at
about 15 times the
cost per table
scored, or about 2
to 8 times per
question, with the
gap narrowing as
schemas grow
(Sections 6.5 and
8).

\textbf{Contributions.}

\begin{enumerate}
\def\labelenumi{\arabic{enumi}.}
\tightlist
\item
  \textbf{NIFT-SQL}
  is a table
  retriever that
  needs only table
  names, column
  names, and
  declared keys. It
  fuses hybrid
  search over a
  retriever
  fine-tuned per
  customer with a
  per-customer
  scorer that reads
  every table in the
  schema, through a
  logistic
  regression fitted
  on held-out
  questions, with no
  generative LLM
  calls. An optional
  high-accuracy mode
  adds two features
  from a zero-shot
  reranker.
\item
  \textbf{A
  pre-specified
  evaluation,
  followed by
  controlled
  experiments and
  exploratory
  analyses,} on
  three public
  benchmarks: four
  random splits
  grouped by
  gold-table set,
  confidence
  intervals
  clustered by
  gold-table set,
  and means over
  three training
  runs. NIFT-SQL
  reaches
  complete@15 of
  69.2\%, 89.5\%,
  and 85.6\%, 39 to
  64 points above
  hybrid search;
  NIFT-SQL-base, the
  same pipeline with
  off-the-shelf
  search, whose
  configuration was
  fixed before the
  test runs (the
  cross-encoder
  scorer was added
  afterwards;
  Section 4),
  already leads
  search by 33 to 58
  points, and by 24
  and 47 points with
  the database given
  on the two
  collections with
  large databases
  (Sections 5, 6.5,
  6.6, and 6.8).
\item
  \textbf{Customer
  supervision is the
  active
  ingredient.} Two
  scorer
  architectures, a
  cross-encoder and
  the Laya decision
  model, both
  deliver large
  gains, and in
  NIFT-SQL-base the
  same cross-encoder
  without
  fine-tuning is
  about 21 points
  lower on
  complete@15. The
  trained scorer
  also adds to a
  retriever
  fine-tuned on the
  same labeled
  questions: with
  the label budget
  matched, NIFT-SQL
  leads that
  retriever at both
  cutoffs on every
  collection
  (complete@15 +8.1,
  +7.7, and +15.1
  points; Section
  6.8). On a newer
  BEAVER release
  scored once, only
  the complete@8
  lead on the
  synthetic test
  replicates; on
  real questions,
  the retriever
  leads at
  complete@15
  (Section 6.9). The
  fine-tuned
  retriever, in
  turn, lifts
  NIFT-SQL 3 to 6
  points above
  NIFT-SQL-base.
  Against a
  training-free rank
  fusion of the same
  two parts,
  grouping by
  database is the
  larger step on
  LiveSQLBench
  (+13.1 points of
  complete@15); on
  LiveSQLBench-Large,
  the learned
  weights add more
  under this
  decomposition
  (+4.3 against
  +2.9), but on no
  collection is
  their gain over
  grouped rank
  fusion resolved.
  Learned fusion
  determines how the
  high-accuracy mode
  weights the
  reranker's scores
  against the other
  two signals
  (Section 6.8).
\item
  \textbf{A
  comparison with a
  strong zero-shot
  reranker}
  (Qwen3-Reranker-8B),
  which needs no
  labels, with its
  routing chosen on
  validation
  questions. After
  reordering the
  search shortlist,
  it trails
  NIFT-SQL-base by
  about 30 points of
  complete@15 on the
  971-table
  collection because
  the shortlist
  misses reference
  tables for about
  half the
  questions. Routed
  or grouped by
  database, with the
  variant chosen per
  collection on
  validation, it
  comes much closer:
  NIFT-SQL leads it
  at complete@15 on
  BEAVER (+16.3
  {[}4.5, 28.4{]})
  and at complete@8
  on
  LiveSQLBench-Large
  (+11.5 {[}3.5,
  20.0{]}), with no
  resolved
  difference
  elsewhere. The two
  are complementary:
  the high-accuracy
  mode, at an
  estimated 1.7 to
  2.8 s more of
  batched scoring
  per question,
  leads the reranker
  at complete@8 on
  all three
  collections and at
  complete@15 on
  BEAVER and
  LiveSQLBench-Large
  (Section 6.8).
\item
  \textbf{Learning
  curves for both
  scorers} (with
  NIFT-SQL-base):
  with 21 to 69
  training questions
  per customer, the
  pipeline with the
  cross-encoder
  improves
  complete@15 over
  hybrid search by
  26 to 55
  percentage points,
  with additional
  labeled validation
  questions used to
  fit the fusion.
\end{enumerate}

\hypertarget{problem-setting-and-metrics}{%
\subsection{2
Problem setting and
metrics}\label{problem-setting-and-metrics}}

\textbf{Task.} Given
a natural-language
question and a
customer's schema,
rank all tables so
that every table the
SQL query needs
appears near the
top. The database is
not given (Section
6.5 also evaluates
the known-database
case): the retriever
searches the whole
schema, across all
databases in a
dataset.

\textbf{Schema
cards.} A schema
card, a single line
of text, represents
each table. In our
\textbf{names-and-keys
setting}, the card
holds only the table
name and its column
names
(\texttt{fclt\_\allowbreak{}org\_\allowbreak{}dlc\_\allowbreak{}key:\ FCLT\_\allowbreak{}ORGANIZATION\_\allowbreak{}KEY,\ DLC\_\allowbreak{}KEY}),
and declared foreign
keys, where they
exist, are passed to
the scorer alongside
it as join hints
(\texttt{mesh.equipref\ =\ equipment.equipregistry}).
We ignore column
types, comments, and
sample rows, even
when the benchmark
provides them. We
can append richer
metadata to the card
when a customer
provides it; we
evaluate this
limited-metadata
setting.

\textbf{Per-customer
setting.} A customer
is one schema plus
its example
questions. We train
and fit models
separately for each
customer; nothing
transfers between
customers. BEAVER is
one customer.
LiveSQLBench and
LiveSQLBench-Large
share databases and
many question texts,
so they form one
customer and are
always split
together.

\textbf{Metrics.}
The headline metric
is
\textbf{complete@K}:
the share of
questions whose
entire reference
table set appears
among the top K
retrieved tables. We
report K = 8 and 15
as two table-context
budgets, with
additional results
at K = 5 and 25; we
have not validated
them as optimal
context sizes for a
downstream SQL
generator.
complete@K measures
reference-table
coverage, not SQL
correctness: extra
tables may distract
a generator, and
alternative correct
queries may use
different table
sets. We also report
\textbf{R@N}, the
fraction of
reference tables
retrieved in the top
N, averaged across
questions. R@N gives
partial credit and
can hide failures on
multi-table
questions (Section
6.2 shows a case
where two methods
tie on R@10 but
differ on
complete@K).

\hypertarget{method}{%
\subsection{3
Method}\label{method}}

NIFT-SQL has three
parts (Figure 1):
hybrid search ranks
all tables, a
per-customer scorer
scores all tables,
and a learned fusion
combines the two
over a candidate
set.

\begin{figure}
\centering
\includegraphics{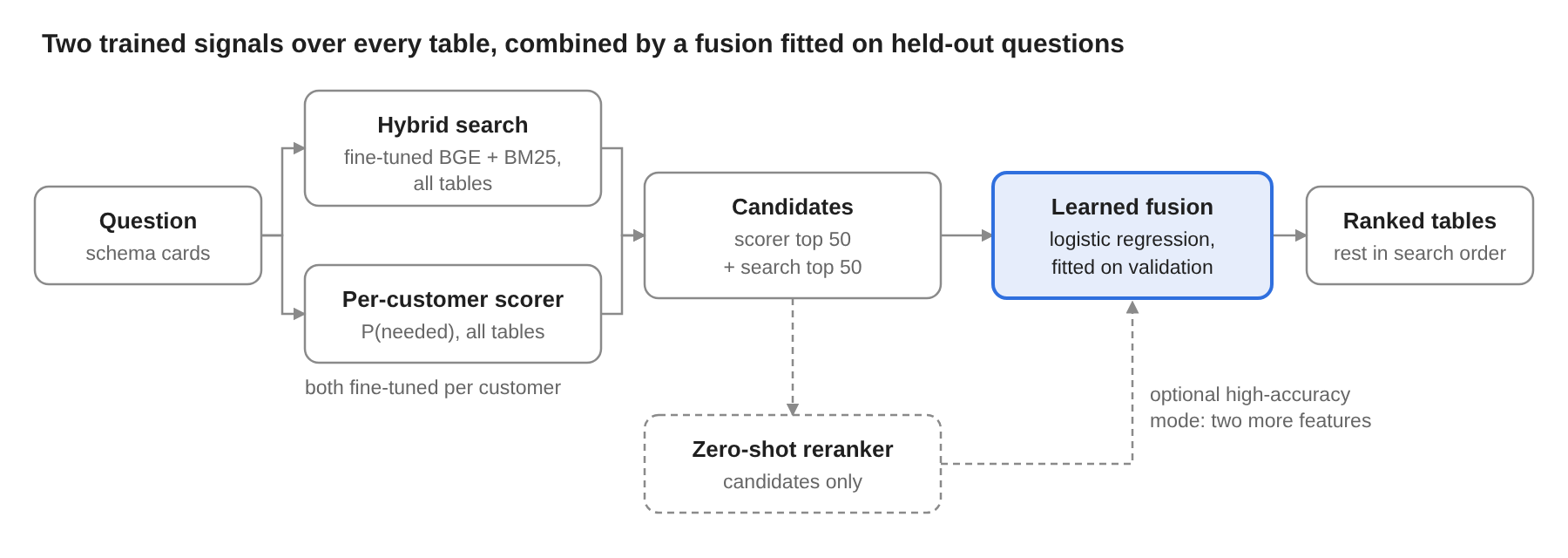}
\caption{NIFT-SQL at
inference: hybrid
search with a
retriever fine-tuned
per customer, a
per-customer scorer
over every table,
and a fusion fitted
on held-out
questions; the
optional
high-accuracy mode
adds two features
from a zero-shot
reranker over the
candidates}
\end{figure}

\textbf{3.1 Hybrid
search.} Each card
is embedded with
BGE-large-en-v1.5,
off the shelf, and
indexed with BM25
over name tokens
(split on
underscores and
camel case). For a
question, the cosine
scores and the BM25
scores are each
min-max normalized
over all tables,
then combined as 0.6
× dense + 0.4 ×
BM25. The weights
are fixed, not
tuned. In NIFT-SQL,
BGE is first
fine-tuned per
customer with a
contrastive loss
(in-batch negatives
plus one hard
negative) on the
scorer's training
rows, each training
question paired with
a gold table and a
hard negative: 2
epochs, learning
rate 2e-5, batch 32
(Appendix A). It
sees only the
training questions,
so its scores on the
validation questions
that fit the fusion
are out-of-sample,
as the scorer's are.
BM25 is unchanged.
NIFT-SQL-base, the
configuration fixed
before the test runs
and used for the
component analyses
of Sections 5 to 7,
uses BGE off the
shelf.

\textbf{3.2
Per-customer
scorer.} For every
(question, table)
pair, the scorer
estimates the
probability that the
table is required to
write the SQL query
that answers the
question: it holds
answer columns,
filter columns or
values, or is needed
to join such tables.
Operationally, we
label a table as
needed if it appears
in the benchmark's
relevance judgments,
which Zeng et
al.~(2026) derived
from the tables the
reference SQL uses;
we do not test
whether another
correct query could
avoid it. This is a
necessity label, not
topical relevance:
positives are the
tables the gold SQL
uses, including
join-only tables
that look unrelated
to the question, and
the hard negatives
are tables that look
relevant but are not
used. Training on
these labels turns a
general relevance
model into a
customer-specific
scorer (Section 5,
finding 1). We do
not test whether
necessity labels
beat relevance-style
labels: the
retrieval release
provides questions
and gold tables but
not the SQL, and we
did not join it back
to the source
benchmarks' SQL to
identify and relabel
join-only tables.
``Necessity''
therefore names the
labeling target, not
a tested advantage.
Its input is the
question, the
table's card, and up
to six of the
table's declared
joins. Training rows
come only from the
customer's training
questions. For each
training question,
we take every gold
table (label yes),
the four
highest-ranked
non-gold tables by
off-the-shelf hybrid
search (hard
negatives, label
no), and one random
non-gold table
(label no). At
inference, the
scorer scores
\textbf{every table
in the schema}, not
a shortlist. We
instantiate the
necessity scorer
with a cross-encoder
(headline) and a
decision model
(pre-specified),
built from the same
rows:

\begin{itemize}
\tightlist
\item
  \textbf{Fine-tuned
  cross-encoder
  (headline).}
  BAAI/bge-reranker-large
  (BAAI, n.d.; 560M
  parameters),
  fine-tuned from
  its public
  checkpoint for 4
  epochs with binary
  cross-entropy
  (learning rate
  2e-5, batch 16;
  inputs truncated
  at 512 tokens, a
  limit no input
  reached because
  cards are capped
  at 700 characters
  and six joins, the
  same caps Laya's
  inputs use). The
  question is paired
  with the card,
  followed by
  \texttt{Joins:}
  and the declared
  joins.
\item
  \textbf{Decision
  model, Laya
  (pre-specified).}
  A decision model
  is a classifier
  that answers a
  typed question,
  here yes or no,
  with a probability
  from a single
  forward pass,
  without generating
  text. We use Laya
  (convaiinnovations/laya;
  421M parameters;
  Python package
  version 0.4.8) and
  fine-tune it from
  its public
  checkpoint for
  four epochs using
  soft
  cross-entropy,
  reserving 10\% of
  the supplied
  training rows for
  temperature
  scaling (Guo et
  al., 2017). At
  inference, we
  apply the fitted
  temperature to
  obtain
  temperature-scaled
  ``yes''
  probabilities.
  Laya therefore
  trains on 90\% of
  the rows, while
  the cross-encoder
  trains on all of
  them. We do not
  treat either
  scorer's output as
  a calibrated
  probability: the
  fusion uses the
  scorer's logit and
  rank and is fitted
  on validation
  questions.
\end{itemize}

\textbf{3.3 Fusion.}
The candidates for a
question are the
union of the
scorer's top 50 and
search's top 50,
about 80 to 90
tables in our
collections. Each
candidate gets eight
features, with ranks
counted from 0 among
all tables of the
collection:

\begin{itemize}
\tightlist
\item
  \textbf{scorer
  logit:} log(p / (1
  - p)) of the
  scorer's
  probability p that
  the table is
  needed
\item
  \textbf{scorer
  rank:} log(1 + the
  table's rank by
  the scorer)
\item
  \textbf{search
  rank:} log(1 + the
  table's rank by
  hybrid search)
\item
  \textbf{dense
  cosine} and
  \textbf{BM25
  score:} search's
  two components for
  the table (BM25
  min-max normalized
  per question)
\item
  \textbf{rank gap:}
  the absolute
  difference between
  the scorer-rank
  and search-rank
  features
\item
  \textbf{database
  best search rank}
  and
  \textbf{database
  best scorer rank:}
  log(1 + rank) of
  the best-ranked
  table, by search
  and by the scorer,
  in the candidate's
  database
\end{itemize}

Every feature is
computed from the
question and the
tables alone; none
uses the gold
tables. The fusion's
training data is one
row per candidate of
every
\textbf{validation}
question of the
customer, with label
1 if the candidate
is one of the
question's gold
tables and 0
otherwise. The
scorer was trained
only on the training
questions, so its
scores on validation
questions are
out-of-sample, just
as on test
questions. On seed
1, this gives 63
validation questions
and 5,092 rows for
BEAVER (6.2\%
positive) and 226
questions and 19,926
rows for the
LiveSQLBench
customer (2.9\%
positive); Appendix
H shows one
question's rows. We
fit a standardized
logistic regression
(C = 1) per customer
on these rows. We
rank candidates by
their probability;
the remaining tables
follow in search
order.

\textbf{Why fit the
fusion on
validation?} Fitting
on training
questions would use
scores the scorer
was trained on,
which are
overconfident.
Validation keeps the
fusion honest, at
the cost of fitting
it on 30\% of the
questions; Section
6.5 shows that much
of what the fitted
fusion adds over
plain rank fusion is
grouping tables by
database, and
Section 7 shows that
20 validation
questions are nearly
enough to fit it.

\textbf{High-accuracy
mode.} Optionally, a
zero-shot reranker
(Qwen3-Reranker-8B
with the prompt in
Appendix E) scores
every candidate, and
two more features
join the eight: its
logit and its log
rank among the
candidates (Section
6.8).

\hypertarget{experimental-setup}{%
\subsection{4
Experimental
setup}\label{experimental-setup}}

\textbf{Datasets.}
We use the BEAVER,
LiveSQLBench, and
LiveSQLBench-Large
collections from
Zeng et al.~(2026),
a public
schema-retrieval
benchmark that
recasts text-to-SQL
datasets as table
retrieval (Table 1).
That work evaluates
every query
zero-shot with
recall@10 and
nDCG@10; we train
per customer and
score held-out test
splits with
complete@K, so the
numbers are not
directly comparable.
Our main setting
searches all
databases of a
collection at once,
which is harder than
retrieving within
the question's
database; Section
6.5 reports the
latter as well.
BEAVER is built from
real enterprise data
warehouses and is
the hardest: it
declares no foreign
keys, its names are
cryptic, and 90\% of
its queries need
three or more
tables.

\needspace{12\baselineskip}

\emph{Table 1.
Datasets, schema
cards in the
names-and-keys
setting.}

\begin{longtable}[]{@{}
  >{\raggedright\arraybackslash}p{(\columnwidth - 6\tabcolsep) * \real{0.2167}}
  >{\raggedright\arraybackslash}p{(\columnwidth - 6\tabcolsep) * \real{0.1500}}
  >{\raggedright\arraybackslash}p{(\columnwidth - 6\tabcolsep) * \real{0.2667}}
  >{\raggedright\arraybackslash}p{(\columnwidth - 6\tabcolsep) * \real{0.3667}}@{}}
\toprule\noalign{}
\begin{minipage}[b]{\linewidth}\raggedright
\end{minipage} &
\begin{minipage}[b]{\linewidth}\raggedright
BEAVER
\end{minipage} &
\begin{minipage}[b]{\linewidth}\raggedright
LiveSQLBench
\end{minipage} &
\begin{minipage}[b]{\linewidth}\raggedright
LiveSQLBench-Large
\end{minipage} \\
\midrule\noalign{}
\endhead
\bottomrule\noalign{}
\endlastfoot
Tables & 463 & 244 &
971 \\
Databases & 6 & 22 &
18 \\
Tables per database
(mean) & 77 & 11 &
54 \\
Questions & 209 &
410 & 332 \\
Columns per table
(mean) & 9.2 & 8.0 &
18.3 \\
Tables per SQL query
(mean) & 4.44 & 2.62
& 2.71 \\
Queries needing 3+
tables & 90\% & 47\%
& 51\% \\
Queries needing 5+
tables & 31\% & 13\%
& 14\% \\
Distinct gold-table
sets & 126 & 277 &
227 \\
Train / val / test
questions (seed 1) &
104 / 63 / 42 & 198
/ 126 / 86 & 160 /
100 / 72 \\
\end{longtable}

\textbf{Splits.} We
split questions
50/30/20 into train,
validation, and test
by question count.
Questions that share
a gold-table set, or
(within a customer)
share the same
question text, are
linked, and every
linked group goes to
exactly one part. So
no test question
repeats a training
table combination or
a training question
text. Hard checks
enforce both rules,
and an additional
check found no
question texts that
differ only in case
or punctuation
across parts.

\textbf{Protocol.}
We used seed 0 for
development only and
fixed the main
configuration, with
Laya as the scorer,
before running the
test evaluations on
seeds 1 to 4. Each
split has separately
trained models. The
evaluation code
prevents repeat test
scoring, and all
four main test runs
used the same code
version. After
inspecting these
results, we added a
supervised
cross-encoder to
investigate whether
the gains required
Laya or could also
be obtained with
conventional
per-customer
supervision. Before
running it, we fixed
its training recipe
to match Laya's
training rows, epoch
count, and fusion
procedure, using
common default
cross-encoder
fine-tuning
hyperparameters
(learning rate 2e-5,
batch 16) set before
the run and not
tuned on any seed.
We used validation
labels to fit the
fusion but not to
select the
cross-encoder's
hyperparameters, and
we evaluated it once
on each of the same
four test splits. We
present it as the
primary
implementation
because its
retrieval
performance is
comparable to Laya's
and its recorded
compute cost is
lower, and we report
both scorers
throughout. We made
this choice after
observing the
results; the
comparison does not
establish
statistical
equivalence, and we
did not collect
timings under fully
matched conditions.
Finally, after the
analyses in Sections
5 to 7, we tested
whether fusing the
scorer with a
retriever fine-tuned
on the same
questions and adding
a zero-shot
reranker's scores
improves the
pipeline (Section
6.8). We fixed their
settings before test
scoring, reported
every variant, and
ran a validation
check first that
predicted their test
order. Because the
fine-tuned retriever
raised complete@15
on every collection
at no extra scoring
cost, we adopted it
into NIFT-SQL after
seeing those
results. The
configuration fixed
before the test runs
is NIFT-SQL-base,
which every
component analysis
uses. Last, we froze
NIFT-SQL and scored
it once on a newer
BEAVER release, with
hypotheses fixed
before scoring in a
dated commit that is
part of the code
release (Section
6.9).

\textbf{Baselines.}
Every method ranks
all tables of the
same schema for the
same test questions.

\begin{itemize}
\tightlist
\item
  \textbf{Search
  alone:} the hybrid
  search of Section
  3.1 with
  off-the-shelf BGE.
\item
  \textbf{Fine-tuned
  retriever alone:}
  hybrid search with
  BGE fine-tuned per
  customer as in
  Section 3.1,
  trained on the
  training
  questions, or, to
  match NIFT-SQL's
  label budget, on
  the training and
  validation
  questions (the
  label-matched
  retriever, or
  matched retriever
  for short).
\item
  \textbf{Scorer
  alone:} the
  fine-tuned
  cross-encoder's or
  Laya's ranking of
  all tables. We
  also run the same
  cross-encoder
  zero-shot, without
  fine-tuning, alone
  and inside
  NIFT-SQL-base, and
  a fine-tuned
  cross-encoder
  without join hints
  in its input.
\item
  \textbf{Rank
  average:}
  reciprocal-rank
  fusion of a
  scorer's ranking
  and the search
  ranking with k =
  60, score = 1/(60
  + scorer rank) +
  1/(60 + search
  rank), for each
  scorer. It uses
  the same inputs as
  NIFT-SQL-base with
  that scorer but
  learns nothing, so
  the gap measures
  what the fitted
  fusion adds.
\item
  \textbf{Search +
  Qwen3-Reranker-8B
  (zero-shot):} the
  search top 100
  reordered by an
  off-the-shelf 8B
  cross-encoder
  reranker that
  reads the question
  with the same
  schema card and
  declared join
  hints that the
  scorers read;
  tables below 100
  keep search order.
  Its prompt is in
  Appendix E.
\end{itemize}

\textbf{Statistics.}
We report means and
standard deviations
over the four seeds.
For differences, we
pool the
per-question
differences of all
four test sets and
compute a 95\%
bootstrap interval,
resampling
gold-table sets
(3,000 resamples).
We cluster by
gold-table set
across seeds because
questions with the
same gold set are
not independent, and
because the four
random splits
overlap: 36 of 130,
83 of 249, and 66 of
200 distinct test
questions appear in
more than one seed's
test set. Tables 2
and 3 weight each
seed equally, while
paired differences
pool questions so
that a difference
can differ from the
gap between two
Table 3 means by a
few tenths of a
point. Pooled tests
across the three
datasets cluster
LiveSQLBench and
LiveSQLBench-Large
questions that share
text. Intervals are
not adjusted for
multiple
comparisons: we
treat complete@15
against hybrid
search as the
primary comparison
and the rest as
secondary. The code
prevents scoring any
trained model on a
second test. Still,
after seeing the
results, we ran
analyses beyond the
pre-specified Laya
comparison. Every
comparison other
than NIFT-SQL-base
with Laya against
its baselines is
exploratory, and its
interval is nominal.
The intervals
resample questions
with each trained
model held fixed, so
they do not include
the variation from
retraining a scorer;
Section 6.6 measures
that variation for
the cross-encoder
and reports
two-level intervals
that include it
(Table G5). Three
training runs give
only a coarse
estimate of training
variance, so those
intervals may still
be too narrow.

\hypertarget{main-results}{%
\subsection{5 Main
results}\label{main-results}}

Table 2 gives
NIFT-SQL's results.
Each of the
cross-encoder's
three training runs
is paired with a
training run of the
fine-tuned
retriever, and
differences are
means over training
runs with 95\%
intervals from a
two-level bootstrap
over training runs
and gold-table-set
clusters (Section
6.6). NIFT-SQL has
the highest mean
complete@15 of any
method except its
own high-accuracy
mode on BEAVER and
LiveSQLBench-Large,
and is within one
point of the
zero-shot reranker
variant selected on
validation on
LiveSQLBench (89.5\%
against 90.3\%). On
LiveSQLBench,
reranking NIFT-SQL's
own candidates,
grouped by database,
reaches 93.4\%, the
highest of any
method there, but it
depends on
NIFT-SQL's trained
candidates (Section
6.8). Against
NIFT-SQL-base, the
fine-tuned retriever
adds 5.6 {[}0.7,
11.1{]}, 3.3 {[}0.6,
6.4{]}, and 5.6
{[}2.7, 8.8{]}
points of
complete@15. Against
the retriever
fine-tuned on the
same training and
validation
questions, NIFT-SQL
leads by 8.1 {[}1.1,
16.3{]}, 7.7 {[}3.5,
12.7{]}, and 15.1
{[}10.0, 21.0{]}
points of
complete@15 and 12.9
{[}2.9, 23.0{]},
14.4 {[}9.4,
19.6{]}, and 23.7
{[}17.6, 30.2{]} of
complete@8. Against
the zero-shot
reranker with its
variant chosen on
validation, it leads
by 16.3 {[}4.5,
28.4{]} points of
complete@15 on
BEAVER and 11.5
{[}3.5, 20.0{]}
points of complete@8
on
LiveSQLBench-Large;
the other
differences remain
unresolved. The
high-accuracy mode
adds 3.2 {[}−1.9,
8.9{]}, 3.0 {[}0.5,
5.9{]}, and 3.1
{[}0.5, 5.9{]}
points of
complete@15 and 8.1
{[}2.5, 15.3{]}, 6.8
{[}2.8, 11.4{]}, and
4.2 {[}0.3, 8.0{]}
points of complete@8
to NIFT-SQL.

\needspace{12\baselineskip}

\emph{Table 2.
NIFT-SQL, test,
complete@8 /
complete@15 in \%,
mean over seeds 1 to
4 and, for trained
methods, over three
training runs. The
reranker variant is
chosen per
collection and seed
on validation
questions (Section
6.5). Intervals for
these differences
are in Table G7 and
its note.}

\begin{longtable}[]{@{}
  >{\raggedright\arraybackslash}p{(\columnwidth - 6\tabcolsep) * \real{0.3333}}
  >{\raggedright\arraybackslash}p{(\columnwidth - 6\tabcolsep) * \real{0.1333}}
  >{\raggedright\arraybackslash}p{(\columnwidth - 6\tabcolsep) * \real{0.2167}}
  >{\raggedright\arraybackslash}p{(\columnwidth - 6\tabcolsep) * \real{0.3167}}@{}}
\toprule\noalign{}
\begin{minipage}[b]{\linewidth}\raggedright
Method
\end{minipage} &
\begin{minipage}[b]{\linewidth}\raggedright
BEAVER
\end{minipage} &
\begin{minipage}[b]{\linewidth}\raggedright
LiveSQLBench
\end{minipage} &
\begin{minipage}[b]{\linewidth}\raggedright
LiveSQLBench-Large
\end{minipage} \\
\midrule\noalign{}
\endhead
\bottomrule\noalign{}
\endlastfoot
Search alone
(off-the-shelf BGE)
& 20.2 / 30.4 & 28.3
/ 41.0 & 19.9 /
21.8 \\
Fine-tuned retriever
alone, training
questions & 29.6 /
58.9 & 52.4 / 73.1 &
38.7 / 58.1 \\
Fine-tuned retriever
alone, training +
validation questions
& 33.5 / 61.1 & 60.5
/ 81.7 & 50.1 /
70.4 \\
Search +
Qwen3-Reranker-8B
(search top 100) &
36.9 / 53.6 & 64.8 /
77.1 & 46.2 /
50.3 \\
Qwen3-Reranker-8B,
variant chosen on
validation & 37.5 /
53.0 & 72.0 / 90.3 &
62.2 / 78.7 \\
Qwen3-Reranker-8B on
NIFT-SQL's
candidates, grouped
(control; needs
NIFT-SQL's trained
candidates) & 38.3 /
54.4 & 75.1 / 93.4 &
61.2 / 73.8 \\
NIFT-SQL-base
(off-the-shelf
search) & 44.4 /
63.7 & 71.0 / 86.2 &
67.5 / 80.0 \\
\textbf{NIFT-SQL} &
\textbf{46.4 / 69.2}
& \textbf{74.9 /
89.5} & \textbf{74.0
/ 85.6} \\
NIFT-SQL,
high-accuracy mode
(+ reranker
features) & 54.6 /
72.4 & 81.6 / 92.5 &
78.0 / 88.6 \\
\end{longtable}

NIFT-SQL's final
form was chosen in
light of earlier
results (Section 4),
so its intervals do
not account for that
choice; Section 6.9
scores it once on a
newer release, whose
synthetic questions
are new and whose
real questions
largely overlap the
development
questions. The rest
of this section and
Sections 6.1 to 6.7
and 7 analyze the
components on
NIFT-SQL-base, the
same pipeline with
off-the-shelf
search, which was
fixed before the
test runs. Its
figures come from
the main training
run; Section 6.6
reports means over
three training runs
(63.7\%, 86.2\%, and
80.0\% complete@15).

NIFT-SQL-base with
the fine-tuned
cross-encoder has
the highest
complete@8 on all
three datasets and
the highest
complete@15 on
LiveSQLBench (Table
3). On BEAVER, rank
averaging with the
same scorer reaches
a similar
complete@15 (68.5 vs
67.3, not resolved),
and on
LiveSQLBench-Large,
NIFT-SQL-base with
Laya is 0.5 points
higher in the main
run. However, the
cross-encoder
pipeline's
three-training mean
is 0.9 points above
Laya's (80.0 against
79.1; pooled paired
difference +0.6
{[}−4.6, 6.3{]};
Section 6.6).
NIFT-SQL-base more
than doubles
search's complete@15
on BEAVER and
LiveSQLBench and
more than triples it
on
LiveSQLBench-Large.

\needspace{12\baselineskip}

\emph{Table 3. Test
results, complete@8
/ complete@15 in \%,
mean (standard
deviation) over
seeds 1 to 4. All K
and N are in
Appendix F.}

\begin{longtable}[]{@{}
  >{\raggedright\arraybackslash}p{(\columnwidth - 6\tabcolsep) * \real{0.3000}}
  >{\raggedright\arraybackslash}p{(\columnwidth - 6\tabcolsep) * \real{0.1833}}
  >{\raggedright\arraybackslash}p{(\columnwidth - 6\tabcolsep) * \real{0.2167}}
  >{\raggedright\arraybackslash}p{(\columnwidth - 6\tabcolsep) * \real{0.3000}}@{}}
\toprule\noalign{}
\begin{minipage}[b]{\linewidth}\raggedright
Method
\end{minipage} &
\begin{minipage}[b]{\linewidth}\raggedright
BEAVER
\end{minipage} &
\begin{minipage}[b]{\linewidth}\raggedright
LiveSQLBench
\end{minipage} &
\begin{minipage}[b]{\linewidth}\raggedright
LiveSQLBench-Large
\end{minipage} \\
\midrule\noalign{}
\endhead
\bottomrule\noalign{}
\endlastfoot
Search alone & 20.2
(8.8) / 30.4 (6.3) &
28.3 (4.1) / 41.0
(3.7) & 19.9 (5.9) /
21.8 (5.4) \\
Search +
Qwen3-Reranker-8B
(zero-shot) & 36.9
(14.1) / 53.6 (12.4)
& 64.8 (2.4) / 77.1
(5.2) & 46.2 (4.4) /
50.3 (4.9) \\
Zero-shot
cross-encoder alone
& 16.7 (9.1) / 28.0
(9.2) & 41.9 (5.4) /
55.1 (2.8) & 25.7
(2.6) / 35.3
(5.4) \\
Laya alone & 13.7
(10.7) / 31.0 (13.3)
& 34.9 (9.8) / 51.6
(7.1) & 35.7 (10.7)
/ 54.7 (7.4) \\
Fine-tuned
cross-encoder alone
& 21.4 (8.0) / 45.2
(15.2) & 58.2 (13.9)
/ 69.1 (12.3) & 58.7
(15.7) / 71.7
(12.8) \\
Rank average, Laya +
search (no learning)
& 36.9 (7.4) / 60.1
(16.4) & 51.0 (5.1)
/ 68.6 (6.4) & 46.9
(2.6) / 62.3
(3.9) \\
Rank average,
fine-tuned
cross-encoder +
search (no learning)
& 38.7 (10.7) / 68.5
(7.9) & 50.4 (8.8) /
64.7 (6.0) & 53.0
(7.8) / 66.8
(8.5) \\
NIFT-SQL-base with
zero-shot
cross-encoder & 20.2
(8.8) / 41.1 (9.6) &
60.1 (6.5) / 77.8
(6.5) & 35.0 (9.0) /
45.6 (6.5) \\
NIFT-SQL-base with
Laya (pre-specified)
& 41.7 (9.2) / 60.1
(11.9) & 70.3 (4.0)
/ 83.2 (3.5) & 63.8
(4.9) / 79.1
(2.7) \\
\textbf{NIFT-SQL-base
with fine-tuned
cross-encoder} &
\textbf{48.8 (13.0)
/ 67.3 (11.1)} &
\textbf{71.5 (6.2) /
86.2 (4.9)} &
\textbf{68.0 (9.3) /
78.6 (9.0)} \\
\end{longtable}

\needspace{12\baselineskip}

\emph{Table 4.
NIFT-SQL-base with
the fine-tuned
cross-encoder minus
other methods,
complete@8 /
complete@15 in
points, all four
test sets pooled,
95\% interval
clustered by
gold-table set
across seeds. Bold:
interval excludes
zero.}

\begin{longtable}[]{@{}
  >{\raggedright\arraybackslash}p{(\columnwidth - 6\tabcolsep) * \real{0.2623}}
  >{\raggedright\arraybackslash}p{(\columnwidth - 6\tabcolsep) * \real{0.2295}}
  >{\raggedright\arraybackslash}p{(\columnwidth - 6\tabcolsep) * \real{0.2295}}
  >{\raggedright\arraybackslash}p{(\columnwidth - 6\tabcolsep) * \real{0.2787}}@{}}
\toprule\noalign{}
\begin{minipage}[b]{\linewidth}\raggedright
NIFT-SQL-base
(cross-encoder)
minus
\end{minipage} &
\begin{minipage}[b]{\linewidth}\raggedright
BEAVER
\end{minipage} &
\begin{minipage}[b]{\linewidth}\raggedright
LiveSQLBench
\end{minipage} &
\begin{minipage}[b]{\linewidth}\raggedright
LiveSQLBench-Large
\end{minipage} \\
\midrule\noalign{}
\endhead
\bottomrule\noalign{}
\endlastfoot
Search alone &
\textbf{+28.6}
{[}13.9, 42.5{]} /
\textbf{+36.9}
{[}20.8, 52.5{]} &
\textbf{+43.2}
{[}35.0, 51.7{]} /
\textbf{+45.3}
{[}36.4, 54.8{]} &
\textbf{+48.0}
{[}39.9, 56.2{]} /
\textbf{+56.8}
{[}47.9, 65.3{]} \\
Search +
Qwen3-Reranker-8B &
\textbf{+11.9}
{[}2.4, 21.5{]} /
\textbf{+13.7}
{[}2.6, 25.5{]} &
+6.8 {[}−1.4,
14.7{]} /
\textbf{+9.1}
{[}2.2, 16.4{]} &
\textbf{+21.6}
{[}13.9, 29.5{]} /
\textbf{+28.2}
{[}18.7, 37.6{]} \\
Rank average, same
scorer + search &
\textbf{+10.1}
{[}2.9, 17.5{]} /
−1.2 {[}−6.0, 3.6{]}
& \textbf{+21.2}
{[}15.3, 27.4{]} /
\textbf{+21.5}
{[}16.0, 27.1{]} &
\textbf{+15.0}
{[}9.8, 20.7{]} /
\textbf{+11.7}
{[}6.1, 17.8{]} \\
NIFT-SQL-base with
the same
cross-encoder,
zero-shot &
\textbf{+28.6}
{[}14.5, 41.8{]} /
\textbf{+26.2}
{[}14.8, 39.1{]} &
\textbf{+11.5}
{[}4.1, 18.9{]} /
\textbf{+8.5}
{[}2.0, 14.7{]} &
\textbf{+33.0}
{[}25.6, 40.7{]} /
\textbf{+33.0}
{[}25.1, 40.9{]} \\
NIFT-SQL-base with
Laya & +7.1 {[}−4.2,
16.9{]} / +7.1
{[}0.0, 14.5{]} &
+1.2 {[}−5.1, 7.7{]}
/ +2.9 {[}−2.6,
8.6{]} & +4.0
{[}−1.3, 9.6{]} /
−0.7 {[}−5.9,
4.5{]} \\
\end{longtable}

Table 4 gives the
paired differences.
Pooled over the
three datasets,
NIFT-SQL-base with
the fine-tuned
cross-encoder leads
the zero-shot
reranker by +16.8
{[}10.6, 22.8{]}
complete@15, leads
the same pipeline
with the zero-shot
cross-encoder by
+20.9 {[}15.6,
25.8{]}, and differs
from NIFT-SQL-base
with Laya by +2.6
{[}−1.1, 6.4{]}.

\needspace{12\baselineskip}

\emph{Table 5.
NIFT-SQL-base with
Laya minus each
baseline, same
format. This
pre-specified
comparison isolates
the method's
components.}

\begin{longtable}[]{@{}
  >{\raggedright\arraybackslash}p{(\columnwidth - 6\tabcolsep) * \real{0.2667}}
  >{\raggedright\arraybackslash}p{(\columnwidth - 6\tabcolsep) * \real{0.2167}}
  >{\raggedright\arraybackslash}p{(\columnwidth - 6\tabcolsep) * \real{0.2333}}
  >{\raggedright\arraybackslash}p{(\columnwidth - 6\tabcolsep) * \real{0.2833}}@{}}
\toprule\noalign{}
\begin{minipage}[b]{\linewidth}\raggedright
NIFT-SQL-base (Laya)
minus
\end{minipage} &
\begin{minipage}[b]{\linewidth}\raggedright
BEAVER
\end{minipage} &
\begin{minipage}[b]{\linewidth}\raggedright
LiveSQLBench
\end{minipage} &
\begin{minipage}[b]{\linewidth}\raggedright
LiveSQLBench-Large
\end{minipage} \\
\midrule\noalign{}
\endhead
\bottomrule\noalign{}
\endlastfoot
Search alone &
\textbf{+21.4}
{[}11.0, 31.6{]} /
\textbf{+29.8}
{[}17.1, 42.0{]} &
\textbf{+42.1}
{[}34.5, 49.7{]} /
\textbf{+42.4}
{[}34.2, 50.8{]} &
\textbf{+44.0}
{[}35.6, 51.8{]} /
\textbf{+57.5}
{[}48.2, 66.3{]} \\
Laya alone &
\textbf{+28.0}
{[}16.9, 38.8{]} /
\textbf{+29.2}
{[}18.9, 39.2{]} &
\textbf{+35.6}
{[}28.2, 42.6{]} /
\textbf{+31.8}
{[}25.3, 38.4{]} &
\textbf{+28.2}
{[}20.6, 36.1{]} /
\textbf{+24.5}
{[}19.1, 30.6{]} \\
Rank average & +4.8
{[}−3.4, 12.4{]} /
0.0 {[}−4.5, 5.3{]}
& \textbf{+19.4}
{[}14.3, 24.5{]} /
\textbf{+14.7}
{[}9.9, 19.7{]} &
\textbf{+16.8}
{[}11.3, 22.7{]} /
\textbf{+16.8}
{[}10.7, 23.4{]} \\
Search +
Qwen3-Reranker-8B &
+4.8 {[}−6.4,
17.2{]} / +6.5
{[}−4.8, 19.4{]} &
+5.6 {[}−0.3,
12.2{]} /
\textbf{+6.2}
{[}0.3, 12.2{]} &
\textbf{+17.6}
{[}10.2, 25.0{]} /
\textbf{+28.9}
{[}20.3, 38.4{]} \\
\end{longtable}

Six findings stand
out.

\begin{enumerate}
\def\labelenumi{\arabic{enumi}.}
\tightlist
\item
  \textbf{Training
  on the customer's
  questions is what
  makes the scorer
  work.} Comparing
  the same
  cross-encoder
  zero-shot and
  fine-tuned on our
  rows estimates the
  benefit of
  supervision while
  holding the
  architecture
  fixed. Inside
  NIFT-SQL-base,
  fine-tuning adds
  8.5 to 33.0 points
  of complete@15
  (pooled +20.9
  {[}15.6, 25.8{]});
  alone, it adds 14
  to 36 points
  (borderline on
  BEAVER).
  Zero-shot, this
  cross-encoder lags
  the zero-shot Qwen
  reranker by 15 to
  26 points;
  fine-tuned, inside
  NIFT-SQL-base, it
  outperforms it.
\item
  \textbf{Both
  trained scorer
  architectures
  support strong
  fused retrieval.}
  The fine-tuned
  cross-encoder
  alone beats Laya
  alone by 14 to 18
  points of
  complete@15
  (intervals in
  Appendix F, all
  above zero).
  Inside
  NIFT-SQL-base, the
  two scorers are
  close: no resolved
  difference favors
  Laya, and the
  largest in the
  cross-encoder's
  favor, +7.1 at
  complete@15 on
  BEAVER, has an
  interval that
  touches zero and
  comes from the
  best of three
  cross-encoder
  trainings (1.8 to
  7.1 across the
  three; Section
  6.6); pooled over
  the three
  datasets, the
  difference is +2.6
  {[}−1.1, 6.4{]}
  complete@15.
  Averaged over
  three
  cross-encoder
  trainings, the
  differences are
  +3.6 {[}−5.2,
  11.5{]}, +2.8
  {[}−1.7, 8.0{]},
  and +0.6 {[}−4.6,
  6.3{]} (Section
  6.6). We use the
  cross-encoder as
  the primary
  implementation
  because its
  observed accuracy
  is comparable and
  its recorded cost
  is lower (Section
  8, measured in
  separate runs);
  the experiment
  does not establish
  that it is the
  better
  architecture.
\item
  \textbf{In the
  main setting, the
  trained
  combination beats
  both of its parts
  on all three
  datasets.} With
  Laya, the gain
  over each part is
  21 to 58 points,
  and every interval
  is far from zero.
  With the
  cross-encoder, the
  gain over the
  scorer alone is
  smaller, 7 to 27
  points at
  complete@8 and
  complete@15,
  because the scorer
  alone is stronger.
  Neither search nor
  a scorer is good
  enough on its own;
  the cross-encoder
  intervals are in
  Appendix F.
\item
  \textbf{In
  NIFT-SQL-base,
  learned fusion
  beats plain rank
  fusion on the
  LiveSQLBench
  datasets, mostly
  by grouping tables
  by database.} With
  Laya,
  NIFT-SQL-base
  beats
  reciprocal-rank
  fusion by 15 to 19
  points on both
  headline metrics
  there; with the
  cross-encoder, by
  12 to 22 points on
  complete@15. On
  BEAVER, the
  intervals do not
  separate rank
  fusion and learned
  fusion at
  complete@15 for
  either scorer, but
  with the
  cross-encoder,
  learned fusion
  leads at
  complete@8 (+10.1
  {[}2.9, 17.5{]}).
  Most of this gain
  comes from the
  database grouping
  that fusion learns
  and rank averaging
  lacks. When rank
  averaging is
  regrouped by
  database without
  training (Section
  6.5), learned
  fusion with the
  cross-encoder is
  within 2 points of
  it at complete@15
  on BEAVER and
  LiveSQLBench
  (intervals include
  zero) and leads by
  4.8 {[}0.4, 9.4{]}
  on
  LiveSQLBench-Large;
  at complete@8 it
  leads by 6.5
  {[}1.1, 12.8{]} on
  BEAVER and 8.1
  {[}3.3, 12.9{]} on
  LiveSQLBench-Large
  in the main run,
  but over three
  trainings only the
  LiveSQLBench-Large
  gains remain
  resolved (+7.1
  {[}2.8, 11.6{]} at
  complete@8 and
  +5.6 {[}1.6,
  9.9{]} at
  complete@15).
  These comparisons
  use NIFT-SQL-base,
  with off-the-shelf
  search; the
  corresponding
  two-level
  comparison for
  NIFT-SQL, with
  fine-tuned search
  on both sides, is
  +4.3 {[}−0.4,
  9.4{]} on
  LiveSQLBench-Large
  and is not
  resolved (Section
  6.8).
\item
  \textbf{With its
  variant chosen on
  validation, a
  strong zero-shot
  reranker shows no
  resolved
  difference from
  NIFT-SQL-base.}
  Reordering the
  search top 100, it
  trails
  NIFT-SQL-base with
  the cross-encoder
  at complete@15 by
  13.7 points on
  BEAVER, 9.1 on
  LiveSQLBench, and
  28.2 on
  LiveSQLBench-Large
  in the main run
  (over three
  trainings: 10.1
  {[}−1.3, 21.8{]},
  9.0 {[}2.7,
  15.6{]}, and 29.5
  {[}20.0, 39.7{]}).
  On the two
  LiveSQLBench sets,
  those leads come
  from the shortlist
  and the missing
  database step, not
  from the
  reranker's scores:
  grouped by
  database, it
  reaches 91.8\% on
  LiveSQLBench, 5.7
  {[}0.7, 10.7{]}
  points above
  NIFT-SQL-base over
  three trainings,
  and routed to the
  two databases its
  own ranking puts
  first, 78.7\% on
  LiveSQLBench-Large.
  With one variant
  chosen per
  collection on
  validation
  questions,
  NIFT-SQL-base's
  differences over
  three trainings
  are +10.7 {[}−0.7,
  22.3{]}, −4.2
  {[}−9.2, 0.8{]},
  and +1.0 {[}−5.7,
  8.5{]} at
  complete@15, none
  resolved. BEAVER
  is different:
  NIFT-SQL-base's
  lead there stays
  near 10 points
  whichever variant
  the reranker uses
  (8.9 to 11.3 at
  complete@15,
  including the
  whole known
  database) and is
  positive in every
  training run, but
  it is resolved
  only against the
  search-routed
  variant, which
  routing made
  slightly weaker
  than the plain
  reranker (Sections
  6.5 and 6.6).
\item
  \textbf{Results on
  the LiveSQLBench
  sets are stable
  across splits}
  (standard
  deviation 3 to 5
  points for
  NIFT-SQL-base with
  Laya, 5 to 9 with
  the
  cross-encoder),
  while BEAVER
  varies more (9 to
  13), as expected
  from its
  42-question test
  sets.
\end{enumerate}

\hypertarget{analysis-of-retrieval-gains}{%
\subsection{6
Analysis of
retrieval
gains}\label{analysis-of-retrieval-gains}}

Sections 6 and 7
report both scorers.
All rankings in this
section were rebuilt
from the runs'
cached scores and
reproduce each run's
own test results
exactly.

\hypertarget{scoring-every-table-escapes-the-search-shortlist}{%
\subsubsection{6.1
Scoring every table
escapes the search
shortlist}\label{scoring-every-table-escapes-the-search-shortlist}}

On the large schema,
search cannot find
the tables and a
trained scorer can.
Table 6 counts how
often a shortlist
contains every gold
table, which bounds
anything that only
reorders that
shortlist.

\needspace{12\baselineskip}

\emph{Table 6. Share
of test questions
whose shortlist
contains every gold
table, \%, mean over
seeds 1 to 4. Bold:
the values discussed
in the text.}

\begin{longtable}[]{@{}
  >{\raggedright\arraybackslash}p{(\columnwidth - 6\tabcolsep) * \real{0.2667}}
  >{\raggedright\arraybackslash}p{(\columnwidth - 6\tabcolsep) * \real{0.1500}}
  >{\raggedright\arraybackslash}p{(\columnwidth - 6\tabcolsep) * \real{0.2333}}
  >{\raggedright\arraybackslash}p{(\columnwidth - 6\tabcolsep) * \real{0.3500}}@{}}
\toprule\noalign{}
\begin{minipage}[b]{\linewidth}\raggedright
Shortlist
\end{minipage} &
\begin{minipage}[b]{\linewidth}\raggedright
BEAVER (463 tables)
\end{minipage} &
\begin{minipage}[b]{\linewidth}\raggedright
LiveSQLBench (244)
\end{minipage} &
\begin{minipage}[b]{\linewidth}\raggedright
LiveSQLBench-Large
(971)
\end{minipage} \\
\midrule\noalign{}
\endhead
\bottomrule\noalign{}
\endlastfoot
Search top 50 & 73.8
& 66.4 & 41.1 \\
Search top 100 (the
reranker's input) &
88.1 & 85.7 &
\textbf{51.7} \\
Search top 200 &
89.3 & 99.1 &
64.9 \\
Zero-shot
cross-encoder top
100 & 82.1 & 90.6 &
63.4 \\
Fine-tuned
cross-encoder top
100 & 86.9 & 84.0 &
\textbf{81.0} \\
Laya top 100 & 86.9
& 91.4 &
\textbf{91.7} \\
NIFT-SQL-base
candidates,
cross-encoder
(scorer 50 + search
50) & 95.2 & 93.6 &
87.6 \\
NIFT-SQL-base
candidates, Laya
(scorer 50 + search
50) & 92.3 & 93.9 &
88.3 \\
\end{longtable}

On
LiveSQLBench-Large,
search's top 100
contains every
reference table for
51.7\% of test
questions. This sets
an upper bound on
complete@K for any
reranker restricted
to that shortlist
when K ≤ 100. At
complete@25, the
reranker's reported
mean matches this
candidate-coverage
ceiling at 51.7\%
(Appendix F). When
scoring all 971
tables, Laya's top
100 covers every
reference table for
91.7\% of questions,
while the fine-tuned
cross-encoder covers
them for 81.0\%. The
same zero-shot
cross-encoder
reaches 63.4\%,
showing that
fine-tuning improves
candidate coverage
beyond full-catalog
scoring alone. The
fine-tuned
cross-encoder also
achieves standalone
complete@25 of
77.2\%, exceeding
the 51.7\% ceiling
imposed by search's
top-100 shortlist.
On the two smaller
catalogs, search's
top 100 already
covers 86 to 88\% of
questions, making
candidate coverage a
less restrictive
ceiling for the
reranker.

This is a cost
argument as much as
an accuracy one. The
8B reranker takes
about 25 to 34 ms
per table, so 100
tables cost about 3
s per question. The
fine-tuned
cross-encoder takes
2.0 ms per table, so
the whole 971-table
collection costs 1.9
s (Laya: 3.7 s).
With the
cross-encoder,
NIFT-SQL-base covers
the full schema at a
lower per-question
cost than the
reranker spends on
about a tenth of it;
with Laya, the cost
is similar.
Extrapolating the
reranker's measured
per-table throughput
linearly, it would
need about 24 to 33
s per question to
score all 971
tables; we did not
benchmark that
configuration. With
the database given,
the reranker can
score every table of
that database, and
the shortlist no
longer limits it
(Section 6.5).

\hypertarget{the-advantage-is-in-completing-multi-table-sets}{%
\subsubsection{6.2
The advantage is in
completing
multi-table
sets}\label{the-advantage-is-in-completing-multi-table-sets}}

On LiveSQLBench,
NIFT-SQL-base with
Laya and the
shortlist reranker
have similar R@10,
with no resolved
difference (R@10
86.7\% vs 86.3\%,
difference +0.4
{[}−3.2, 4.1{]}),
yet NIFT-SQL-base
more often finds all
of them (complete@15
+6.2 {[}0.3,
12.2{]}). The
cross-encoder shows
the same pattern:
its R@10 is 86.8\%,
against the
reranker's 86.3\%,
and its complete@15
lead is +9.1 {[}2.2,
16.4{]}. Table 7
breaks complete@15
down by the number
of gold tables,
against the
shortlist reranker
and against the two
strongest baselines:
the reranker with
its variant chosen
on validation
(Section 6.5) and
the retriever
fine-tuned on the
training and
validation questions
(Section 6.7).

\needspace{12\baselineskip}

\emph{Table 7.
Complete@15 by
number of gold
tables, \%, test
questions of seeds 1
to 4 pooled. Trained
methods are means
over their training
runs (three for the
cross-encoder
pipelines;
NIFT-SQL-base with
Laya, one). Question
counts in the
dataset rows. Bold:
highest on five or
more tables.}

\begin{longtable}[]{@{}
  >{\raggedright\arraybackslash}p{(\columnwidth - 8\tabcolsep) * \real{0.4262}}
  >{\raggedright\arraybackslash}p{(\columnwidth - 8\tabcolsep) * \real{0.1475}}
  >{\raggedright\arraybackslash}p{(\columnwidth - 8\tabcolsep) * \real{0.1639}}
  >{\raggedright\arraybackslash}p{(\columnwidth - 8\tabcolsep) * \real{0.1311}}
  >{\raggedright\arraybackslash}p{(\columnwidth - 8\tabcolsep) * \real{0.1311}}@{}}
\toprule\noalign{}
\begin{minipage}[b]{\linewidth}\raggedright
Dataset, method
\end{minipage} &
\begin{minipage}[b]{\linewidth}\raggedright
1 table
\end{minipage} &
\begin{minipage}[b]{\linewidth}\raggedright
2 tables
\end{minipage} &
\begin{minipage}[b]{\linewidth}\raggedright
3 to 4
\end{minipage} &
\begin{minipage}[b]{\linewidth}\raggedright
5 or more
\end{minipage} \\
\midrule\noalign{}
\endhead
\bottomrule\noalign{}
\endlastfoot
\textbf{BEAVER} (4 /
17 / 92 / 55
questions) & & &
& \\
Search alone & 50.0
& 35.3 & 40.2 &
10.9 \\
Reranker, search top
100 & 100.0 & 94.1 &
63.0 & 21.8 \\
Reranker, variant
chosen on validation
& 100.0 & 94.1 &
60.9 & 23.6 \\
Fine-tuned
retriever, train +
validation & 50.0 &
88.2 & 73.6 &
32.7 \\
NIFT-SQL-base (Laya)
& 75.0 & 94.1 & 66.3
& 38.2 \\
NIFT-SQL-base
(cross-encoder) &
100.0 & 90.2 & 71.0
& 40.6 \\
NIFT-SQL & 83.3 &
90.2 & 74.6 &
\textbf{52.7} \\
NIFT-SQL,
high-accuracy mode &
100.0 & 100.0 & 78.3
& 52.1 \\
\textbf{LiveSQLBench}
(93 / 85 / 119 / 43
questions) & & &
& \\
Search alone & 68.8
& 48.2 & 23.5 &
14.0 \\
Reranker, search top
100 & 94.6 & 82.4 &
70.6 & 46.5 \\
Reranker, variant
chosen on validation
& 97.8 & 89.4 & 87.4
& 83.7 \\
Fine-tuned
retriever, train +
validation & 95.0 &
88.2 & 73.7 &
62.0 \\
NIFT-SQL-base (Laya)
& 90.3 & 87.1 & 79.8
& 69.8 \\
NIFT-SQL-base
(cross-encoder) &
93.5 & 91.4 & 79.8 &
76.7 \\
NIFT-SQL & 93.5 &
93.7 & 83.8 &
\textbf{87.6} \\
NIFT-SQL,
high-accuracy mode &
97.5 & 96.5 & 88.0 &
86.0 \\
\textbf{LiveSQLBench-Large}
(76 / 61 / 98 / 38
questions) & & &
& \\
Search alone & 57.9
& 18.0 & 4.1 &
0.0 \\
Reranker, search top
100 & 80.3 & 63.9 &
30.6 & 18.4 \\
Reranker, variant
chosen on validation
& 92.1 & 95.1 & 69.4
& 50.0 \\
Fine-tuned
retriever, train +
validation & 90.4 &
81.4 & 55.4 &
50.0 \\
NIFT-SQL-base (Laya)
& 96.1 & 86.9 & 70.4
& 55.3 \\
NIFT-SQL-base
(cross-encoder) &
93.4 & 86.3 & 69.7 &
67.5 \\
NIFT-SQL & 95.6 &
91.3 & 77.2 &
76.3 \\
NIFT-SQL,
high-accuracy mode &
100.0 & 94.0 & 80.3
& \textbf{77.2} \\
\end{longtable}

Most methods do well
on one- and
two-table questions,
and differences
widen as sets grow.
NIFT-SQL's advantage
over the matched
retriever is largest
on questions
requiring five or
more tables: it
completes the set
for 52.7\%, 87.6\%,
and 76.3\% of them,
against 32.7\%,
62.0\%, and 50.0\%
(+20.0 {[}7.8,
34.8{]}, +25.6
{[}8.3, 43.9{]}, and
+26.3 {[}12.0,
42.9{]}), while on
three- or four-table
questions the
difference is +1.1
{[}−7.7, 11.2{]}
(not resolved),
+10.1 {[}2.6,
19.1{]}, and 21.8
{[}12.9, 32.0{]}.
The fine-tuned
retriever in
NIFT-SQL also helps:
NIFT-SQL leads
NIFT-SQL-base by
12.1 {[}3.6,
24.8{]}, 10.9
{[}2.7, 20.6{]}, and
8.8 {[}0.9, 20.5{]}
points on five or
more tables. Against
the
validation-chosen
reranker, the
five-or-more lead is
+26.3 {[}9.8,
44.8{]} on
LiveSQLBench-Large
and +29.1 {[}0.0,
57.4{]} on BEAVER,
where only 55 such
questions across
seeds put the lower
bound at zero; on
LiveSQLBench the
routed reranker is
within 4.3 points of
NIFT-SQL at every
set size (−4.3 on
one table, +4.3 on
two, −3.6 {[}−10.9,
3.2{]} on three to
four, +3.9 {[}−10.3,
16.3{]} on five or
more), consistent
with Section 6.5:
once the right small
database is found,
the zero-shot
reranker completes
its sets. The
high-accuracy mode
is close to NIFT-SQL
on five or more
tables and gains
mainly on two to
four. BEAVER's
single-table
category holds only
four test questions
across seeds and
should be read
cautiously. One
explanation for the
multi-table pattern
is that a bridge
table containing
only join keys looks
irrelevant on its
own. At the same
time, a per-customer
scorer is trained on
that customer's gold
sets, which include
such tables. Section
6.3 tests this as
far as our data
allow. These
breakdowns reuse the
rankings of Section
6.8
(experiments/final\_\allowbreak{}breakdowns.py
checks that every
test question
reproduces).

\hypertarget{tables-that-do-not-look-relevant}{%
\subsubsection{6.3
Tables that do not
look
relevant}\label{tables-that-do-not-look-relevant}}

We use a simple
proxy for tables
that do not look
relevant: gold
tables whose name
shares no word with
the question (364,
706, and 570 such
gold tables over the
four test sets).
Table 8 counts how
often each method
ranks them in the
top 15. We also
retrained the
cross-encoder
without join hints,
with the same rows
and settings.

\needspace{12\baselineskip}

\emph{Table 8. Gold
tables whose names
share no word with
the question: share
ranked in the top
15, \%, test
questions of seeds 1
to 4 pooled.
``Runs'': training
runs averaged (main
= the main training
run only). Bold:
highest per
collection.}

\begin{longtable}[]{@{}
  >{\raggedright\arraybackslash}p{(\columnwidth - 8\tabcolsep) * \real{0.2542}}
  >{\raggedright\arraybackslash}p{(\columnwidth - 8\tabcolsep) * \real{0.1017}}
  >{\raggedright\arraybackslash}p{(\columnwidth - 8\tabcolsep) * \real{0.1186}}
  >{\raggedright\arraybackslash}p{(\columnwidth - 8\tabcolsep) * \real{0.2203}}
  >{\raggedright\arraybackslash}p{(\columnwidth - 8\tabcolsep) * \real{0.3051}}@{}}
\toprule\noalign{}
\begin{minipage}[b]{\linewidth}\raggedright
Method
\end{minipage} &
\begin{minipage}[b]{\linewidth}\raggedright
Runs
\end{minipage} &
\begin{minipage}[b]{\linewidth}\raggedright
BEAVER
\end{minipage} &
\begin{minipage}[b]{\linewidth}\raggedright
LiveSQLBench
\end{minipage} &
\begin{minipage}[b]{\linewidth}\raggedright
LiveSQLBench-Large
\end{minipage} \\
\midrule\noalign{}
\endhead
\bottomrule\noalign{}
\endlastfoot
Search alone & none
& 41.2 & 55.8 &
25.6 \\
Reranker, search top
100 (zero-shot) &
none & 62.6 & 85.0 &
60.9 \\
Reranker, variant
chosen on validation
(zero-shot) & none &
60.4 & 91.2 &
86.0 \\
Zero-shot
cross-encoder alone
& none & 42.3 & 67.0
& 43.2 \\
Fine-tuned
cross-encoder alone
& main & 54.1 & 79.6
& 81.6 \\
Fine-tuned
cross-encoder alone,
no join hints & main
& 56.3 & 70.5 &
72.6 \\
Fine-tuned
retriever, train +
validation & three &
64.8 & 88.7 &
78.5 \\
NIFT-SQL-base (Laya)
& main & 66.2 & 88.2
& 84.7 \\
NIFT-SQL-base
(cross-encoder) &
three & 69.2 & 89.1
& 85.0 \\
NIFT-SQL & three &
76.1 & 93.1 &
89.6 \\
NIFT-SQL,
high-accuracy mode &
three &
\textbf{78.4} &
\textbf{95.1} &
\textbf{92.0} \\
\end{longtable}

Fine-tuning raises
the cross-encoder's
top-15 retrieval
rate for these
tables on
LiveSQLBench-Large
from 43.2\% to
81.6\%, and
NIFT-SQL-base
retrieves 85.0\% of
them, compared with
60.9\% for the
shortlist reranker.
Against the
strongest baselines,
NIFT-SQL retrieves
more of these tables
than the matched
retriever on every
collection (+11.3
{[}3.3, 19.7{]},
+4.3 {[}1.4, 7.6{]},
and +11.1 {[}7.5,
15.0{]}). Still, its
lead over the
reranker with its
variant chosen on
validation (Section
6.5) is not resolved
(+15.7 {[}−2.4,
28.3{]}, +1.8
{[}−2.0, 5.7{]}, and
+3.6 {[}−1.0,
8.1{]}); the
high-accuracy mode
leads that reranker
on all three (+17.9
{[}2.7, 29.4{]},
+3.9 {[}0.8, 7.4{]},
and +6.1 {[}2.6,
9.8{]}). So
unrelated-looking
tables are found
both by per-customer
training and by a
large zero-shot
reranker that can
read the right
database, and
combining the two
finds the most.
Removing join hints
reduces the
standalone rate from
79.6\% to 70.5\% on
LiveSQLBench and
from 81.6\% to
72.6\% on
LiveSQLBench-Large.
BEAVER declares no
foreign keys, so its
two runs read the
same inputs, and its
change (54.1\% to
56.3\%) cannot come
from join hints. In
the full pipeline,
removing join hints
costs 6.2 points of
complete@15 on
LiveSQLBench
(interval 2.7 to
10.2), 4.8 on
LiveSQLBench-Large
(interval −0.3 to
10.2), and nothing
measurable on BEAVER
(1.2, interval −2.5
to 5.0). A hash of
every training and
scoring input
confirms that
BEAVER's two runs
read identical text
across all four
seeds, so its
differences (up to
2.4 points) come
from retraining
alone; they fall
within the
retraining range
measured in Section
6.6. These results
are consistent with
per-customer
training teaching
the scorer which
unrelated-looking
tables a query
needs, with declared
joins helping it do
so. They do not show
that the scorer
reasons over joins
or selects table
sets jointly: every
pair is still scored
separately, and word
overlap is only a
proxy for a bridge
table.

\hypertarget{fusion-with-a-weak-standalone-scorer}{%
\subsubsection{6.4
Fusion with a weak
standalone
scorer}\label{fusion-with-a-weak-standalone-scorer}}

Laya alone performs
poorly on BEAVER at
complete@8 (13.7\%,
compared with
search's 20.2\%),
while NIFT-SQL-base
with Laya reaches
41.7\%. A scorer
with weak standalone
performance can
still add useful
complementary
ranking information.
Learned fusion
improves over rank
averaging on the
LiveSQLBench
datasets by 15 to 19
points (Table 5),
much of it from
grouping tables by
database (Section
6.5), whereas the
two have the same
mean complete@15 on
BEAVER. These
results do not
guarantee that
fusion protects
against an
unreliable scorer in
other settings. With
the stronger
cross-encoder,
fusion still gains 7
to 27 points over
the scorer alone,
but less.

\hypertarget{how-much-of-the-gain-is-finding-the-right-database}{%
\subsubsection{6.5
How much of the gain
is finding the right
database?}\label{how-much-of-the-gain-is-finding-the-right-database}}

The main setting
searches all
databases in a
collection at once
(Section 2), and two
of the eight fusion
features describe a
candidate's
database. Three
further analyses
separate finding the
right database from
finding the right
tables inside it
(Tables F4 and G1).
All use the cached
scores from the same
runs, except that
the reranker also
scored every table
in each question's
database.

\textbf{Database
given.} Every test
question's reference
tables lie in one
database, so we
restrict each
ranking to that
database. On
LiveSQLBench, this
nearly solves the
task: search alone
reaches 98.8\%
complete@15, and
complete@8 still
separates the
methods (87.1\% for
NIFT-SQL-base with
the cross-encoder
against 77.5\% for
search). Nearly all
of NIFT-SQL-base's
complete@15 gain on
LiveSQLBench
therefore comes from
ranking the right
database's tables
first. On BEAVER and
LiveSQLBench-Large
it does not: with
the database given,
NIFT-SQL-base with
the cross-encoder
still leads search
by 27.4 {[}14.8,
40.0{]} and 46.2
{[}37.5, 54.6{]}
points of
complete@15 in the
main run (23.8 and
46.8 over three
trainings, Table
G5). On
LiveSQLBench-Large,
most of that lead
comes from the
trained scorer: the
fine-tuned
cross-encoder alone
reaches 83.9\% with
the database given,
versus 85.0\% for
the full pipeline.
On BEAVER, the
fusion adds more
(50.0\% for the
cross-encoder alone
against 67.3\%).
With the database
given, the two
scorers' pipelines
are within 5.4
points of each
other, and no
difference is
resolved (the
largest is +5.4
{[}−1.5, 12.4{]} on
BEAVER). Restricting
the reranker's main
list to the database
keeps its search
shortlist, so we
also let it score
every table in the
known database. It
then reaches 54.8\%,
99.4\%, and 85.3\%
complete@15. In the
main training run,
NIFT-SQL-base with
the cross-encoder
leads it by 12.5
{[}1.2, 24.4{]}
points on BEAVER;
averaged over the
three trainings, the
lead is 8.9 {[}−2.6,
20.6{]} and not
resolved. At
complete@15, the
difference is not
resolved on
LiveSQLBench (+0.6
{[}0.0, 1.9{]}; at
complete@8
NIFT-SQL-base leads
by 5.7 {[}0.1,
11.5{]} over three
trainings) or
LiveSQLBench-Large
(+0.2 {[}−6.0,
7.0{]} over three
trainings). The
fine-tuned
cross-encoder alone
also shows no
resolved difference
from it on either
LiveSQLBench set,
while on BEAVER it
trails the reranker
at complete@8 (−7.7
{[}−14.5, −0.7{]}),
so NIFT-SQL-base's
lead comes from
fusing the trained
scorer with search.
On the two
benchmark-built
collections, we find
no statistically
resolved complete@15
difference between a
zero-shot 8B
reranker that reads
every table of the
right database and
the trained 560M
pipeline, which does
not establish
equivalence; the
reranker costs 15 to
17 times as much per
table (22 to 30 ms
against 1.5 to 2.0
ms, Table 10) but
needs no labels or
training. On
BEAVER's real
warehouse schemas,
NIFT-SQL-base leads
every training run
(by 7.1 to 12.5
points), but the
interval over
training runs
includes zero. With
the database given,
the reranker scores
every table of one
database, 11 to 77
tables on average
depending on the
collection (Table
1), or about 0.2 to
2.3 s per question
at its measured
per-table cost,
while the
cross-encoder would
need about 0.02 to
0.14 s for the same
tables (both
extrapolated from
per-table throughput
and average database
size); for
NIFT-SQL-base, the
trade-off is
therefore some 40 to
330 labeled
questions and a
minute or two of
training against
roughly 15 times the
scoring compute per
question, for a gain
that is consistent
on BEAVER but not
statistically
resolved.

\textbf{Routing
without labels.} The
reranker's
main-setting deficit
could reflect our
search shortlist and
the missing database
step rather than the
reranker's scores.
We tried two
zero-label routes,
each letting the
reranker score every
table of the top two
databases and
appending the
remaining databases
in grouped order:
databases ordered by
their best
search-ranked table,
and databases
ordered by the
reranker's own
grouped ranking of
the search top 100.
We fixed two
databases before
running either, as
the smallest number
that still recovers
a question's
database when it is
ranked second, and
did not tune that
number. Routing by
search helps on
LiveSQLBench-Large
(from 50.3\% to
63.1\% complete@15)
but hurts on
LiveSQLBench
(72.9\%), because
search puts the
right database first
for only 55\% and
58\% of questions on
the two LiveSQLBench
sets (Appendix F).
Routing by the
reranker's own
ranking is much
stronger: 53.6\%,
89.1\%, and 78.7\%.
On BEAVER, this
equals the plain
reranker's 53.6\%
for a simple reason:
the plain ranking
already puts the
gold database first
for 97.6\% of
questions, and
routing, which
scores about 216
tables per question,
changes the top 15
for 63\% of
questions but
complete@15 for only
two (one gained, one
lost). The gold
tables outside the
search top 100
(11.9\% of BEAVER
questions) all lie
in the routed
databases, but the
reranker rarely
ranks them in the
top 15. To avoid
choosing the
reranker's variant
claim by claim, we
also chose one
variant per
collection and seed
by validation
complete@15 among
the four (as is,
grouped by database,
routed by search,
and routed by its
own ranking).
Validation picks the
grouped variant on
BEAVER for every
seed and on
LiveSQLBench for
three of four seeds,
and the own-ranking
routing on
LiveSQLBench-Large
for every seed. The
chosen reranker
reaches 37.5 / 53.0,
72.0 / 90.3, and
62.2 / 78.7
complete@8 /
complete@15, and
over three trainings
NIFT-SQL-base's
differences from it
are +6.9 {[}−2.4,
17.0{]} / +10.7
{[}−0.7, 22.3{]},
−1.1 {[}−8.5, 6.2{]}
/ −4.2 {[}−9.2,
0.8{]}, and +5.0
{[}−2.6, 13.5{]} /
+1.0 {[}−5.7,
8.5{]}, none
resolved (note to
Table G5). On the
large collection,
the reranker's
30-point deficit is
therefore a
shortlist effect,
not a scoring one.
Routed, the reranker
scores the search
top 100 plus at most
two databases, about
100 to 250 tables
per question; at its
measured 21 to 32 ms
per table, that is
roughly 2 to 7 s,
against 0.4 to 1.9 s
for the
cross-encoder over
the whole collection
(Table 10;
extrapolated from
per-table throughput
and average database
size).

\textbf{Fusion
without the database
features.} Refitting
the fusion without
the two database
features lowers
complete@15 with the
cross-encoder by 8.8
{[}5.5, 12.9{]}
points on
LiveSQLBench, 3.3
{[}0.4, 6.8{]} on
LiveSQLBench-Large,
and 1.8 {[}−0.8,
5.5{]} on BEAVER
(with Laya, by 10.6,
9.2, and 0.0).
Without them,
NIFT-SQL-base with
the cross-encoder
still leads search
on every dataset
(65.5\%, 77.4\%, and
75.3\% against
30.4\%, 41.0\%, and
21.8\%) and leads
the reranker on
BEAVER and
LiveSQLBench-Large
(against 53.6\% and
50.3\%), while on
LiveSQLBench the two
means are nearly
identical (77.4\%
against 77.1\%; we
did not test
equivalence).
Removing the
database features
thus removes nearly
all of
NIFT-SQL-base's lead
over the reranker on
LiveSQLBench.

\textbf{Database
grouping for the
baselines.} The
fusion's database
features let it rank
whole databases, and
the baselines do not
have this step. We
therefore regroup
every ranking by
database without
training: databases
are ordered by their
best-ranked table,
and tables keep
their order within
each database (a
second rule, using
each database's top
three tables, agrees
within about one
point). Grouping
lifts search on
LiveSQLBench from
41.0\% to 61.1\%
complete@15 and the
reranker from 77.1\%
to 91.8\%, changes
BEAVER and
LiveSQLBench-Large
by at most 7 points
for any method, and
changes
NIFT-SQL-base
itself, with either
scorer, by at most
3.3 points. Against
grouped baselines,
NIFT-SQL-base with
the cross-encoder
still leads search
by 33.9, 25.0, and
53.8 points and the
reranker by 14.3 and
33.7 points on
BEAVER and
LiveSQLBench-Large
in the main run
(over three
trainings, 10.7
{[}−0.7, 22.3{]},
not resolved, and
35.0 {[}26.7,
43.9{]}), but the
grouped reranker
scores 5.6 points
above it on
LiveSQLBench, a loss
that is resolved
over three trainings
(−5.7 {[}−10.7,
−0.7{]}). Grouped
rank averaging of
the cross-encoder
and search comes
within 2 points of
NIFT-SQL-base at
complete@15 on
BEAVER and
LiveSQLBench, so on
these collections
most of what learned
fusion adds over
rank averaging is
the database
grouping; on
LiveSQLBench-Large
learned fusion still
leads by 4.8 {[}0.4,
9.4{]}. In the main
setting, combining
the trained scorer
with search still
improves complete@15
over the grouped
scorer alone by 23.8
{[}16.0, 32.8{]},
8.2 {[}3.7, 12.9{]},
and 8.4 {[}3.9,
13.4{]} points; with
the database given,
its added value is
large on BEAVER but
small on the two
LiveSQLBench sets
(+0.6 and +1.1).
When labeled
questions are
scarce, grouped rank
averaging of the
fine-tuned scorer
and search needs no
validation labels
and matches learned
fusion at
complete@15 on
BEAVER and
LiveSQLBench;
fitting the fusion
(on all validation
questions; Section 7
shows that 20 lose
at most 1.6 points)
pays off mainly on
the large
collection: over
three trainings,
+7.1 {[}2.8, 11.6{]}
at complete@8 and
+5.6 {[}1.6, 9.9{]}
at complete@15 on
LiveSQLBench-Large,
while the main run's
+6.5 at complete@8
on BEAVER is not
resolved over runs.

\hypertarget{how-much-does-retraining-move-the-results}{%
\subsubsection{6.6
How much does
retraining move the
results?}\label{how-much-does-retraining-move-the-results}}

We retrained the
cross-encoder twice
more on every split,
with the same rows
and settings but
different dropout
and data-order
seeds, and refitted
the fusion each time
(Table G2). Across
the three training
runs, the four-seed
mean complete@15 of
NIFT-SQL-base with
the cross-encoder
ranges from 61.9\%
to 67.3\% on BEAVER,
85.7\% to 86.6\% on
LiveSQLBench, and
78.6\% to 81.9\% on
LiveSQLBench-Large.
The main BEAVER run
is the best of the
three; averaged over
the three runs,
NIFT-SQL-base with
the cross-encoder
reaches 63.7\%,
86.2\%, and 80.0\%.
The comparison with
Laya remains
uncertain: across
the three trainings,
the cross-encoder
pipeline's mean lead
over the Laya
pipeline at
complete@15 ranges
from 1.8 to 7.1
points on BEAVER,
2.4 to 3.4 on
LiveSQLBench, and
−0.6 to 2.7 on
LiveSQLBench-Large,
and we did not
retrain Laya. Two
runs with PyTorch's
deterministic
algorithms still
differed, so we
could not make
training bit-for-bit
repeatable on our
GPU setup or
attribute the
variation to a
specific source;
hashing confirms the
inputs are identical
(Section 6.3).

Intervals that
include this
variation come from
a two-level
bootstrap that
resamples training
runs as well as
gold-table-set
clusters (Table G5).
With it, the gains
over hybrid search
(33.3 {[}16.8,
48.7{]}, 45.2
{[}36.4, 53.6{]},
and 58.1 {[}49.5,
67.2{]} points of
complete@15), over
grouped search, and
over the reranker
reordering the
search shortlist on
the two LiveSQLBench
sets (9.0 {[}2.7,
15.6{]} and 29.5
{[}20.0, 39.7{]})
remain resolved, as
does learned
fusion's lead over
grouped rank
averaging on
LiveSQLBench-Large
(5.6 {[}1.6,
9.9{]}). The leads
over the reranker on
BEAVER (10.1
{[}−1.3, 21.8{]} in
the main setting and
8.9 {[}−2.6, 20.6{]}
with the database
given) and over Laya
(+3.6, +2.8, and
+0.6, all intervals
including zero) are
not, and against the
reranker with its
routing chosen on
validation, no
difference is
resolved on any
collection (Section
6.5). The grouped
reranker's lead on
LiveSQLBench is
resolved over three
trainings (−5.7
{[}−10.7, −0.7{]}
for NIFT-SQL-base).

\hypertarget{is-a-supervised-scorer-better-than-a-supervised-retriever}{%
\subsubsection{6.7
Is a supervised
scorer better than a
supervised
retriever?}\label{is-a-supervised-scorer-better-than-a-supervised-retriever}}

We fine-tuned the
BGE retriever per
customer on the same
training rows
(contrastive loss,
two epochs, as in
Appendix A) and used
it in hybrid search
on all four splits,
three times with
different training
seeds (Table G6 and
the note to Table
G5). Supervision
also helps the
retriever: averaged
over its three
trainings,
complete@15 rises
from 30.4\%, 41.0\%,
and 21.8\% to
58.9\%, 73.1\%, and
58.1\% (standard
deviation over
trainings 0.6 to 1.8
points).
NIFT-SQL-base with
the cross-encoder,
which fuses its
scorer with the base
search, differs from
it by +4.8 {[}−3.1,
13.3{]} (not
resolved), +13.0
{[}7.4, 19.5{]}, and
+21.7 {[}15.0,
28.8{]} points of
complete@15 and by
14.9 {[}3.3,
26.9{]}, 18.7
{[}12.4, 25.6{]},
and 28.6 {[}21.9,
35.8{]} at
complete@8, with
intervals that
resample both
methods' training
runs. This retriever
sees fewer labels
than NIFT-SQL-base,
whose fusion also
uses the validation
questions. Trained
on the training and
validation questions
together, every
labeled question
NIFT-SQL-base uses,
it reaches 61.1\%,
81.7\%, and 70.4\%
complete@15, and
NIFT-SQL-base's lead
shrinks to 2.6
{[}−4.7, 10.7{]},
4.4 {[}−0.4, 9.7{]},
and 9.5 {[}3.9,
15.7{]}; at
complete@8 it stays
resolved on all
three collections
(10.9 {[}0.9,
21.0{]}, 10.6
{[}4.8, 16.9{]}, and
17.2 {[}11.3,
23.9{]}). With the
label budget
matched,
NIFT-SQL-base's
advantage over a
supervised retriever
is therefore mainly
at the top of the
ranking: it places
the full reference
set in the top 8
more often, but at
complete@15 it is
clearly ahead only
on the 971-table
collection, and with
the database given,
the two show no
resolved complete@15
difference (+1.2,
0.0, and +1.5
points, intervals
including zero). The
scorer alone is not
what wins: against
the first retriever
training, it trails
on BEAVER (−14.1
{[}−24.2, −4.9{]} at
complete@15), shows
no resolved
difference on
LiveSQLBench (−4.9
{[}−12.3, 2.6{]}),
and leads only on
LiveSQLBench-Large
(+13.2 {[}5.7,
21.3{]}). The gain
comes from combining
a scorer that reads
every table with
search. Replacing
off-the-shelf search
with the fine-tuned
retriever inside the
cross-encoder
pipeline combines
the two; Appendix A
found that it adds
about 5 points with
Laya on two seeds,
and Section 6.8
shows that with the
cross-encoder it
does better than
either alone, which
is why NIFT-SQL uses
it. The fine-tuned
retriever is also
much cheaper at
query time: like
base search, it
embeds each card
once (1.0 to 5.6 s
per schema after
training) and then
costs a few
milliseconds per
question (4 to 13 ms
for base search,
Table 10; we did not
time the fine-tuned
retriever
separately, but its
inference is
identical), against
0.37 to 1.94 s for
the cross-encoder to
score every table.
Between the two
components on their
own, a fine-tuned
retriever is
therefore the
cheaper choice for
similar complete@15
coverage when the
collection is not
large, and the
scorer pipeline is
the better choice
when the context
budget is tight
(complete@8), or the
collection is large;
combined, as
NIFT-SQL, they lead
the matched
retriever at both
cutoffs on every
collection (Section
6.8).

\hypertarget{the-fine-tuned-retriever-and-the-high-accuracy-mode}{%
\subsubsection{6.8
The fine-tuned
retriever and the
high-accuracy
mode}\label{the-fine-tuned-retriever-and-the-high-accuracy-mode}}

This section gives
the experiment
behind NIFT-SQL's
final form. Sections
6.5 and 6.7 found
two strong
alternatives to
NIFT-SQL-base: a
retriever fine-tuned
on the same
questions, and a
zero-shot reranker
routed by database.
We tested whether
they complement the
trained scorer
rather than replace
it, with three
variants whose
settings were fixed
before their test
scoring
(experiments/fusion\_\allowbreak{}extensions.py).
(A) NIFT-SQL-base
with the fine-tuned
retriever in place
of base search, with
the same eight
features; that
retriever is trained
on the training
questions only, so
its validation
scores, which fit
the fusion, come
from a model that
never saw them. (B)
NIFT-SQL-base plus
two reranker
features for every
candidate: the
reranker's logit and
the log rank of its
score among the
candidates. (C)
Both. A is NIFT-SQL
as defined in
Section 3, and C is
its high-accuracy
mode. We recomputed
the reranker's
scores for each
question's
candidates because
earlier caches
covered the full
gold database for
some questions, so
whether a table had
a score could reveal
it. Every
cross-encoder
training run is
paired with a
fine-tuned-retriever
run, and the
intervals resample
both runs and
gold-table-set
clusters (Table G7).
A validation-only
check, run before
the test part, split
each customer's
validation questions
into two halves,
fitted each
variant's fusion on
one half and scored
it on the other, in
both directions; it
ranked the variants
at complete@15 as
the test part later
did: C, A, B, then
NIFT-SQL-base.

Both additions help.
A uses the same
labeled questions as
NIFT-SQL-base and
adds only a few
milliseconds of
search per question.
It raises
complete@15 from
63.7\%, 86.2\%, and
80.0\% to 69.2\%,
89.5\%, and 85.6\%
(+5.6 {[}0.7,
11.1{]}, +3.3
{[}0.6, 6.4{]}, and
+5.6 {[}2.7,
8.8{]}), and it now
leads the retriever
fine-tuned on the
same training and
validation questions
at both cutoffs on
every collection
(complete@15 +8.1
{[}1.1, 16.3{]},
+7.7 {[}3.5,
12.7{]}, and +15.1
{[}10.0, 21.0{]}),
where NIFT-SQL-base
alone led at
complete@15 only on
LiveSQLBench-Large
(Section 6.7). C
reaches 54.6\%,
81.6\%, and 78.0\%
complete@8 and
72.4\%, 92.5\%, and
88.6\% complete@15,
8.7 {[}2.9, 15.3{]},
6.4 {[}2.8, 10.1{]},
and 8.7 {[}4.9,
12.8{]} points of
complete@15 above
NIFT-SQL-base.

The reranker adds
information the
supervised pipeline
lacks. C leads A at
complete@8 on all
three collections
(+8.1 {[}2.5,
15.3{]}, +6.8
{[}2.8, 11.4{]}, and
+4.2 {[}0.3, 8.0{]})
and at complete@15
on the two
LiveSQLBench sets
(+3.0 {[}0.5, 5.9{]}
and +3.1 {[}0.5,
5.9{]}), and the
fusion gives the
reranker substantial
weight: on BEAVER
its standardized
logit weight (0.85
in B) is close to
the trained scorer's
(1.03), and on the
LiveSQLBench
customer the
reranker's rank
carries the largest
weight of any
feature (−1.11; a
lower rank is
better). Against the
zero-shot reranker
with its variant
chosen on
validation, C leads
at complete@8 on all
three collections
(+17.1 {[}8.3,
27.3{]}, +9.5
{[}4.2, 15.0{]}, and
+15.6 {[}8.9,
22.6{]}) and at
complete@15 on
BEAVER and
LiveSQLBench-Large
(+19.4 {[}8.4,
30.8{]} and +9.6
{[}3.8, 16.0{]}); on
LiveSQLBench at
complete@15 the
difference is not
resolved (+2.2
{[}−1.4, 6.1{]}),
where both are close
to the ceiling that
routing to the right
database sets. With
the database given,
C also leads the
reranker over the
whole database at
complete@8 on all
three collections
and at complete@15
on BEAVER and
LiveSQLBench-Large
(+17.7 {[}6.9,
28.8{]} and +6.5
{[}1.0, 12.6{]}).

\textbf{Controls.}
Two further controls
separate what the
learned fusion and
the reranker
features contribute;
they fit nothing and
score nothing new
(experiments/fusion\_\allowbreak{}controls.py,
note to Table G7).
First, reciprocal
rank fusion of the
same two trained
parts, the
fine-tuned
cross-encoder and
the fine-tuned
search, reaches
66.3\%, 76.2\%, and
78.3\% complete@15.
NIFT-SQL's learned
fusion leads it on
the two LiveSQLBench
sets (+13.2 {[}9.1,
17.6{]} and +7.3
{[}3.2, 11.6{]}) but
not on BEAVER (+3.0
{[}−1.3, 7.3{]}).
Grouped by database,
the same rank fusion
reaches 67.5\%,
89.3\%, and 81.2\%,
and no difference
from the learned
fusion is resolved
at either cutoff
(complete@15 +1.8
{[}−2.3, 5.9{]},
+0.1 {[}−3.1,
3.0{]}, and +4.3
{[}−0.4, 9.4{]}).
Splitting NIFT-SQL's
gain over plain rank
fusion into its two
steps gives a
different picture on
each collection.
Grouping by database
adds 1.2, 13.1, and
2.9 points of
complete@15 (from
66.3\%, 76.2\%, and
78.3\% to 67.5\%,
89.3\%, and 81.2\%),
and the learned
weights add a
further 1.8 {[}−2.3,
5.9{]}, 0.1 {[}−3.1,
3.0{]}, and 4.3
{[}−0.4, 9.4{]}.
Grouping is the
dominant step on
LiveSQLBench, whose
databases are small;
on
LiveSQLBench-Large,
the largest schema,
the learned weights
add more than
grouping under this
decomposition, which
applies grouping
first and is not a
unique attribution;
both are small on
BEAVER. None of the
learned-weight gains
is statistically
resolved; with 42 to
86 test questions
per seed, this may
reflect low power,
and it does not show
equivalence. Rank
fusion can combine
any number of
rankings but weights
them equally; the
learned fusion
weights them
unequally, which is
how the
high-accuracy mode
weights the
reranker's two
features against the
scorer and search
signals. Grouped
rank averaging needs
no fusion labels;
whether those labels
would do more as
extra training data
for the two trained
parts is untested.
Second, the reranker
alone ordering
NIFT-SQL's own
candidates reaches
54.4\%, 84.0\%, and
77.7\% complete@15
(grouped by
database: 54.4\%,
93.4\%, and 73.8\%),
close to its best
variant chosen on
validation, so a
better candidate
pool does not by
itself explain the
high-accuracy mode.
That mode leads it
at both cutoffs on
every collection
(complete@15 +18.1
{[}8.5, 28.0{]},
+8.4 {[}4.4,
13.3{]}, and +10.7
{[}5.7, 16.7{]});
against the grouped
version it leads on
BEAVER and
LiveSQLBench-Large
and shows no
resolved difference
on LiveSQLBench at
complete@15 (−1.0
{[}−4.9, 2.8{]}).
The candidate pool
alone therefore does
not explain the
high-accuracy mode's
results.

The two variants
trade cost
differently. A costs
what NIFT-SQL-base
costs (0.4 to 1.9 s
per question for the
cross-encoder, Table
10). B and C also
run the reranker on
the fusion's
candidates, about 80
to 90 per question,
roughly 1.7 to 2.8 s
at 21 to 32 ms per
table (Table 10),
which is comparable
to the routed
reranker on its own
(Section 8). The
per-pair time
recorded in this run
for BEAVER (100 ms)
included model
loading and a
restart after a
memory error, so we
use Table 10. A
combines
NIFT-SQL-base with
the fine-tuned
retriever from
Appendix A, which
Appendix A tested
only with Laya, on
two seeds, and with
join completion;
here we test it with
the cross-encoder,
on four seeds and
three training runs,
without join
completion.

\hypertarget{frozen-method-holdout-on-a-newer-beaver-release}{%
\subsubsection{6.9
Frozen-method
holdout on a newer
BEAVER
release}\label{frozen-method-holdout-on-a-newer-beaver-release}}

After the analyses
above, we froze
NIFT-SQL and scored
it once on the
current public
BEAVER release
(Chen, Yang, et al.,
2024), with three
databases and 381
tables searched
together, 7,857
synthetic questions,
and 121 real
questions from query
logs. The
configuration is
that of Section 3
with only the data
changed; declared
foreign keys from
the release's
database dumps are
passed as join hints
(two databases
declare some, the
data warehouse
none). We split the
synthetic questions
50/30/20 as in
Section 4 (seed 1:
3,869 / 2,359 /
1,629 questions) and
kept the real
questions for
testing only. Its
synthetic questions
are new to this
study: none matches
any of the older
release's 209
questions exactly,
and one of 7,857
nearly does (word
overlap of at least
0.8). Its real
questions are not:
112 of the 121
appear verbatim
among those 209
questions on which
NIFT-SQL was
developed, 120
nearly so (word
overlap of at least
0.8), and all 121 at
0.6. All 121 real
questions come from
the data warehouse;
the synthetic
questions of the
other two databases
have no overlap with
the older release's
questions
(results/paper/OVERLAP\_\allowbreak{}CHECK.txt).
Because most real
questions overlap
those used during
method development,
this evaluation is
not an independent
confirmation on new
real questions. Both
NIFT-SQL's design
and the fine-tuned
retriever's recipe
(Appendix A) were
developed on these
questions, so their
reuse does not
obviously favor
either method. We
also explored this
release the day
before in separate
work (search
baselines and a
check of join
signals on its
synthetic and real
questions), so we
wrote the hypotheses
knowing how search
performs on it; we
did not change
NIFT-SQL's
configuration.
Before scoring, we
fixed two hypotheses
at complete@15 on
the synthetic test
in a dated commit,
part of the code
release: NIFT-SQL
beats off-the-shelf
hybrid search (H1),
and it beats the
retriever fine-tuned
on the same training
and validation
questions (H2;
experiments/confirm\_\allowbreak{}beaver\_\allowbreak{}v3.py).

\needspace{12\baselineskip}

\emph{Table 9.
Frozen-method
holdout on the
current BEAVER
release, seed 1,
scored once,
complete@8 /
complete@15 in \%.}

\begin{longtable}[]{@{}
  >{\raggedright\arraybackslash}p{(\columnwidth - 4\tabcolsep) * \real{0.4167}}
  >{\raggedright\arraybackslash}p{(\columnwidth - 4\tabcolsep) * \real{0.3000}}
  >{\raggedright\arraybackslash}p{(\columnwidth - 4\tabcolsep) * \real{0.2833}}@{}}
\toprule\noalign{}
\begin{minipage}[b]{\linewidth}\raggedright
Method
\end{minipage} &
\begin{minipage}[b]{\linewidth}\raggedright
Synthetic test
(1,629 questions)
\end{minipage} &
\begin{minipage}[b]{\linewidth}\raggedright
Real questions (121)
\end{minipage} \\
\midrule\noalign{}
\endhead
\bottomrule\noalign{}
\endlastfoot
Search alone
(off-the-shelf BGE)
& 25.1 / 55.4 & 28.9
/ 42.1 \\
Fine-tuned
retriever, training
+ validation
questions & 70.0 /
95.5 & 31.4 /
62.0 \\
Fine-tuned
cross-encoder alone
& 47.2 / 85.3 & 11.6
/ 29.8 \\
NIFT-SQL-base
(off-the-shelf
search) & 73.5 /
94.0 & 25.6 /
44.6 \\
\textbf{NIFT-SQL} &
\textbf{79.4 / 96.4}
& \textbf{34.7 /
54.5} \\
\end{longtable}

Table 9 gives the
results. H1 holds:
NIFT-SQL leads
search by 41.0
{[}35.3, 47.6{]}
points of
complete@15. H2 is
not supported: the
estimated
complete@15
improvement over the
label-matched
retriever is 1.0
percentage point,
with a 95\% interval
spanning zero
{[}−0.5, 2.5{]}.
Both methods are
near the ceiling: no
synthetic test
question requires a
table absent from
the training gold
sets, although new
table combinations
are held out. At
complete@8, NIFT-SQL
leads the retriever
by 9.4 {[}6.3,
12.9{]} points; this
is a secondary
result, not
confirmation of H2.
The synthetic test
evaluates new
questions from the
training
distribution. Most
real questions
overlap those used
during method
development, but
neither the scorer
nor the retrievers
in this run were
trained on them: all
learned only from
synthetic questions.
This evaluation
therefore examines
synthetic-to-real
transfer and does
not constitute an
independent test on
new real questions.
NIFT-SQL leads
search by 12.4
{[}2.5, 22.6{]}
points at
complete@15, but
trails the
label-matched
retriever by 7.4
points (−7.4
{[}−14.0, −0.9{]}),
with no resolved
difference at
complete@8 (+3.3
{[}−2.7, 9.6{]}).
The cross-encoder
alone reaches 29.8\%
complete@15 on real
questions, below
search, while the
fine-tuned retriever
reaches 62.0\%;
within NIFT-SQL, the
fine-tuned retriever
adds 9.9 {[}3.2,
17.5{]} points over
NIFT-SQL-base. These
results support
gains over search
and a complete@8
advantage over the
label-matched
retriever on the
synthetic test. The
synthetic
complete@15
advantage was not
resolved, and the
real-question
evaluation favors
the retriever at
that cutoff. The
older-release
experiments show
gains when training
on real questions
(Section 6.8),
whereas this
synthetic-trained
run shows weaker
transfer for the
scorer than for the
retriever.

\hypertarget{labeling-requirements-and-learning-curves}{%
\subsection{7
Labeling
requirements and
learning
curves}\label{labeling-requirements-and-learning-curves}}

We measure how
retrieval changes as
the number of
questions used to
train the scorer
increases. We
retrained both
scorers on 10\%,
20\%, and 30\% of
all questions
(dropping whole
gold-set groups from
the training part)
and kept validation
and test exactly as
in the main runs.
The
training-question
counts below exclude
the labeled
validation questions
used to fit the
fusion, and the
total supervision
includes both: the
fusion is still
fitted on the full
validation part (64
questions for BEAVER
and 125 unique
question texts for
the LiveSQLBench
customer, whose two
sets share many
texts). Counted as
unique texts and
averaged over seeds,
the 10\% setting
uses 86 labeled
questions in total
(components and
totals averaged over
seeds and rounded
separately) on
BEAVER (21 for
training) and 194 on
the LiveSQLBench
customer (69 for
training); the main
50\% setting uses
167 and 326.
Cross-fitting, which
would let the fusion
reuse the training
questions instead of
a separate
validation part, is
left for future
work.

\begin{figure}
\centering
\includegraphics{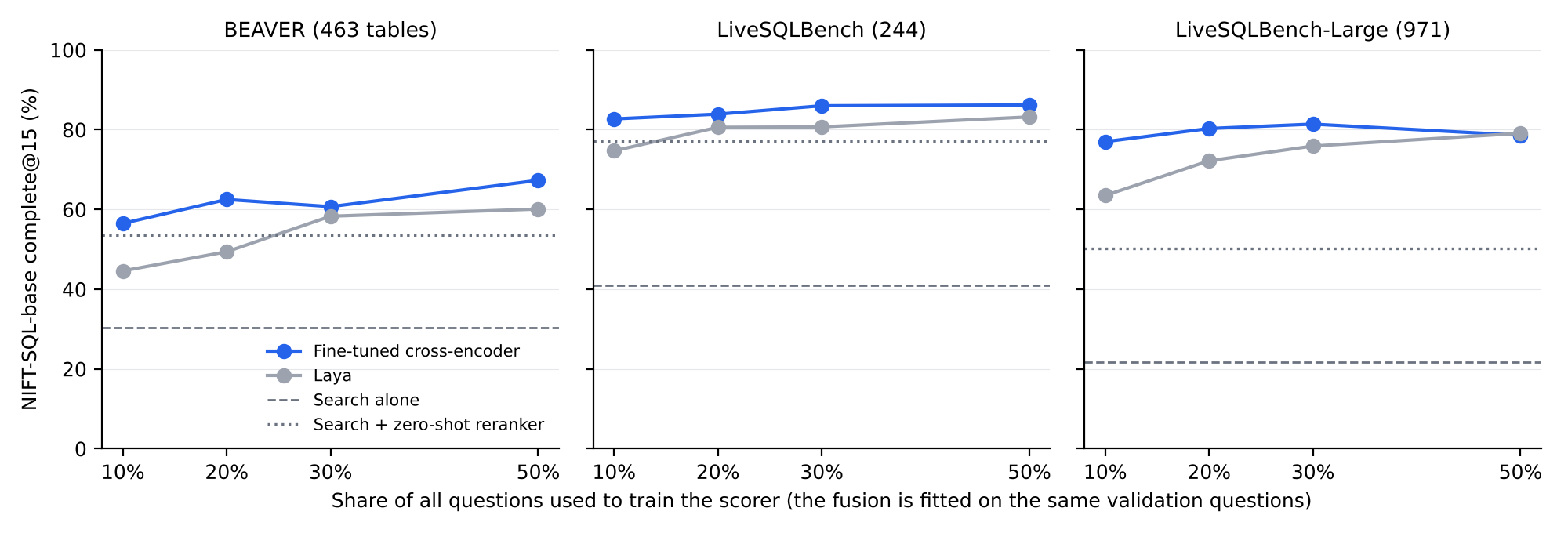}
\caption{NIFT-SQL-base
complete@15 by
training size, test,
seeds 1 to 4 (values
in Table F5)}
\end{figure}

With Laya and the
smallest training
set (10\%, the same
21 and 69 training
questions per
customer),
NIFT-SQL-base
already lifts
complete@15 by 14 to
42 points over
search, and it beats
the zero-shot 8B
reranker on
LiveSQLBench-Large.
At 20\%, its mean is
also above the
reranker's on
LiveSQLBench.
Observed
improvements
diminish after 30\%:
at 30\%,
NIFT-SQL-base is
within 2 to 3 points
of the main result
at complete@15, and
the intervals
include zero,
although this does
not establish
equivalence or show
that more training
data has no benefit
(BEAVER −1.8
{[}−7.5, 3.0{]},
LiveSQLBench −2.6
{[}−5.8, 0.6{]},
LiveSQLBench-Large
−3.3 {[}−7.1,
0.7{]}). At 10\%,
every dataset is
below the main
result by a resolved
margin of 8.5 to
15.8 points.

On
LiveSQLBench-Large,
increasing the
scorer-training
fraction from 10\%
to 50\% raises
Laya's standalone
complete@15 from
21.5\% to 54.7\%,
while NIFT-SQL-base
increases from
63.5\% to 79.1\%.
The fusion therefore
gives substantial
gains even with the
smaller
scorer-training set,
given the fixed
validation labels
used to fit it.

At the small
training sizes we
tested, the
cross-encoder
pipeline scores
higher than the Laya
pipeline (Figure 2):
at 10\%,
NIFT-SQL-base
reaches 56.5, 82.7,
and 77.0 complete@15
with the
cross-encoder
against 44.6, 74.7,
and 63.5 with Laya.
At that size,
NIFT-SQL-base with
the cross-encoder
already improves
complete@15 over
search by 26.2
{[}10.5, 41.4{]},
41.8 {[}34.2,
49.7{]}, and 55.3
{[}46.8, 63.4{]}
points, and leads
the zero-shot
reranker by 26.7
{[}18.0, 35.8{]}
points on
LiveSQLBench-Large;
on the two smaller
schemas its lead
over the reranker at
10\% is not
resolved. On the
LiveSQLBench sets,
it is within 4
points of the 50\%
result already at
10\%, with intervals
that include zero;
on BEAVER it stays
below the 50\%
result even at 30\%
(−6.5 {[}−13.1,
−1.2{]}), although
the 50\% BEAVER run
is the best of three
trainings (Section
6.6), so this gap is
probably overstated.
Figure 2 shows the
mean over seeds;
Table F5 lists the
exact values for
each size. At 10\%,
the cross-encoder
pipeline leads the
Laya pipeline by
11.9 {[}1.9,
23.2{]}, 7.9 {[}2.9,
13.0{]}, and 13.6
{[}8.1, 19.2{]}
points of
complete@15, against
a single Laya
training. A
difference between
two independently
trained models
varies more than
either model's own
retraining range
(Section 6.6);
Laya's variation was
not measured, and
variation with 21 to
69 training
questions may exceed
what we measured at
50\%, so the
LiveSQLBench-Large
difference is the
largest relative to
the training
variation measured
at the main training
size. Still, Laya's
training variation
and the variation at
smaller training
sizes remain
unmeasured. The
fusion also needs a
few validation
questions. Fitted on
20 randomly drawn
validation questions
per customer instead
of all of them,
NIFT-SQL-base with
the cross-encoder
loses at most 1.6
points of
complete@15 in the
main runs. With the
scorer trained at
10\% and the fusion
fitted on 20
validation
questions, about 41
labeled questions
for BEAVER and 89
for the LiveSQLBench
customer, it still
improves on search
by 25.0 {[}10.3,
39.2{]}, 39.6
{[}32.2, 47.3{]},
and 53.3 {[}45.1,
61.2{]} points
(Table G3).

\hypertarget{cost-speed-and-scoring-budgets}{%
\subsection{8 Cost,
speed, and scoring
budgets}\label{cost-speed-and-scoring-budgets}}

Costs below estimate
GPU compute at an
assumed \$1.50 per
GPU hour and exclude
annotation, data
preparation,
engineering, model
loading, and
deployment overhead.
Times are means over
the four seeds on
one GPU. Training
the cross-encoder
takes about 0.7
minutes for BEAVER
and 1.9 minutes for
the shared
LiveSQLBench
customer, and
batched scoring
averages 1.8, 1.5,
and 2.0 ms per table
on the three
datasets. Table 10
also reports Laya's
timings; we timed
the two scorers in
separate runs, so
the comparison is
indicative.

\needspace{12\baselineskip}

\emph{Table 10. Cost
and speed.}

\begin{longtable}[]{@{}
  >{\raggedright\arraybackslash}p{(\columnwidth - 6\tabcolsep) * \real{0.3000}}
  >{\raggedright\arraybackslash}p{(\columnwidth - 6\tabcolsep) * \real{0.2000}}
  >{\raggedright\arraybackslash}p{(\columnwidth - 6\tabcolsep) * \real{0.2000}}
  >{\raggedright\arraybackslash}p{(\columnwidth - 6\tabcolsep) * \real{0.3000}}@{}}
\toprule\noalign{}
\begin{minipage}[b]{\linewidth}\raggedright
Stage
\end{minipage} &
\begin{minipage}[b]{\linewidth}\raggedright
BEAVER
\end{minipage} &
\begin{minipage}[b]{\linewidth}\raggedright
LiveSQLBench
\end{minipage} &
\begin{minipage}[b]{\linewidth}\raggedright
LiveSQLBench-Large
\end{minipage} \\
\midrule\noalign{}
\endhead
\bottomrule\noalign{}
\endlastfoot
Embed all cards with
BGE-large (once per
schema) & 2.8 s &
1.0 s & 5.6 s \\
Search, per question
& 12.7 ms & 4.1 ms &
4.6 ms \\
Train cross-encoder
(once per customer)
& 0.7 min, \$0.02 &
1.9 min, \$0.05 (one
model for both
LiveSQLBench sets) &
same model \\
Cross-encoder, per
table & 1.8 ms & 1.5
ms & 2.0 ms \\
Cross-encoder, all
tables, per question
& 0.86 s & 0.37 s &
1.94 s \\
Cross-encoder, GPU
cost per 1,000
questions & \$0.36 &
\$0.15 & \$0.81 \\
Train Laya (once per
customer) & 1.7 min,
\$0.04 & 3.5 min,
\$0.09 & same
model \\
Laya, per table &
2.8 ms & 2.9 ms &
3.8 ms \\
Laya, all tables,
per question & 1.29
s & 0.71 s & 3.74
s \\
Laya, GPU cost per
1,000 questions &
\$0.54 & \$0.30 &
\$1.56 \\
Qwen3-Reranker-8B,
per table (top 100
only) & about 25 to
34 ms & about 25 to
34 ms & about 25 to
34 ms \\
Qwen3-Reranker-8B,
per table (whole
known database,
Section 6.5) & 30.0
ms & 22.2 ms & 30.0
ms \\
Qwen3-Reranker-8B,
per table (routed to
two databases,
Section 6.5) & 21.3
ms & 25.4 ms & 31.6
ms \\
NIFT-SQL: fine-tune
BGE (once per
customer) & about 2
to 4 GPU-minutes per
split, both
customers together &
(included in
BEAVER's figure) &
same model \\
NIFT-SQL, per
question (search
plus cross-encoder
over all tables) &
about 0.87 s & about
0.37 s & about 1.94
s \\
High-accuracy mode,
per question (adds
the reranker on
about 80 to 90
candidates;
estimated) & about
2.6 to 3.6 s & about
2.2 to 2.7 s & about
4.3 to 4.8 s \\
\end{longtable}

The scorer
dominates, and its
cost grows linearly
with schema size.
The last two rows
are computed from
the rows above
(high-accuracy mode:
cross-encoder over
all tables plus the
reranker on the
candidates at 21 to
32 ms per table); we
did not time them
end-to-end. NIFT-SQL
also fine-tunes BGE
once per customer
(about 2 to 4
GPU-minutes per
split for both
customers together)
and re-embeds the
cards once; at query
time, its search
costs match
off-the-shelf
search, so its
per-question cost
equals
NIFT-SQL-base's. The
two reranker rows
come from separate
runs, so their
per-table times
differ. The reranker
routed by database
(Section 6.5) scores
the search top 100
plus at most two
databases, so at the
per-table times
measured in that run
(21.3, 25.4, and
31.6 ms) it needs an
estimated 2.1 to
5.4, 2.5 to 3.1, and
3.2 to 6.6 s per
question, against
0.86, 0.37, and 1.94
s for the
cross-encoder over
the whole
collection. Appendix
B evaluates
restricted scoring
budgets from cached
full-schema
predictions: scoring
only search's top
50\% of tables meets
the pre-set
validation rule. It
halves the number of
question-table pairs
scored. However, we
still need to
measure the
end-to-end gain,
which will depend on
batching and fixed
overhead. These
findings use Laya
and the
fine-tuned-search
variant in Appendix
A; the same
simulation applies
to the
cross-encoder, whose
table score also
does not depend on
other tables. In a
deployment where the
user's database is
known, the scorer
would score only
that database
(LiveSQLBench-Large
averages 54 tables
per database instead
of 971). Section 6.5
evaluates retrieval
with the database
given, using
full-schema scores;
it did not measure
the cost of scoring
only that database.

These are throughput
figures from batched
scoring.
Single-question
latency on an idle
GPU is higher per
table because
batches are smaller.

\emph{All runs used
one NVIDIA
A100-SXM4-40GB GPU,
as recorded in each
run's manifest.}

\hypertarget{related-work}{%
\subsection{9
Related
work}\label{related-work}}

\textbf{Enterprise
text-to-SQL
benchmarks.} BEAVER
(Chen, Yang, et al.,
2024) collects
questions over real
enterprise data
warehouses. In its
current release,
state-of-the-art
agentic frameworks
reach 10.8\%
accuracy with
GPT-5.2, and oracle
schema-linking hints
roughly double
execution accuracy.
LiveSQLBench (BIRD
Team, 2025) provides
contamination-resistant
databases, including
a large-schema
variant. We use the
table-retrieval
versions of both
from Zeng et
al.~(2026), whose
benchmark also
recasts Spider (Yu
et al., 2018) and
BIRD (Li et al.,
2023). We leave
those two out
because their
databases are small
and their table and
column names are
descriptive, which
is not the
enterprise setting
we study.

\textbf{Schema
linking and table
retrieval.}
Text-to-SQL
pipelines usually
place schema linking
before generation,
either with the LLM
itself, as in
DIN-SQL (Pourreza \&
Rafiei, 2023),
DAIL-SQL (Gao et
al., 2024), CHESS
(Talaei et al.,
2024), and ReFoRCE
(Deng et al., 2025),
or with a separate
retriever over
tables and columns,
such as CRUSH4SQL
(Kothyari et al.,
2023) or dense table
retrieval (Herzig et
al., 2021). The
BEAVER baselines
retrieve with
Qwen3-Embedding-8B
followed by
Qwen3-Reranker-8B
(Zhang et al.,
2025). These
approaches are
zero-shot with
respect to the
customer: a general
model reads a schema
it has never seen.
We start from the
observation that
many enterprise
deployments have
labeled questions,
and test whether a
small model trained
on them per customer
retrieves better and
more cheaply.

\textbf{Closest
work.} Zeng et
al.~(2026) built the
schema-retrieval
benchmark we use.
They fine-tuned a
305M embedding model
per corpus on
synthetic queries
with hard negatives,
raising average
recall@10 across
five corpora from
60.4 to 75.6. We
differ in three
ways: we use real
question-to-table
labels from the
customer, a
supervised scorer
that reads every
table, and a fusion
fitted on held-out
questions. The
closest analog to
their adapted
embedder in our
setup is the
per-customer
fine-tuned BGE of
Appendix A, which
NIFT-SQL now
includes as its
retriever (Sections
3.1 and 6.8);
Section 6.7
evaluates it as a
stand-alone baseline
on all four splits.
REaR (Agarwal et
al., 2026) and
CORE-T (Soliman et
al., 2026) also
target multi-table
retrieval; CORE-T is
training-free and
evaluates BEAVER
with database
identifiers removed,
as in our main
setting. PEARL (Ho
et al., 2026) is
also training-free:
offline, it
generates multi-hop
queries over
pre-identified join
paths so that a
standard dense
retriever can find
multi-table
evidence; it
evaluates on BIRD,
Spider, and MMQA.
JoinGR (De et al.,
2026) is the closest
supervised method:
it trains a
lightweight scorer
on gold tables to
traverse join graphs
and reports recall
gains over dense
retrieval and
reranking on BEAVER.
CORE-T evaluates the
same 463-table,
209-question BEAVER
collection, but it
writes an LLM
description of each
table and reads five
sample rows, which
our names-and-keys
setting excludes,
and it reports
set-selection
metrics rather than
complete@K. Its code
is public, but as
published it needs
the descriptions and
sample rows our
setting excludes;
without them, it
would no longer be
CORE-T. JoinGR
evaluates a smaller
BEAVER
data-warehouse
subset (97 tables,
121 questions) with
197 join edges; its
paper does not say
how those edges were
obtained. The BEAVER
release we use
declares no foreign
keys, so they would
have to be inferred
(for instance, with
the name-matching
join completion of
Appendix A). No
JoinGR code is
released, so a
comparison would
require a
reimplementation.
Chen, Zhang, and
Roth (2024) score
query-table and
table-table join
relevance jointly
and select a
coherent table set
with a mixed-integer
program; NIFT-SQL
scores tables
independently and
does not optimize a
join plan or a table
set, which may
contribute to its
misses on questions
that need five or
more tables.
Differences in
inputs, supervision,
candidate selection,
and metrics prevent
a direct numerical
comparison under
their reported
protocols, and we
leave a controlled
comparison under
matched inputs to
future work.

\textbf{Retrieval
and reranking.} We
build on dense
retrieval with BGE
(Xiao et al., 2024),
BM25 (Robertson \&
Zaragoza, 2009),
cross-encoder
reranking (Nogueira
\& Cho, 2019; Zhang
et al., 2025), and
reciprocal rank
fusion (Cormack et
al., 2009).
Fine-tuning a
cross-encoder on
in-domain relevance
labels is standard
in passage ranking;
our contribution is
not a new scorer
architecture but
evidence on when
per-customer
supervision pays,
with the recipe
around it: schema
cards with joins,
scoring the whole
schema instead of a
shortlist, and a
fusion with search
fitted on held-out
questions. Our
fusion is a learned
version of rank
fusion with a
handful of rank and
score features, and
Section 6.5 shows
that much of what it
adds over plain rank
fusion is grouping
tables by database.

\textbf{Decision
models.} Laya
(Convai Innovations,
n.d.) and similar
``decision'' models
answer typed
questions (yes/no,
choice, score) with
temperature-scaled
probabilities in one
forward pass,
without generating
text. We use one as
the pre-specified
per-customer scorer.
Appendix C compares
it with a larger
zero-shot decision
model and with the
same larger model
fine-tuned on our
rows.

\hypertarget{limitations}{%
\subsection{10
Limitations}\label{limitations}}

\begin{itemize}
\tightlist
\item
  \textbf{Retrieval
  only.} We measure
  coverage of the
  reference query's
  tables, not
  downstream SQL
  execution
  accuracy.
  Reproducing the
  reference query
  when SQL
  generation is
  restricted to the
  retrieved tables
  requires
  retrieving every
  table the
  reference query
  uses. However,
  retrieval coverage
  does not guarantee
  correct or fully
  executable SQL:
  further generation
  errors can arise
  in mapping
  question intent to
  SQL logic, and
  alternative
  correct queries
  may use different
  table sets. As
  such, while strong
  retrieval
  performance is an
  important step, it
  is not sufficient
  on its own for
  correct end-to-end
  text-to-SQL
  deployment.
\item
  \textbf{BEAVER
  releases.} The
  main analyses use
  the older
  table-retrieval
  release (209
  questions over 463
  tables). Section
  6.9 uses the
  current public
  release (7,857
  synthetic and 121
  real questions,
  one split seed);
  we do not use the
  benchmark's
  private test set,
  so our numbers are
  not comparable to
  the BEAVER
  leaderboard.
\item
  \textbf{Synthetic-to-real
  transfer.} On the
  current release,
  the trained
  cross-encoder
  alone falls below
  search on real
  logged questions,
  while the
  fine-tuned
  retriever
  transfers well
  (Section 6.9). In
  deployment,
  training questions
  are often curated
  or synthetic and
  differ from what
  users ask. Trained
  on real questions,
  as in the main
  analyses, the
  scorer works well
  on them. We do not
  isolate the cause:
  a shift in
  phrasing,
  different
  relevance
  patterns, or other
  dataset
  differences. Only
  BEAVER comes from
  real enterprise
  data warehouses;
  the LiveSQLBench
  databases were
  built for a
  benchmark, with
  enterprise-style
  schemas.
\item
  \textbf{Small and
  overlapping test
  sets.} BEAVER's
  test sets have 42
  questions per
  seed, so its
  results vary by 9
  to 13 points
  between seeds. The
  four random splits
  overlap, which we
  account for by
  clustering
  intervals by
  gold-table set
  across seeds; the
  seeds are not four
  independent
  samples.
\item
  \textbf{Two
  LiveSQLBench sets
  are not two
  confirmations.}
  LiveSQLBench and
  LiveSQLBench-Large
  share databases
  and many
  questions, so our
  protocol treats
  them as one
  customer, and
  their results are
  correlated. In
  total, the study
  has two customer
  environments, so
  it does not yet
  show how reliably
  the approach
  adapts across many
  separate
  enterprise
  customers.
\item
  \textbf{The method
  evolved with the
  results.} We chose
  the cross-encoder
  as the headline
  scorer after the
  pre-specified Laya
  runs, and added
  the fine-tuned
  retriever and the
  high-accuracy
  mode's reranker
  features after
  analyzing Sections
  5 to 7. We fixed
  each addition's
  settings before
  its own test
  scoring and scored
  it once per split;
  the pre-specified
  Laya results
  independently
  support gains over
  search, but the
  intervals for the
  final
  configuration do
  not account for
  these choices.
  Most component
  analyses (scorer
  comparison,
  learning curves,
  database grouping,
  join hints, the
  breakdowns of
  Tables 7 and 8)
  were run with
  NIFT-SQL-base and
  not repeated with
  the fine-tuned
  retriever.
\item
  \textbf{Indirect
  mechanism
  evidence.} The
  bridge-table
  evidence rests on
  a word-overlap
  proxy and a
  join-hint
  ablation; it does
  not show that any
  scorer reasons
  over joins. The
  training-size
  curves keep the
  validation labels
  fixed.
\item
  \textbf{The
  reranker's
  routing.} In the
  main setting, it
  reorders the BGE
  hybrid top 100.
  Section 6.5 also
  groups it by
  database and
  routes it to two
  databases, by
  search or by its
  own ranking, and
  chooses one
  variant per
  collection on
  validation
  questions; with
  that choice, no
  difference from
  NIFT-SQL-base is
  resolved on any
  collection, and
  NIFT-SQL leads it
  only at
  complete@15 on
  BEAVER and
  complete@8 on
  LiveSQLBench-Large.
  Other zero-label
  routes, a stronger
  first-stage
  retriever such as
  the
  Qwen3-Embedding-8B
  model the BEAVER
  authors use, or
  few-shot prompts
  could make it
  stronger still; we
  did not test them.
  A matched
  supervised
  retriever comes
  close to
  NIFT-SQL-base:
  trained on the
  same labeled
  questions, it is
  within 2.6 and 4.4
  points of
  NIFT-SQL-base at
  complete@15 on
  BEAVER and
  LiveSQLBench, with
  intervals that
  include zero
  (Section 6.7).
  NIFT-SQL, which
  includes it, leads
  it at both cutoffs
  on every
  collection
  (Section 6.8); we
  did not tune its
  training.
\item
  \textbf{Fixed
  settings.} We
  fixed the search
  weights, both
  scorers' training
  settings, and the
  fusion's
  regularization
  before the test
  runs and did not
  tune them. This is
  fair to the
  baselines but may
  understate what
  tuning could add.
\item
  \textbf{Retraining
  variation.} The
  intervals hold
  each trained model
  fixed. We measured
  retraining
  variation for the
  cross-encoder
  (Section 6.6) but
  not for Laya.
  Hence, the
  comparison between
  the two scorers is
  less certain than
  its intervals
  suggest, and the
  BEAVER headline
  run is the best of
  three
  cross-encoder
  trainings. The
  two-level
  intervals in Table
  G5 include the
  cross-encoder's
  retraining
  variation but not
  Laya's.
\item
  \textbf{Capped
  inputs.} Cards are
  cut at 700
  characters and six
  joins. This
  shortens 3.8\% of
  BEAVER's
  gold-table cards
  (by 14 columns on
  average when cut)
  and drops some
  joins for 23.5\%
  of
  LiveSQLBench-Large's
  gold tables and
  8.6\% of
  LiveSQLBench's,
  which may
  understate both
  scorers (Table
  G4).
\item
  \textbf{Throughput,
  not latency.}
  Speeds are
  measured with
  batched scoring.
  Interactive
  single-question
  latency is higher
  per table.
\item
  \textbf{Untested
  alternatives.} We
  ran the reranker
  zero-shot with its
  standard
  instruction;
  adding few-shot
  customer examples
  to the prompt
  might help, at the
  cost of longer
  inputs for every
  table. We rank a
  fixed number of
  tables rather than
  thresholding the
  scorer's
  probability, so we
  do not optimize
  the number of
  distractor tables
  passed to a
  generator.
\item
  \textbf{Database
  not given.} The
  main setting
  searches all
  databases in a
  dataset at once.
  Section 6.5
  evaluates a known
  database using the
  same scores, but
  not a pipeline
  trained and fitted
  for that setting,
  which would also
  be cheaper to
  score.
\end{itemize}

\hypertarget{conclusion}{%
\subsection{11
Conclusion}\label{conclusion}}

In our experiments
on enterprise table
retrieval from names
and keys alone,
models trained
separately on each
customer's example
questions
substantially
improve table
retrieval. NIFT-SQL,
which fuses a
fine-tuned 560M
cross-encoder with
hybrid search over a
retriever fine-tuned
on the same
questions, retrieves
every reference
table within the top
15 for 69.2\%,
89.5\%, and 85.6\%
of test questions,
averaged over three
trainings, against
22 to 41\% for
hybrid search and 61
to 82\% for the
fine-tuned retriever
alone with the same
labels, which
NIFT-SQL leads at
both cutoffs on
every collection.
Scored once on a
newer BEAVER
release, NIFT-SQL
again beats search
by a wide margin
(+41.0 complete@15)
and leads the
matched retriever at
complete@8 on
synthetic questions
(+9.4), but not at
complete@15, and on
121 real logged
questions the
matched retriever is
ahead at complete@15
(−7.4); a fine-tuned
retriever is
therefore a strong
and cheaper
baseline, and the
scorer's value is
clearest at a tight
context budget. Both
trained parts
matter: in
NIFT-SQL-base, the
zero-shot
cross-encoder is 21
points lower in
pooled complete@15,
and without the
fine-tuned retriever
(NIFT-SQL-base),
complete@15 is 3 to
6 points lower. For
a training-free rank
fusion of the two
trained parts,
grouping by database
is the dominant step
on LiveSQLBench, and
learned weights are
the larger step on
LiveSQLBench-Large
(+4.3 points, not
resolved); learned
fusion is how the
reranker is weighted
as a third signal.
Grouped rank fusion
of the two trained
parts is a
competitive
alternative that
needs no fusion
labels; we did not
test it at small
label budgets. A
zero-shot 8B
reranker needs no
labels and trails
NIFT-SQL-base by
about 30 points on
the 971-table
collection when it
reorders a search
shortlist; routed or
grouped by database,
with the variant
chosen on
validation, it comes
close, and NIFT-SQL
leads it by a
resolved margin only
at complete@15 on
BEAVER and
complete@8 on
LiveSQLBench-Large.
The reranker
complements the
trained pipeline
rather than
replacing it: adding
its scores as two
fusion features
raises complete@15
to 72.4\%, 92.5\%,
and 88.6\%, ahead of
the
validation-selected
reranker variant on
BEAVER and
LiveSQLBench-Large
with no resolved
difference from it
on LiveSQLBench, at
an estimated 1.7 to
2.8 s more of
batched scoring per
question; ordering
NIFT-SQL's own
candidates by the
reranker alone does
not reproduce that
gain. NIFT-SQL uses
167 and 326 labeled
questions per
customer and a few
minutes of training;
NIFT-SQL-base,
tested at smaller
budgets, already
improves on search
by 26 to 55 points
with 21 to 69
training questions.
On LiveSQLBench,
whose databases are
small, nearly all of
the gain comes from
finding the right
database; on BEAVER
and
LiveSQLBench-Large,
NIFT-SQL-base leads
search by 24 and 47
points when the
database is given,
and the
high-accuracy mode
leads the reranker
over the whole known
database at
complete@8 on all
three collections
(Sections 6.5 and
6.8). Retraining the
cross-encoder moves
the four-seed means
by up to 5.4 points
of complete@15 (7.7
at complete@8), so
we report means over
three trainings.
These results hold
for held-out
questions and table
combinations within
known schemas in two
customer
environments; they
do not establish
generalization to
unseen customers or
gains in downstream
SQL execution.

Several directions
follow. First,
retrieval failures
concentrate in
questions requiring
three or more
tables. On BEAVER,
questions requiring
five or more tables
account for about
half of NIFT-SQL's
failures; on the two
LiveSQLBench
datasets, questions
requiring three or
four tables account
for the largest
share. For questions
requiring five or
more tables,
NIFT-SQL reaches
complete@15 of 53 to
88\% across the
three datasets,
leaving substantial
room to improve
retrieval of
complete table sets.
Second, Appendix C
suggests that a
larger fine-tuned
scorer could improve
retrieval over Laya
in the exploratory
validation setting.
Its recorded
per-table scoring
time is 10 to 17
times Laya's and 20
to 30 times that of
the headline
cross-encoder;
whether it improves
on the
cross-encoder, and
whether any gain
justifies the
additional compute,
requires evaluation
under matched
conditions. Third,
schema cards can
carry richer
metadata where
customers have it,
such as descriptions
or column types; how
much that adds
beyond names and
keys is open.
Finally, evaluating
more customer
environments,
stronger first-stage
retrieval under
matched compute, and
downstream SQL
execution would
establish the
approach's broader
deployment value.

\hypertarget{data-and-code-availability}{%
\subsection{Data and
code
availability}\label{data-and-code-availability}}

The three benchmark
collections are
public (Zeng et al.,
2026). All code,
configurations, and
per-question results
needed to reproduce
every table will be
released upon
journal publication.
File paths cited in
the text (configs/,
experiments/,
results/) refer to
that release. We do
not distribute model
weights trained on
the benchmarks; the
code retrains them:
about 0.7 to 2
minutes per customer
for the
cross-encoder and
two to four
GPU-minutes per
split for the
fine-tuned BGE.

\hypertarget{references}{%
\subsection{References}\label{references}}

Agarwal, R.,
Singhal, H., Chen,
P. B., Roy
Choudhury, M., Roth,
D., \& Gupta, V.
(2026). REaR:
Retrieve, expand and
refine for effective
multitable
retrieval. In
Proceedings of the
64th Annual Meeting
of the Association
for Computational
Linguistics (Volume
1: Long Papers)
(pp.~39360--39374).
\url{https://doi.org/10.18653/v1/2026.acl-long.1826}

BAAI. (n.d.).
bge-reranker-large
{[}model card{]}.
Retrieved October
29, 2026, from
\url{https://huggingface.co/BAAI/bge-reranker-large}

BIRD Team. (2025).
LiveSQLBench: A
dynamic and
contamination-free
benchmark for
text-to-SQL.
Retrieved October
29, 2026, from
\url{https://livesqlbench.ai/}

Burges, C. J. C.
(2010). From RankNet
to LambdaRank to
LambdaMART: An
overview (Technical
Report
MSR-TR-2010-82).
Microsoft Research.

Chen, P. B., Yang,
D., Li, W., Wenz,
F., Zhang, Y.,
Tatbul, N.,
Cafarella, M.,
Demiralp, Ç., \&
Stonebraker, M.
(2024). BEAVER: An
enterprise benchmark
for text-to-SQL.
arXiv preprint
arXiv:2409.02038
(version 3, May
2026).
\url{https://doi.org/10.48550/arXiv.2409.02038}

Chen, P. B., Zhang,
Y., \& Roth, D.
(2024). Is table
retrieval a solved
problem? Exploring
join-aware
multi-table
retrieval. In
Proceedings of the
62nd Annual Meeting
of the Association
for Computational
Linguistics (Volume
1: Long Papers)
(pp.~2687--2699).
\url{https://doi.org/10.18653/v1/2024.acl-long.148}

Convai Innovations.
(n.d.). Laya
{[}model card{]}.
Retrieved October
29, 2026, from
\url{https://huggingface.co/convaiinnovations/laya}

Cormack, G. V.,
Clarke, C. L. A., \&
Büttcher, S. (2009).
Reciprocal rank
fusion outperforms
Condorcet and
individual rank
learning methods. In
Proceedings of the
32nd International
ACM SIGIR Conference
on Research and
Development in
Information
Retrieval
(pp.~758--759).
\url{https://doi.org/10.1145/1571941.1572114}

De, S., Chakraborty,
A., Bandyopadhyay,
S., \& Gupta, V.
(2026). JoinGR:
Learning to traverse
join graphs for
table retrieval.
arXiv preprint
arXiv:2610.01064.
\url{https://doi.org/10.48550/arXiv.2610.01064}

Deng, M.,
Ramachandran, A.,
Xu, C., Hu, L., Yao,
Z., Datta, A., \&
Zhang, H. (2025).
ReFoRCE: A
text-to-SQL agent
with
self-refinement,
consensus
enforcement, and
column exploration.
arXiv preprint
arXiv:2502.00675
(version 5; version
1 was titled
``\ldots{} format
restriction, and
column
exploration'').
\url{https://doi.org/10.48550/arXiv.2502.00675}

Gao, D., Wang, H.,
Li, Y., Sun, X.,
Qian, Y., Ding, B.,
\& Zhou, J. (2024).
Text-to-SQL
empowered by large
language models: A
benchmark
evaluation.
Proceedings of the
VLDB Endowment,
17(5), 1132--1145.
\url{https://doi.org/10.14778/3641204.3641221}

Guo, C., Pleiss, G.,
Sun, Y., \&
Weinberger, K. Q.
(2017). On
calibration of
modern neural
networks. In
Proceedings of the
34th International
Conference on
Machine Learning
(pp.~1321--1330).
PMLR.

Herzig, J., Müller,
T., Krichene, S., \&
Eisenschlos, J.
(2021). Open domain
question answering
over tables via
dense retrieval. In
Proceedings of the
2021 Conference of
the North American
Chapter of the
Association for
Computational
Linguistics: Human
Language
Technologies
(pp.~512--519).
\url{https://doi.org/10.18653/v1/2021.naacl-main.43}

Ho, S., Kang, H.,
Kang, S., \& Yoon,
S. (2026). PEARL:
Front-loading
relational chains
for multi-hop table
retrieval. arXiv
preprint
arXiv:2608.30291.
\url{https://doi.org/10.48550/arXiv.2608.30291}

Hu, E. J., Shen, Y.,
Wallis, P.,
Allen-Zhu, Z., Li,
Y., Wang, S., Wang,
L., \& Chen, W.
(2022). LoRA:
Low-rank adaptation
of large language
models. In
International
Conference on
Learning
Representations.
\url{https://doi.org/10.48550/arXiv.2106.09685}

Ke, G., Meng, Q.,
Finley, T., Wang,
T., Chen, W., Ma,
W., Ye, Q., \& Liu,
T.-Y. (2017).
LightGBM: A highly
efficient gradient
boosting decision
tree. In Advances in
Neural Information
Processing Systems
30 (pp.~3146--3154).

Kothyari, M.,
Dhingra, D.,
Sarawagi, S., \&
Chakrabarti, S.
(2023). CRUSH4SQL:
Collective retrieval
using schema
hallucination for
Text2SQL. In
Proceedings of the
2023 Conference on
Empirical Methods in
Natural Language
Processing
(pp.~14054--14066).
\url{https://doi.org/10.18653/v1/2023.emnlp-main.868}

Li, J., Hui, B., Qu,
G., Yang, J., Li,
B., Li, B., Wang,
B., Qin, B., Cao,
R., Geng, R., Huo,
N., Zhou, X., Ma,
C., Li, G., Chang,
K. C. C., Huang, F.,
Cheng, R., \& Li, Y.
(2023). Can LLM
already serve as a
database interface?
A BIg bench for
large-scale database
grounded
text-to-SQLs. In
Advances in Neural
Information
Processing Systems
36 (Datasets and
Benchmarks Track).
\url{https://doi.org/10.48550/arXiv.2305.03111}

Nogueira, R., \&
Cho, K. (2019).
Passage re-ranking
with BERT. arXiv
preprint
arXiv:1901.04085.
\url{https://doi.org/10.48550/arXiv.1901.04085}

Perplexity AI.
(n.d.).
pplx-decider-flash-v1.2
{[}model card{]}.
Retrieved October
29, 2026, from
\url{https://huggingface.co/perplexity-ai/pplx-decider-flash-v1.2}

Pourreza, M., \&
Rafiei, D. (2023).
DIN-SQL: Decomposed
in-context learning
of text-to-SQL with
self-correction. In
Advances in Neural
Information
Processing Systems
36.
\url{https://doi.org/10.48550/arXiv.2304.11015}

Robertson, S., \&
Zaragoza, H. (2009).
The probabilistic
relevance framework:
BM25 and beyond.
Foundations and
Trends in
Information
Retrieval, 3(4),
333--389.
\url{https://doi.org/10.1561/1500000019}

Soliman, H., Gupta,
V., Roth, D., \&
Gurevych, I. (2026).
CORE-T: Coherent
retrieval of tables
for text-to-SQL.
arXiv preprint
arXiv:2601.13111
(version 3, August
2026).
\url{https://doi.org/10.48550/arXiv.2601.13111}

Talaei, S.,
Pourreza, M., Chang,
Y.-C., Mirhoseini,
A., \& Saberi, A.
(2024). CHESS:
Contextual
harnessing for
efficient SQL
synthesis. arXiv
preprint
arXiv:2405.16755.
\url{https://doi.org/10.48550/arXiv.2405.16755}

Xiao, S., Liu, Z.,
Zhang, P.,
Muennighoff, N.,
Lian, D., \& Nie,
J.-Y. (2024).
C-Pack: Packed
resources for
general Chinese
embeddings. In
Proceedings of the
47th International
ACM SIGIR Conference
on Research and
Development in
Information
Retrieval
(pp.~641--649).
\url{https://doi.org/10.1145/3626772.3657878}

Yu, T., Zhang, R.,
Yang, K., Yasunaga,
M., Wang, D., Li,
Z., Ma, J., Li, I.,
Yao, Q., Roman, S.,
Zhang, Z., \& Radev,
D. (2018). Spider: A
large-scale
human-labeled
dataset for complex
and cross-domain
semantic parsing and
text-to-SQL task. In
Proceedings of the
2018 Conference on
Empirical Methods in
Natural Language
Processing
(pp.~3911--3921).
\url{https://doi.org/10.18653/v1/D18-1425}

Zeng, Q., Yu, P.,
Mehta, A., Zhao, F.,
\& Samdani, R.
(2026). Finding the
right tables and
columns: A benchmark
and corpus-adaptive
embeddings for SQL
schema retrieval.
arXiv preprint
arXiv:2607.13311.
\url{https://doi.org/10.48550/arXiv.2607.13311}

Zhang, Y., Li, M.,
Long, D., Zhang, X.,
Lin, H., Yang, B.,
Xie, P., Yang, A.,
Liu, D., Lin, J.,
Huang, F., \& Zhou,
J. (2025). Qwen3
Embedding: Advancing
text embedding and
reranking through
foundation models.
arXiv preprint
arXiv:2506.05176.
\url{https://doi.org/10.48550/arXiv.2506.05176}

\hypertarget{appendix}{%
\subsection{Appendix}\label{appendix}}

Appendices A, F, and
G are test results;
Appendix H
illustrates the
fusion's training
data on one
validation question.
Appendices B, C, and
D are exploratory
development studies
on validation
questions and should
be read as such:
seed 0 was used for
many comparisons,
and they were run
with Laya as the
scorer on the
variant of Appendix
A (fine-tuned BGE
and join
completion), not on
the main
configuration.

\hypertarget{a.-fine-tuned-search-and-join-completion}{%
\subsubsection{A.
Fine-tuned search
and join
completion}\label{a.-fine-tuned-search-and-join-completion}}

Two additions to the
main method with
Laya. (1) BGE
fine-tuned per
customer with a
contrastive loss on
the same training
rows the scorer uses
(question, gold
table, hard
negative), 2 epochs.
(2) Names-only join
completion: two
tables in one
database are linked
when they share a
key-like column name
(ending in key, id,
code, no, num, nbr,
ref, or registry,
not a generic name,
carried by at most 8
tables), plus three
join features in the
fusion. The
combination passed
our pre-set
development rule on
seed-0 validation
(mean complete@8
+2.7, no dataset
worse).

\needspace{12\baselineskip}

\emph{Table A1.
Test, seeds 1 and 2,
complete@8 /
complete@15 in \%,
and the paired
change against
NIFT-SQL-base with
Laya {[}95\%
interval{]}.}

\begin{longtable}[]{@{}
  >{\raggedright\arraybackslash}p{(\columnwidth - 6\tabcolsep) * \real{0.2667}}
  >{\raggedright\arraybackslash}p{(\columnwidth - 6\tabcolsep) * \real{0.1333}}
  >{\raggedright\arraybackslash}p{(\columnwidth - 6\tabcolsep) * \real{0.2500}}
  >{\raggedright\arraybackslash}p{(\columnwidth - 6\tabcolsep) * \real{0.3500}}@{}}
\toprule\noalign{}
\begin{minipage}[b]{\linewidth}\raggedright
\end{minipage} &
\begin{minipage}[b]{\linewidth}\raggedright
BEAVER
\end{minipage} &
\begin{minipage}[b]{\linewidth}\raggedright
LiveSQLBench
\end{minipage} &
\begin{minipage}[b]{\linewidth}\raggedright
LiveSQLBench-Large
\end{minipage} \\
\midrule\noalign{}
\endhead
\bottomrule\noalign{}
\endlastfoot
NIFT-SQL-base with
Laya & 46.4 / 66.7 &
70.8 / 83.9 & 64.0 /
79.9 \\
+ fine-tuned BGE +
join completion &
54.8 / 72.6 & 73.2 /
88.7 & 73.4 /
84.9 \\
Change, complete@8 &
+8.3 {[}−6.3,
21.2{]} & +2.4
{[}−4.8, 9.6{]} &
\textbf{+9.4}
{[}2.0, 17.4{]} \\
Change, complete@15
& +6.0 {[}−6.2,
17.8{]} &
\textbf{+4.8}
{[}0.6, 9.4{]} &
+5.0 {[}0.0,
11.0{]} \\
\end{longtable}

The gain is about 5
points in
complete@15,
resolved on
LiveSQLBench, with
an interval touching
zero on
LiveSQLBench-Large,
from two seeds only.

\hypertarget{b.-scoring-budget-laya-on-part-of-the-schema}{%
\subsubsection{B.
Scoring budget: Laya
on part of the
schema}\label{b.-scoring-budget-laya-on-part-of-the-schema}}

Laya scores each
question-table pair
independently, so we
simulate restricted
scoring budgets by
keeping cached
scores only for
tables within each
search shortlist.
This reproduces the
scorer's predictions
for those pairs but
does not measure
runtime savings. We
refit and evaluate
the fusion with two
folds inside
validation. The
selection rule,
fixed before the
seed-0 comparison,
chooses the smallest
budget whose mean is
no more than one
point below
full-schema scoring
at both complete@8
and complete@15,
with no dataset
significantly worse.

\needspace{12\baselineskip}

\emph{Table B1.
Validation, mean
change against
scoring all tables,
complete@8 /
complete@15 in
points. Reduction in
scored pairs for
BEAVER /
LiveSQLBench /
LiveSQLBench-Large.}

\begin{longtable}[]{@{}
  >{\raggedright\arraybackslash}p{(\columnwidth - 8\tabcolsep) * \real{0.1864}}
  >{\raggedright\arraybackslash}p{(\columnwidth - 8\tabcolsep) * \real{0.2712}}
  >{\raggedright\arraybackslash}p{(\columnwidth - 8\tabcolsep) * \real{0.2373}}
  >{\raggedright\arraybackslash}p{(\columnwidth - 8\tabcolsep) * \real{0.1525}}
  >{\raggedright\arraybackslash}p{(\columnwidth - 8\tabcolsep) * \real{0.1525}}@{}}
\toprule\noalign{}
\begin{minipage}[b]{\linewidth}\raggedright
Laya scores
\end{minipage} &
\begin{minipage}[b]{\linewidth}\raggedright
Reduction in scored
pairs
\end{minipage} &
\begin{minipage}[b]{\linewidth}\raggedright
Seed 0 (rule)
\end{minipage} &
\begin{minipage}[b]{\linewidth}\raggedright
Seed 1
\end{minipage} &
\begin{minipage}[b]{\linewidth}\raggedright
Seed 2
\end{minipage} \\
\midrule\noalign{}
\endhead
\bottomrule\noalign{}
\endlastfoot
Top 50\% & 2× / 2× /
2× & −0.3 / −0.2,
passes & +0.6 / +0.8
& +0.6 / −1.0 \\
Top 25\% & 4× / 4× /
4× & −2.1 / +0.9,
fails & +0.2 / +1.4
& −0.4 / −0.3 \\
Top 100 & 4.6× /
2.4× / 9.7× & −1.6 /
−0.9, fails & +0.5 /
+0.3 & −1.6 /
−0.7 \\
Top 50 & 9.3× / 4.9×
/ 19.4× & −4.1 /
−3.4, fails & −1.7 /
−1.7 & −1.6 /
−1.7 \\
\end{longtable}

The top 50\% passes
the rule and holds
on seeds 1 and 2.
Smaller budgets lose
recall where the
search shortlist
starts to miss gold
tables, which again
points to
first-stage recall
on large schemas as
the limit.

\hypertarget{c.-does-model-size-or-training-matter-more}{%
\subsubsection{C.
Does model size or
training matter
more?}\label{c.-does-model-size-or-training-matter-more}}

We replaced Laya
with
pplx-decider-flash-v1.2
(Perplexity AI,
n.d.; about 4B
parameters,
Qwen3.5-4B
backbone), first
zero-shot and then
fine-tuned with LoRA
(Hu et al., 2022;
rank 16, 2 epochs,
learning rate 1e-4)
on exactly Laya's
training rows.
Settings were fixed
before the run. All
scorers score the
same search top 100
per question.

\needspace{12\baselineskip}

\emph{Table C1.
Seed-0 validation,
complete@8 /
complete@15 in \%.}

\begin{longtable}[]{@{}
  >{\raggedright\arraybackslash}p{(\columnwidth - 8\tabcolsep) * \real{0.2333}}
  >{\raggedright\arraybackslash}p{(\columnwidth - 8\tabcolsep) * \real{0.2500}}
  >{\raggedright\arraybackslash}p{(\columnwidth - 8\tabcolsep) * \real{0.2000}}
  >{\raggedright\arraybackslash}p{(\columnwidth - 8\tabcolsep) * \real{0.1500}}
  >{\raggedright\arraybackslash}p{(\columnwidth - 8\tabcolsep) * \real{0.1667}}@{}}
\toprule\noalign{}
\begin{minipage}[b]{\linewidth}\raggedright
Scorer
\end{minipage} &
\begin{minipage}[b]{\linewidth}\raggedright
Alone, complete@15
(BEAVER / LSB /
LSB-L)
\end{minipage} &
\begin{minipage}[b]{\linewidth}\raggedright
In the fusion
(BEAVER)
\end{minipage} &
\begin{minipage}[b]{\linewidth}\raggedright
In the fusion (LSB)
\end{minipage} &
\begin{minipage}[b]{\linewidth}\raggedright
In the fusion
(LSB-L)
\end{minipage} \\
\midrule\noalign{}
\endhead
\bottomrule\noalign{}
\endlastfoot
Laya, 421M,
fine-tuned & 26.2 /
61.6 / 68.7 & 47.7 /
66.2 & 70.4 / 84.0 &
67.7 / 82.8 \\
pplx 4B, zero-shot &
26.2 / 32.0 / 17.2 &
44.6 / 70.8 & 74.4 /
89.6 & 54.5 /
74.7 \\
pplx 4B, fine-tuned
(LoRA) & 60.0 / 88.0
/ 82.8 & 66.2 / 76.9
& 83.2 / 93.6 & 77.8
/ 85.9 \\
\end{longtable}

Zero-shot, the 4B
model ties
fine-tuned Laya
alone on BEAVER and
is far weaker on the
two LiveSQLBench
sets; in the fusion
it is ahead of
fine-tuned Laya at
complete@15 on
BEAVER and at both
cutoffs on
LiveSQLBench, but
worse by a resolved
margin on
LiveSQLBench-Large,
suggesting that a
strong zero-shot
model can add useful
signal to the fusion
on smaller schemas.
Fine-tuned on the
same rows, it beats
Laya in the fusion
by 10 to 19 points
of complete@8 (all
resolved on seed 0).
Fine-tuning helps
both models, most on
the large schema;
size helps once the
model is trained.
The 4B model scores
at about 38 to 48 ms
per table, 10 to 17
times Laya and 20 to
30 times the
fine-tuned
cross-encoder, and
trains in 10 to 16
minutes per
customer. This
result is from one
validation split and
needs test
confirmation.

\hypertarget{d.-fusers-and-features}{%
\subsubsection{D.
Fusers and
features}\label{d.-fusers-and-features}}

On seed-0
validation, we
compared
logistic-regression
fusion with an MLP
(64 and 32 units),
gradient-boosted
trees, and
LambdaRank (Burges,
2010) as implemented
in LightGBM (Ke et
al., 2017), using
fixed settings.
Alternative fusers
produced mixed
results: with Laya,
LambdaRank increased
mean
complete@8/complete@15
by 2.8/0.3
percentage points
relative to logistic
regression, while
trees reduced them
by 2.0/0.7 points
and the MLP by
3.7/3.3 points.
Adding 15 features
to the 11 used with
join completion
reduced mean
complete@8 by 1.1
points with Laya and
4.4 points with the
fine-tuned 4B
scorer; spline terms
reduced it further.
We fitted fusion on
half the validation
questions,
approximately 30 to
110 per customer.
This limited fitting
set may constrain
more flexible
models, although the
experiments do not
isolate that
explanation.
Cross-fitting the
scorer to provide
more out-of-sample
predictions for
fusion training is a
direction for future
work.

\hypertarget{e.-reranker-prompt-and-reproducibility}{%
\subsubsection{E.
Reranker prompt and
reproducibility}\label{e.-reranker-prompt-and-reproducibility}}

The
Qwen3-Reranker-8B
baseline uses the
model card's format.
The system message
reads: \emph{Judge
whether the Document
meets the
requirements based
on the Query and the
Instruct provided.
Note that the answer
can only be ``yes''
or ``no''.} The user
message is
\texttt{\textless{}Instruct\textgreater{}:\ Given\ a\ question\ about\ a\ database,\ judge\ whether\ this\ table\ is\ required\ to\ write\ the\ SQL\ query\ that\ answers\ the\ question:\ it\ holds\ answer\ columns,\ filter\ columns,\ or\ is\ needed\ to\ join\ such\ tables.}
followed by
\texttt{\textless{}Query\textgreater{}:\ \textless{}question\textgreater{}}
and
\texttt{\textless{}Document\textgreater{}:\ \textless{}card\textgreater{}}
plus \texttt{Joins:}
and the declared
joins when the table
has any. The score
is the probability
of ``yes'' against
``no'' at the last
position. Inputs are
truncated at 2,048
tokens.

All code is in the
repository described
under Data and code
availability. Main
runs Laya using
configs/default.yaml
with seeds 1 to 4;
cross-encoder runs
use
experiments/supervised\_\allowbreak{}crossencoder.py
on the same seeds
and splits;
training-size runs
use the train10,
train20, and train30
configurations;
tables come from
experiments/paper\_\allowbreak{}tables.py
and
experiments/additional\_\allowbreak{}analyses.py.
The repository's
README lists the
code version and
configuration of
every run.

\hypertarget{f.-full-main-results}{%
\subsubsection{F.
Full main
results}\label{f.-full-main-results}}

\needspace{12\baselineskip}

\emph{Table F1.
Test, mean over
seeds 1 to 4, \%;
c@K is complete@K.}

\begin{longtable}[]{@{}
  >{\raggedright\arraybackslash}p{(\columnwidth - 16\tabcolsep) * \real{0.2540}}
  >{\raggedright\arraybackslash}p{(\columnwidth - 16\tabcolsep) * \real{0.1905}}
  >{\raggedright\arraybackslash}p{(\columnwidth - 16\tabcolsep) * \real{0.0794}}
  >{\raggedright\arraybackslash}p{(\columnwidth - 16\tabcolsep) * \real{0.0794}}
  >{\raggedright\arraybackslash}p{(\columnwidth - 16\tabcolsep) * \real{0.0794}}
  >{\raggedright\arraybackslash}p{(\columnwidth - 16\tabcolsep) * \real{0.0794}}
  >{\raggedright\arraybackslash}p{(\columnwidth - 16\tabcolsep) * \real{0.0794}}
  >{\raggedright\arraybackslash}p{(\columnwidth - 16\tabcolsep) * \real{0.0794}}
  >{\raggedright\arraybackslash}p{(\columnwidth - 16\tabcolsep) * \real{0.0794}}@{}}
\toprule\noalign{}
\begin{minipage}[b]{\linewidth}\raggedright
Dataset
\end{minipage} &
\begin{minipage}[b]{\linewidth}\raggedright
Method
\end{minipage} &
\begin{minipage}[b]{\linewidth}\raggedright
c@5
\end{minipage} &
\begin{minipage}[b]{\linewidth}\raggedright
c@8
\end{minipage} &
\begin{minipage}[b]{\linewidth}\raggedright
c@15
\end{minipage} &
\begin{minipage}[b]{\linewidth}\raggedright
c@25
\end{minipage} &
\begin{minipage}[b]{\linewidth}\raggedright
R@5
\end{minipage} &
\begin{minipage}[b]{\linewidth}\raggedright
R@10
\end{minipage} &
\begin{minipage}[b]{\linewidth}\raggedright
R@15
\end{minipage} \\
\midrule\noalign{}
\endhead
\bottomrule\noalign{}
\endlastfoot
BEAVER & Search
alone & 14.3 & 20.2
& 30.4 & 44.6 & 44.3
& 55.6 & 65.0 \\
BEAVER & Search +
reranker & 21.4 &
36.9 & 53.6 & 69.6 &
59.3 & 74.9 &
81.4 \\
BEAVER & Laya alone
& 6.5 & 13.7 & 31.0
& 47.0 & 33.1 & 52.3
& 63.5 \\
BEAVER & Rank
average & 24.4 &
36.9 & 60.1 & 76.2 &
59.6 & 76.9 &
83.4 \\
BEAVER &
NIFT-SQL-base with
Laya & 21.4 & 41.7 &
60.1 & 74.4 & 58.8 &
76.0 & 83.3 \\
LiveSQLBench &
Search alone & 20.4
& 28.3 & 41.0 & 51.6
& 45.8 & 59.3 &
66.6 \\
LiveSQLBench &
Search + reranker &
53.3 & 64.8 & 77.1 &
80.7 & 77.3 & 86.3 &
89.9 \\
LiveSQLBench & Laya
alone & 22.1 & 34.9
& 51.6 & 64.2 & 45.9
& 62.0 & 71.1 \\
LiveSQLBench & Rank
average & 40.1 &
51.0 & 68.6 & 78.6 &
64.9 & 77.0 &
84.0 \\
LiveSQLBench &
NIFT-SQL-base with
Laya & 50.6 & 70.3 &
83.2 & 89.5 & 72.7 &
86.7 & 90.6 \\
LiveSQLBench-Large &
Search alone & 13.6
& 19.9 & 21.8 & 28.0
& 27.6 & 39.9 &
43.5 \\
LiveSQLBench-Large &
Search + reranker &
40.3 & 46.2 & 50.3 &
51.7 & 65.5 & 71.4 &
72.4 \\
LiveSQLBench-Large &
Laya alone & 25.8 &
35.7 & 54.7 & 65.2 &
48.5 & 65.3 &
73.9 \\
LiveSQLBench-Large &
Rank average & 37.4
& 46.9 & 62.3 & 72.2
& 64.0 & 75.6 &
81.7 \\
LiveSQLBench-Large &
NIFT-SQL-base with
Laya & 48.5 & 63.8 &
79.1 & 85.0 & 72.9 &
86.4 & 90.2 \\
\end{longtable}

At complete@5 and
R@5, the reranker
beats NIFT-SQL-base
with Laya on
LiveSQLBench (53.3
vs 50.6 and 77.3 vs
72.7); on smaller
schemas, it ranks
higher at the top.
The fine-tuned
cross-encoder
narrows this
(complete@5: 52.1,
Table F2).

\needspace{12\baselineskip}

\emph{Table F2.
Cross-encoder runs
(fine-tuned,
zero-shot, and
fine-tuned without
join hints), test,
mean over seeds 1 to
4, \%; c@K is
complete@K.}

\begin{longtable}[]{@{}
  >{\raggedright\arraybackslash}p{(\columnwidth - 12\tabcolsep) * \real{0.3051}}
  >{\raggedright\arraybackslash}p{(\columnwidth - 12\tabcolsep) * \real{0.2712}}
  >{\raggedright\arraybackslash}p{(\columnwidth - 12\tabcolsep) * \real{0.0847}}
  >{\raggedright\arraybackslash}p{(\columnwidth - 12\tabcolsep) * \real{0.0847}}
  >{\raggedright\arraybackslash}p{(\columnwidth - 12\tabcolsep) * \real{0.0847}}
  >{\raggedright\arraybackslash}p{(\columnwidth - 12\tabcolsep) * \real{0.0847}}
  >{\raggedright\arraybackslash}p{(\columnwidth - 12\tabcolsep) * \real{0.0847}}@{}}
\toprule\noalign{}
\begin{minipage}[b]{\linewidth}\raggedright
Dataset
\end{minipage} &
\begin{minipage}[b]{\linewidth}\raggedright
Method
\end{minipage} &
\begin{minipage}[b]{\linewidth}\raggedright
c@5
\end{minipage} &
\begin{minipage}[b]{\linewidth}\raggedright
c@8
\end{minipage} &
\begin{minipage}[b]{\linewidth}\raggedright
c@15
\end{minipage} &
\begin{minipage}[b]{\linewidth}\raggedright
c@25
\end{minipage} &
\begin{minipage}[b]{\linewidth}\raggedright
R@10
\end{minipage} \\
\midrule\noalign{}
\endhead
\bottomrule\noalign{}
\endlastfoot
BEAVER & Zero-shot
cross-encoder alone
& 12.5 & 16.7 & 28.0
& 40.5 & 52.2 \\
BEAVER & Fine-tuned
cross-encoder alone
& 11.9 & 21.4 & 45.2
& 63.1 & 62.3 \\
BEAVER & Fine-tuned,
no join hints, alone
& 11.3 & 22.0 & 42.9
& 64.3 & 63.0 \\
BEAVER & Rank
average, fine-tuned
cross-encoder +
search & 25.6 & 38.7
& 68.5 & 78.0 &
77.5 \\
BEAVER &
NIFT-SQL-base with
zero-shot
cross-encoder & 16.1
& 20.2 & 41.1 & 58.9
& 63.9 \\
BEAVER &
NIFT-SQL-base with
fine-tuned, no join
hints & 28.6 & 46.4
& 66.1 & 79.8 &
80.2 \\
BEAVER &
NIFT-SQL-base with
fine-tuned
cross-encoder & 26.8
& 48.8 & 67.3 & 77.4
& 80.2 \\
LiveSQLBench &
Zero-shot
cross-encoder alone
& 31.9 & 41.9 & 55.1
& 67.7 & 67.6 \\
LiveSQLBench &
Fine-tuned
cross-encoder alone
& 41.2 & 58.2 & 69.1
& 74.4 & 78.0 \\
LiveSQLBench &
Fine-tuned, no join
hints, alone & 31.8
& 43.0 & 53.6 & 65.6
& 69.2 \\
LiveSQLBench & Rank
average, fine-tuned
cross-encoder +
search & 40.1 & 50.4
& 64.7 & 79.1 &
76.8 \\
LiveSQLBench &
NIFT-SQL-base with
zero-shot
cross-encoder & 44.5
& 60.1 & 77.8 & 84.2
& 82.2 \\
LiveSQLBench &
NIFT-SQL-base with
fine-tuned, no join
hints & 48.9 & 65.7
& 80.1 & 84.2 &
83.4 \\
LiveSQLBench &
NIFT-SQL-base with
fine-tuned
cross-encoder & 52.1
& 71.5 & 86.2 & 89.2
& 86.8 \\
LiveSQLBench-Large &
Zero-shot
cross-encoder alone
& 20.2 & 25.7 & 35.3
& 44.8 & 48.2 \\
LiveSQLBench-Large &
Fine-tuned
cross-encoder alone
& 40.6 & 58.7 & 71.7
& 77.2 & 80.3 \\
LiveSQLBench-Large &
Fine-tuned, no join
hints, alone & 36.8
& 44.9 & 58.8 & 68.4
& 71.7 \\
LiveSQLBench-Large &
Rank average,
fine-tuned
cross-encoder +
search & 38.7 & 53.0
& 66.8 & 76.7 &
78.2 \\
LiveSQLBench-Large &
NIFT-SQL-base with
zero-shot
cross-encoder & 24.2
& 35.0 & 45.6 & 56.1
& 58.4 \\
LiveSQLBench-Large &
NIFT-SQL-base with
fine-tuned, no join
hints & 50.4 & 63.2
& 73.8 & 81.1 &
81.6 \\
LiveSQLBench-Large &
NIFT-SQL-base with
fine-tuned
cross-encoder & 52.3
& 68.0 & 78.6 & 83.2
& 85.6 \\
\end{longtable}

\needspace{12\baselineskip}

\emph{Table F3.
Cross-encoder
comparisons, A minus
B, complete@8 /
complete@15, pooled
over seeds 1 to 4,
95\% interval
clustered by
gold-table set.
Bold: the interval
excludes zero.}

\begin{longtable}[]{@{}
  >{\raggedright\arraybackslash}p{(\columnwidth - 6\tabcolsep) * \real{0.2787}}
  >{\raggedright\arraybackslash}p{(\columnwidth - 6\tabcolsep) * \real{0.2295}}
  >{\raggedright\arraybackslash}p{(\columnwidth - 6\tabcolsep) * \real{0.2295}}
  >{\raggedright\arraybackslash}p{(\columnwidth - 6\tabcolsep) * \real{0.2623}}@{}}
\toprule\noalign{}
\begin{minipage}[b]{\linewidth}\raggedright
A minus B
\end{minipage} &
\begin{minipage}[b]{\linewidth}\raggedright
BEAVER
\end{minipage} &
\begin{minipage}[b]{\linewidth}\raggedright
LiveSQLBench
\end{minipage} &
\begin{minipage}[b]{\linewidth}\raggedright
LiveSQLBench-Large
\end{minipage} \\
\midrule\noalign{}
\endhead
\bottomrule\noalign{}
\endlastfoot
Fine-tuned minus
zero-shot
cross-encoder, alone
& +4.8 {[}−7.6,
17.8{]} / +17.3
{[}−0.6, 33.5{]} &
\textbf{+16.5}
{[}8.4, 24.3{]} /
\textbf{+14.1}
{[}7.1, 21.6{]} &
\textbf{+32.6}
{[}23.0, 42.1{]} /
\textbf{+36.3}
{[}27.8, 45.2{]} \\
Fine-tuned minus
zero-shot
cross-encoder, in
NIFT-SQL-base &
\textbf{+28.6}
{[}14.5, 41.8{]} /
\textbf{+26.2}
{[}14.8, 39.1{]} &
\textbf{+11.5}
{[}4.1, 18.9{]} /
\textbf{+8.5}
{[}2.0, 14.7{]} &
\textbf{+33.0}
{[}25.6, 40.7{]} /
\textbf{+33.0}
{[}25.1, 40.9{]} \\
Cross-encoder alone
minus Laya alone &
+7.7 {[}−1.9,
17.3{]} /
\textbf{+14.3}
{[}1.9, 27.8{]} &
\textbf{+23.5}
{[}16.9, 30.5{]} /
\textbf{+17.6}
{[}11.5, 23.8{]} &
\textbf{+22.7}
{[}14.7, 31.3{]} /
\textbf{+16.8}
{[}10.0, 24.3{]} \\
NIFT-SQL-base
(cross-encoder)
minus cross-encoder
alone &
\textbf{+27.4}
{[}16.6, 37.6{]} /
\textbf{+22.0}
{[}15.3, 30.4{]} &
\textbf{+13.2}
{[}7.4, 18.9{]} /
\textbf{+17.1}
{[}12.2, 22.3{]} &
\textbf{+9.5}
{[}3.4, 15.5{]} /
\textbf{+7.0}
{[}3.8, 10.6{]} \\
NIFT-SQL-base
(cross-encoder)
minus search alone &
\textbf{+28.6}
{[}13.9, 42.5{]} /
\textbf{+36.9}
{[}20.8, 52.5{]} &
\textbf{+43.2}
{[}35.0, 51.7{]} /
\textbf{+45.3}
{[}36.4, 54.8{]} &
\textbf{+48.0}
{[}39.9, 56.2{]} /
\textbf{+56.8}
{[}47.9, 65.3{]} \\
No join hints minus
join hints, alone &
+0.6 {[}−4.5, 6.1{]}
/ −2.4 {[}−12.2,
6.5{]} &
\textbf{−15.3}
{[}−21.3, −9.6{]} /
\textbf{−15.6}
{[}−21.9, −10.0{]} &
\textbf{−13.6}
{[}−19.5, −7.7{]} /
\textbf{−12.8}
{[}−18.8, −7.3{]} \\
No join hints minus
join hints, in
NIFT-SQL-base & −2.4
{[}−7.6, 2.3{]} /
−1.2 {[}−5.0, 2.5{]}
& \textbf{−5.9}
{[}−11.9, −0.3{]} /
\textbf{−6.2}
{[}−10.2, −2.7{]} &
−4.8 {[}−10.4,
0.7{]} / −4.8
{[}−10.2, 0.3{]} \\
\end{longtable}

Pooled over the
three datasets, with
LiveSQLBench and
LiveSQLBench-Large
questions that share
a text clustered
together,
complete@15:
NIFT-SQL-base with
the cross-encoder
minus the reranker
+16.8 {[}10.6,
22.8{]}, minus the
same pipeline with
the zero-shot
cross-encoder +20.9
{[}15.6, 25.8{]},
minus NIFT-SQL-base
with Laya +2.6
{[}−1.1, 6.4{]}.

\needspace{12\baselineskip}

\emph{Table F4.
Finding the database
versus finding the
tables (Section
6.5), complete@15,
test, mean over
seeds 1 to 4, \%.
``All'' searches
every database in
the collection (the
main setting);
``given'' restricts
each ranking to the
question's
database.}

\begin{longtable}[]{@{}
  >{\raggedright\arraybackslash}p{(\columnwidth - 12\tabcolsep) * \real{0.1803}}
  >{\raggedright\arraybackslash}p{(\columnwidth - 12\tabcolsep) * \real{0.0820}}
  >{\raggedright\arraybackslash}p{(\columnwidth - 12\tabcolsep) * \real{0.0820}}
  >{\raggedright\arraybackslash}p{(\columnwidth - 12\tabcolsep) * \real{0.1311}}
  >{\raggedright\arraybackslash}p{(\columnwidth - 12\tabcolsep) * \real{0.1311}}
  >{\raggedright\arraybackslash}p{(\columnwidth - 12\tabcolsep) * \real{0.1967}}
  >{\raggedright\arraybackslash}p{(\columnwidth - 12\tabcolsep) * \real{0.1967}}@{}}
\toprule\noalign{}
\begin{minipage}[b]{\linewidth}\raggedright
Method
\end{minipage} &
\begin{minipage}[b]{\linewidth}\raggedright
BEAVER all
\end{minipage} &
\begin{minipage}[b]{\linewidth}\raggedright
BEAVER given
\end{minipage} &
\begin{minipage}[b]{\linewidth}\raggedright
LiveSQLBench all
\end{minipage} &
\begin{minipage}[b]{\linewidth}\raggedright
LiveSQLBench given
\end{minipage} &
\begin{minipage}[b]{\linewidth}\raggedright
LiveSQLBench-Large
all
\end{minipage} &
\begin{minipage}[b]{\linewidth}\raggedright
LiveSQLBench-Large
given
\end{minipage} \\
\midrule\noalign{}
\endhead
\bottomrule\noalign{}
\endlastfoot
Search alone & 30.4
& 39.9 & 41.0 & 98.8
& 21.8 & 38.9 \\
Search +
Qwen3-Reranker-8B &
53.6 & 54.2 & 77.1 &
98.8 & 50.3 &
52.8 \\
Fine-tuned
cross-encoder alone
& 45.2 & 50.0 & 69.1
& 99.4 & 71.7 &
83.9 \\
NIFT-SQL-base with
Laya & 60.1 & 61.9 &
83.2 & 100.0 & 79.1
& 85.4 \\
NIFT-SQL-base with
fine-tuned
cross-encoder & 67.3
& 67.3 & 86.2 &
100.0 & 78.6 &
85.0 \\
NIFT-SQL-base with
Laya, no database
features & 60.1 & &
72.7 & & 70.0 & \\
NIFT-SQL-base with
cross-encoder, no
database features &
65.5 & & 77.4 & &
75.3 & \\
NIFT-SQL-base with
cross-encoder, no
join hints, no
database features &
67.3 & & 66.8 & &
64.9 & \\
\end{longtable}

Share of test
questions whose
top-ranked table is
in the right
database, all
databases searched:
search 92.3\%,
55.2\%, and 58.0\%;
the reranker 97.6\%,
90.3\%, and 85.0\%;
NIFT-SQL-base with
the cross-encoder
97.0\%, 81.2\%, and
85.6\% (BEAVER,
LiveSQLBench,
LiveSQLBench-Large).
Database-given
complete@8: search
25.6\%, 77.5\%, and
29.1\%;
NIFT-SQL-base with
the cross-encoder:
49.4\%, 87.1\%, and
72.7\%.

\needspace{12\baselineskip}

\emph{Table F5.
Training-size curve
(Section 7),
NIFT-SQL-base
complete@8 /
complete@15, test,
mean over seeds 1 to
4, \%. Training size
is the share of all
questions used to
train the scorer;
fusion is always
fitted on the same
validation
questions. Search
alone and the
reranker do not
depend on the
training size.}

\begin{longtable}[]{@{}
  >{\raggedright\arraybackslash}p{(\columnwidth - 10\tabcolsep) * \real{0.3390}}
  >{\raggedright\arraybackslash}p{(\columnwidth - 10\tabcolsep) * \real{0.2542}}
  >{\raggedright\arraybackslash}p{(\columnwidth - 10\tabcolsep) * \real{0.1017}}
  >{\raggedright\arraybackslash}p{(\columnwidth - 10\tabcolsep) * \real{0.1017}}
  >{\raggedright\arraybackslash}p{(\columnwidth - 10\tabcolsep) * \real{0.1017}}
  >{\raggedright\arraybackslash}p{(\columnwidth - 10\tabcolsep) * \real{0.1017}}@{}}
\toprule\noalign{}
\begin{minipage}[b]{\linewidth}\raggedright
Dataset
\end{minipage} &
\begin{minipage}[b]{\linewidth}\raggedright
Scorer
\end{minipage} &
\begin{minipage}[b]{\linewidth}\raggedright
10\%
\end{minipage} &
\begin{minipage}[b]{\linewidth}\raggedright
20\%
\end{minipage} &
\begin{minipage}[b]{\linewidth}\raggedright
30\%
\end{minipage} &
\begin{minipage}[b]{\linewidth}\raggedright
50\%
\end{minipage} \\
\midrule\noalign{}
\endhead
\bottomrule\noalign{}
\endlastfoot
BEAVER & Laya & 23.8
/ 44.6 & 25.6 / 49.4
& 35.1 / 58.3 & 41.7
/ 60.1 \\
BEAVER & Fine-tuned
cross-encoder & 37.5
/ 56.5 & 38.7 / 62.5
& 38.1 / 60.7 & 48.8
/ 67.3 \\
LiveSQLBench & Laya
& 58.6 / 74.7 & 64.8
/ 80.6 & 65.4 / 80.7
& 70.3 / 83.2 \\
LiveSQLBench &
Fine-tuned
cross-encoder & 66.8
/ 82.7 & 69.2 / 83.9
& 71.5 / 86.0 & 71.5
/ 86.2 \\
LiveSQLBench-Large &
Laya & 44.4 / 63.5 &
56.9 / 72.2 & 59.4 /
75.9 & 63.8 /
79.1 \\
LiveSQLBench-Large &
Fine-tuned
cross-encoder & 62.7
/ 77.0 & 65.0 / 80.3
& 68.6 / 81.4 & 68.0
/ 78.6 \\
\end{longtable}

For reference,
search alone: 20.2 /
30.4, 28.3 / 41.0,
and 19.9 / 21.8;
search +
Qwen3-Reranker-8B:
36.9 / 53.6, 64.8 /
77.1, and 46.2 /
50.3 (BEAVER,
LiveSQLBench,
LiveSQLBench-Large).
Laya alone,
complete@15 at 10\%
/ 20\% / 30\% /
50\%: 12.5 / 20.8 /
26.2 / 31.0, 19.5 /
34.0 / 39.8 / 51.6,
and 21.5 / 33.8 /
46.3 / 54.7.

\hypertarget{g.-additional-analyses}{%
\subsubsection{G.
Additional
analyses}\label{g.-additional-analyses}}

All from
experiments/additional\_\allowbreak{}analyses.py,
test, seeds 1 to 4,
means over seeds of
each seed's test
mean, \%.

\needspace{12\baselineskip}

\emph{Table G1.
Database grouping
and the known
database (Section
6.5), complete@8 /
complete@15.
``Grouped'' orders
databases by their
best-ranked table
and keeps each
method's order
within a database.
``Database given''
restricts the
ranking to the
question's database;
for the reranker,
``whole database''
scores every table
in it instead of the
search top 100.}

\begin{longtable}[]{@{}
  >{\raggedright\arraybackslash}p{(\columnwidth - 6\tabcolsep) * \real{0.3167}}
  >{\raggedright\arraybackslash}p{(\columnwidth - 6\tabcolsep) * \real{0.1333}}
  >{\raggedright\arraybackslash}p{(\columnwidth - 6\tabcolsep) * \real{0.2333}}
  >{\raggedright\arraybackslash}p{(\columnwidth - 6\tabcolsep) * \real{0.3167}}@{}}
\toprule\noalign{}
\begin{minipage}[b]{\linewidth}\raggedright
Method
\end{minipage} &
\begin{minipage}[b]{\linewidth}\raggedright
BEAVER
\end{minipage} &
\begin{minipage}[b]{\linewidth}\raggedright
LiveSQLBench
\end{minipage} &
\begin{minipage}[b]{\linewidth}\raggedright
LiveSQLBench-Large
\end{minipage} \\
\midrule\noalign{}
\endhead
\bottomrule\noalign{}
\endlastfoot
Search alone & 20.2
/ 30.4 & 28.3 / 41.0
& 19.9 / 21.8 \\
Search alone,
grouped & 20.2 /
33.3 & 42.7 / 61.1 &
18.8 / 24.6 \\
Search +
Qwen3-Reranker-8B &
36.9 / 53.6 & 64.8 /
77.1 & 46.2 /
50.3 \\
Search +
Qwen3-Reranker-8B,
grouped & 37.5 /
53.0 & 72.6 / 91.8 &
41.4 / 44.8 \\
Rank average,
cross-encoder +
search & 38.7 / 68.5
& 50.4 / 64.7 & 53.0
/ 66.8 \\
Rank average,
cross-encoder +
search, grouped &
42.3 / 69.0 & 69.7 /
86.2 & 59.8 /
73.7 \\
Cross-encoder alone,
grouped & 25.0 /
43.5 & 62.1 / 77.9 &
60.6 / 70.0 \\
NIFT-SQL-base with
cross-encoder & 48.8
/ 67.3 & 71.5 / 86.2
& 68.0 / 78.6 \\
NIFT-SQL-base with
cross-encoder,
grouped & 47.6 /
65.5 & 71.2 / 84.7 &
65.8 / 76.3 \\
Search alone,
database given &
25.6 / 39.9 & 77.5 /
98.8 & 29.1 /
38.9 \\
Reranker, database
given, search top
100 & 37.5 / 54.2 &
80.6 / 98.8 & 49.1 /
52.8 \\
Reranker, database
given, whole
database & 37.5 /
54.8 & 81.2 / 99.4 &
69.5 / 85.3 \\
Cross-encoder alone,
database given &
29.8 / 50.0 & 79.7 /
99.4 & 72.3 /
83.9 \\
NIFT-SQL-base with
cross-encoder,
database given &
49.4 / 67.3 & 87.1 /
100.0 & 72.7 /
85.0 \\
\end{longtable}

NIFT-SQL-base with
the cross-encoder
minus each grouped
baseline,
complete@15: search
+33.9 {[}20.0,
47.7{]}, +25.0
{[}16.6, 33.4{]},
+53.8 {[}45.9,
61.7{]}; reranker
+14.3 {[}3.1,
26.4{]}, −5.6
{[}−11.3, 0.0{]},
+33.7 {[}25.5,
42.1{]}; rank
average −1.8
{[}−6.5, 2.9{]}, 0.0
{[}−4.3, 4.3{]},
+4.8 {[}0.4, 9.4{]}.
Database given,
NIFT-SQL-base (main
run) minus the
reranker over the
whole database:
+12.5 {[}1.2,
24.4{]}, +0.6
{[}0.0, 1.9{]}, −0.4
{[}−6.1, 5.8{]};
NIFT-SQL-base minus
the cross-encoder
alone: +17.3
{[}10.4, 25.6{]},
+0.6 {[}0.0, 1.5{]},
+1.1 {[}−3.4, 5.8{]}
(BEAVER,
LiveSQLBench,
LiveSQLBench-Large).

Further paired
differences behind
Sections 5 and 6.5
(BEAVER,
LiveSQLBench,
LiveSQLBench-Large).
NIFT-SQL-base with
the cross-encoder
minus grouped rank
averaging,
complete@8: +6.5
{[}1.1, 12.8{]},
+1.8 {[}−3.1,
6.9{]}, +8.1 {[}3.3,
12.9{]}.
NIFT-SQL-base minus
the grouped
cross-encoder alone,
complete@15: +23.8
{[}16.0, 32.8{]},
+8.2 {[}3.7,
12.9{]}, +8.4
{[}3.9, 13.4{]}.
Database given,
complete@15:
NIFT-SQL-base minus
search +27.4
{[}14.8, 40.0{]},
+1.2 {[}0.0, 3.8{]},
+46.2 {[}37.5,
54.6{]};
NIFT-SQL-base with
the cross-encoder
minus NIFT-SQL-base
with Laya +5.4
{[}−1.5, 12.4{]},
0.0 {[}0.0, 0.0{]},
−0.4 {[}−4.8,
4.6{]}; the
cross-encoder alone
minus the
whole-database
reranker −4.8
{[}−15.5, 5.6{]},
0.0 {[}−0.9, 0.8{]},
−1.5 {[}−7.9,
5.0{]}, and at
complete@8 −7.7
{[}−14.5, −0.7{]},
−1.5 {[}−7.7,
4.5{]}, +2.6
{[}−5.3, 10.6{]};
NIFT-SQL-base minus
the whole-database
reranker at
complete@8 +11.9
{[}2.4, 21.4{]},
+5.9 {[}−0.6,
12.4{]}, +2.9
{[}−5.3, 11.1{]}.

NIFT-SQL-base with
Laya, grouped,
complete@8 /
complete@15: 40.5 /
57.7, 73.2 / 86.5,
and 62.7 / 76.5
(ungrouped: 41.7 /
60.1, 70.3 / 83.2,
and 63.8 / 79.1).

\needspace{12\baselineskip}

\emph{Table G2.
Retraining the
cross-encoder
(Section 6.6):
NIFT-SQL-base with
the cross-encoder,
four-seed mean
complete@8 /
complete@15 for the
main training run
and two retrains
with different
dropout and
data-order seeds.}

\begin{longtable}[]{@{}
  >{\raggedright\arraybackslash}p{(\columnwidth - 6\tabcolsep) * \real{0.2000}}
  >{\raggedright\arraybackslash}p{(\columnwidth - 6\tabcolsep) * \real{0.1500}}
  >{\raggedright\arraybackslash}p{(\columnwidth - 6\tabcolsep) * \real{0.2667}}
  >{\raggedright\arraybackslash}p{(\columnwidth - 6\tabcolsep) * \real{0.3833}}@{}}
\toprule\noalign{}
\begin{minipage}[b]{\linewidth}\raggedright
Training run
\end{minipage} &
\begin{minipage}[b]{\linewidth}\raggedright
BEAVER
\end{minipage} &
\begin{minipage}[b]{\linewidth}\raggedright
LiveSQLBench
\end{minipage} &
\begin{minipage}[b]{\linewidth}\raggedright
LiveSQLBench-Large
\end{minipage} \\
\midrule\noalign{}
\endhead
\bottomrule\noalign{}
\endlastfoot
Main run & 48.8 /
67.3 & 71.5 / 86.2 &
68.0 / 78.6 \\
Retrain 1 & 41.1 /
61.9 & 69.2 / 85.7 &
66.9 / 81.9 \\
Retrain 2 & 43.4 /
61.9 & 72.4 / 86.6 &
67.6 / 79.5 \\
Mean of the three &
44.4 / 63.7 & 71.0 /
86.2 & 67.5 /
80.0 \\
\end{longtable}

Standard deviation
of a single seed's
mean across the
three runs, averaged
over seeds,
complete@15: 4.9,
2.2, and 5.1 points.
Input hashes of
every training row
and table card are
identical with and
without join hints
on BEAVER for all
four seeds, and
differ on the
LiveSQLBench
customer. Two BEAVER
trainings with
PyTorch's
deterministic
algorithms enabled
(seeds 1 and 2) did
not reproduce each
other exactly
(largest score
difference 0.96 and
0.99).

\needspace{12\baselineskip}

\emph{Table G3.
Fusion fitted on
fewer validation
questions per
customer (10 random
draws per seed,
metrics averaged
over draws),
NIFT-SQL-base with
the cross-encoder,
complete@8 /
complete@15.}

\begin{longtable}[]{@{}
  >{\raggedright\arraybackslash}p{(\columnwidth - 6\tabcolsep) * \real{0.2667}}
  >{\raggedright\arraybackslash}p{(\columnwidth - 6\tabcolsep) * \real{0.1333}}
  >{\raggedright\arraybackslash}p{(\columnwidth - 6\tabcolsep) * \real{0.2500}}
  >{\raggedright\arraybackslash}p{(\columnwidth - 6\tabcolsep) * \real{0.3500}}@{}}
\toprule\noalign{}
\begin{minipage}[b]{\linewidth}\raggedright
Scorer training
size, fusion fitted
on
\end{minipage} &
\begin{minipage}[b]{\linewidth}\raggedright
BEAVER
\end{minipage} &
\begin{minipage}[b]{\linewidth}\raggedright
LiveSQLBench
\end{minipage} &
\begin{minipage}[b]{\linewidth}\raggedright
LiveSQLBench-Large
\end{minipage} \\
\midrule\noalign{}
\endhead
\bottomrule\noalign{}
\endlastfoot
50\%, all validation
questions (main) &
48.8 / 67.3 & 71.5 /
86.2 & 68.0 /
78.6 \\
50\%, 40 validation
questions & 49.3 /
67.1 & 69.2 / 84.3 &
67.7 / 78.5 \\
50\%, 20 validation
questions & 48.0 /
67.5 & 68.9 / 84.6 &
66.1 / 77.4 \\
50\%, 10 validation
questions & 48.5 /
66.4 & 68.5 / 83.4 &
65.6 / 77.3 \\
10\%, all validation
questions & 37.5 /
56.5 & 66.8 / 82.7 &
62.7 / 77.0 \\
10\%, 20 validation
questions & 35.0 /
55.4 & 66.6 / 80.5 &
58.8 / 74.9 \\
\end{longtable}

Cross-encoder minus
Laya pipeline,
complete@15, at 10\%
/ 20\% / 30\%
training: BEAVER
+11.9 {[}1.9,
23.2{]} / +13.1
{[}2.4, 24.3{]} /
+2.4 {[}−7.6,
11.4{]};
LiveSQLBench +7.9
{[}2.9, 13.0{]} /
+3.2 {[}−1.4, 7.7{]}
/ +5.3 {[}1.1,
9.6{]};
LiveSQLBench-Large
+13.6 {[}8.1,
19.2{]} / +8.1
{[}3.7, 12.6{]} /
+5.5 {[}1.6, 9.7{]}.
With the scorer at
10\% and the fusion
on 20 validation
questions,
NIFT-SQL-base minus
search alone,
complete@15: +25.0
{[}10.3, 39.2{]},
+39.6 {[}32.2,
47.3{]}, and +53.3
{[}45.1, 61.2{]}.

\needspace{12\baselineskip}

\emph{Table G4.
Inputs cut by the
caps applied before
tokenization (700
characters per card,
six joins per
table). A table is
``gold'' if some
question needs it.
No input exceeded
the cross-encoder's
512-token limit
after the caps.}

\begin{longtable}[]{@{}
  >{\raggedright\arraybackslash}p{(\columnwidth - 8\tabcolsep) * \real{0.3667}}
  >{\raggedright\arraybackslash}p{(\columnwidth - 8\tabcolsep) * \real{0.1500}}
  >{\raggedright\arraybackslash}p{(\columnwidth - 8\tabcolsep) * \real{0.1500}}
  >{\raggedright\arraybackslash}p{(\columnwidth - 8\tabcolsep) * \real{0.1667}}
  >{\raggedright\arraybackslash}p{(\columnwidth - 8\tabcolsep) * \real{0.1667}}@{}}
\toprule\noalign{}
\begin{minipage}[b]{\linewidth}\raggedright
Dataset
\end{minipage} &
\begin{minipage}[b]{\linewidth}\raggedright
Tables
\end{minipage} &
\begin{minipage}[b]{\linewidth}\raggedright
Cards cut, \% (all /
gold)
\end{minipage} &
\begin{minipage}[b]{\linewidth}\raggedright
Columns lost when
cut, mean (all /
gold)
\end{minipage} &
\begin{minipage}[b]{\linewidth}\raggedright
Tables with joins
cut, \% (all / gold)
\end{minipage} \\
\midrule\noalign{}
\endhead
\bottomrule\noalign{}
\endlastfoot
BEAVER & 463 & 1.3 /
3.8 & 17.7 / 14.4 &
0.0 / 0.0 \\
LiveSQLBench & 244 &
0.0 / 0.0 & none &
7.0 / 8.6 \\
LiveSQLBench-Large &
971 & 1.8 / 0.0 &
5.4 / none & 5.7 /
23.5 \\
\end{longtable}

\needspace{12\baselineskip}

\emph{Table G5.
Replicate-aware
comparisons (Section
6.6): NIFT-SQL-base
with the
cross-encoder
averaged over its
three training runs
minus each baseline,
complete@8 /
complete@15, with
95\% intervals from
a two-level
bootstrap over
training runs and
gold-table-set
clusters. ``Same
run'' pairs each
training run with
the baseline built
from that run's
scores. Bold: the
interval excludes
zero.}

\begin{longtable}[]{@{}
  >{\raggedright\arraybackslash}p{(\columnwidth - 6\tabcolsep) * \real{0.2667}}
  >{\raggedright\arraybackslash}p{(\columnwidth - 6\tabcolsep) * \real{0.2333}}
  >{\raggedright\arraybackslash}p{(\columnwidth - 6\tabcolsep) * \real{0.2333}}
  >{\raggedright\arraybackslash}p{(\columnwidth - 6\tabcolsep) * \real{0.2667}}@{}}
\toprule\noalign{}
\begin{minipage}[b]{\linewidth}\raggedright
NIFT-SQL-base
(cross-encoder, mean
of three runs) minus
\end{minipage} &
\begin{minipage}[b]{\linewidth}\raggedright
BEAVER
\end{minipage} &
\begin{minipage}[b]{\linewidth}\raggedright
LiveSQLBench
\end{minipage} &
\begin{minipage}[b]{\linewidth}\raggedright
LiveSQLBench-Large
\end{minipage} \\
\midrule\noalign{}
\endhead
\bottomrule\noalign{}
\endlastfoot
Search alone &
\textbf{+24.2}
{[}9.4, 38.8{]} /
\textbf{+33.3}
{[}16.8, 48.7{]} &
\textbf{+42.7}
{[}34.3, 51.1{]} /
\textbf{+45.2}
{[}36.4, 53.6{]} &
\textbf{+47.5}
{[}40.0, 55.8{]} /
\textbf{+58.1}
{[}49.5, 67.2{]} \\
Search alone,
grouped &
\textbf{+24.2}
{[}9.1, 38.9{]} /
\textbf{+30.4}
{[}16.4, 43.6{]} &
\textbf{+28.3}
{[}19.7, 36.6{]} /
\textbf{+24.9}
{[}16.7, 32.8{]} &
\textbf{+48.6}
{[}41.1, 56.8{]} /
\textbf{+55.2}
{[}47.5, 63.2{]} \\
Search +
Qwen3-Reranker-8B &
+7.5 {[}−1.7,
17.6{]} / +10.1
{[}−1.3, 21.8{]} &
+6.3 {[}−1.4,
14.2{]} /
\textbf{+9.0}
{[}2.7, 15.6{]} &
\textbf{+21.1}
{[}13.8, 29.4{]} /
\textbf{+29.5}
{[}20.0, 39.7{]} \\
Search +
Qwen3-Reranker-8B,
grouped & +6.9
{[}−2.4, 17.0{]} /
+10.7 {[}−0.7,
22.3{]} & −1.7
{[}−9.2, 6.0{]} /
\textbf{−5.7}
{[}−10.7, −0.7{]} &
\textbf{+25.9}
{[}18.4, 33.8{]} /
\textbf{+35.0}
{[}26.7, 43.9{]} \\
NIFT-SQL-base with
Laya & +2.8 {[}−8.1,
13.0{]} / +3.6
{[}−5.2, 11.5{]} &
+0.7 {[}−4.9, 6.5{]}
/ +2.8 {[}−1.7,
8.0{]} & +3.5
{[}−2.0, 9.4{]} /
+0.6 {[}−4.6,
6.3{]} \\
Rank average,
grouped (same run) &
+2.2 {[}−3.5, 8.5{]}
/ −2.0 {[}−6.4,
2.6{]} & +0.8
{[}−4.7, 5.8{]} /
−0.1 {[}−4.6, 4.3{]}
& \textbf{+7.1}
{[}2.8, 11.6{]} /
\textbf{+5.6}
{[}1.6, 9.9{]} \\
Fine-tuned search
alone (first
training run) &
\textbf{+16.5}
{[}4.7, 28.0{]} /
+6.5 {[}−0.8,
14.6{]} &
\textbf{+19.5}
{[}13.1, 26.3{]} /
\textbf{+12.8}
{[}7.0, 19.5{]} &
\textbf{+27.7}
{[}21.5, 35.0{]} /
\textbf{+21.5}
{[}14.7, 29.0{]} \\
Database given:
reranker over the
whole database &
+7.9 {[}−1.2,
17.7{]} / +8.9
{[}−2.6, 20.6{]} &
\textbf{+5.7}
{[}0.1, 11.5{]} /
+0.6 {[}0.0, 1.9{]}
& +3.2 {[}−4.6,
11.4{]} / +0.2
{[}−6.0, 7.0{]} \\
Database given:
search alone &
\textbf{+19.8}
{[}6.3, 34.4{]} /
\textbf{+23.8}
{[}11.9, 35.8{]} &
\textbf{+9.5}
{[}4.7, 14.9{]} /
+1.2 {[}0.0, 3.8{]}
& \textbf{+43.8}
{[}36.2, 52.3{]} /
\textbf{+46.8}
{[}38.3, 55.8{]} \\
\end{longtable}

Reranker routed by
its own ranking
(every table of the
top two databases of
its grouped ranking
of the search top
100), complete@8 /
complete@15: 36.9 /
53.6, 72.1 / 89.1,
and 62.2 / 78.7;
NIFT-SQL-base minus
it: +7.5 {[}−1.7,
17.6{]} / +10.1
{[}−1.7, 22.2{]},
−1.1 {[}−8.7, 7.0{]}
/ −3.0 {[}−8.0,
2.1{]}, and +5.0
{[}−2.6, 13.5{]} /
+1.0 {[}−5.7,
8.5{]}. Reranker
variant chosen per
collection and seed
on validation
complete@15: 37.5 /
53.0, 72.0 / 90.3,
and 62.2 / 78.7;
NIFT-SQL-base minus
it: +6.9 {[}−2.4,
17.0{]} / +10.7
{[}−0.7, 22.3{]},
−1.1 {[}−8.5, 6.2{]}
/ −4.2 {[}−9.2,
0.8{]}, and +5.0
{[}−2.6, 13.5{]} /
+1.0 {[}−5.7,
8.5{]}. Fine-tuned
retriever over three
trainings,
complete@8 /
complete@15: 29.6 /
58.9, 52.4 / 73.1,
and 38.7 / 58.1;
NIFT-SQL-base minus
it: +14.9 {[}3.3,
26.9{]} / +4.8
{[}−3.1, 13.3{]},
+18.7 {[}12.4,
25.6{]} / +13.0
{[}7.4, 19.5{]}, and
+28.6 {[}21.9,
35.8{]} / +21.7
{[}15.0, 28.8{]}.
Fine-tuned retriever
trained on training
and validation
questions, three
trainings: 33.5 /
61.1, 60.5 / 81.7,
and 50.1 / 70.4;
NIFT-SQL-base minus
it: +10.9 {[}0.9,
21.0{]} / +2.6
{[}−4.7, 10.7{]},
+10.6 {[}4.8,
16.9{]} / +4.4
{[}−0.4, 9.7{]}, and
+17.2 {[}11.3,
23.9{]} / +9.5
{[}3.9, 15.7{]};
with the database
given, complete@15:
+1.2 {[}−5.8,
8.6{]}, 0.0 {[}0.0,
0.0{]}, and +1.5
{[}−4.2, 7.2{]}.
These intervals
resample both
methods' training
runs and
gold-table-set
clusters. Routed
reranker (every
table of the top two
databases by search,
the rest in grouped
search order),
complete@8 /
complete@15: 35.7 /
52.4, 57.6 / 72.9,
and 49.5 / 63.1;
NIFT-SQL-base (mean
of three runs) minus
it: +8.7 {[}−0.8,
18.9{]} / +11.3
{[}0.7, 22.6{]},
+13.3 {[}5.3,
21.5{]} / +13.1
{[}6.1, 20.0{]}, and
+17.8 {[}9.5,
26.0{]} / +16.7
{[}8.8, 24.5{]}. The
cross-encoder alone,
averaged over its
three runs, minus
the fine-tuned
search alone: −7.5
{[}−17.3, 1.4{]} /
−14.1 {[}−24.2,
−4.9{]}, +4.5
{[}−4.0, 13.4{]} /
−4.9 {[}−12.3,
2.6{]}, and +18.4
{[}9.9, 27.7{]} /
+13.2 {[}5.7,
21.3{]}. Four-seed
means of
NIFT-SQL-base with
the cross-encoder
over the three runs:
44.4 / 63.7, 71.0 /
86.2, and 67.5 /
80.0; with the
database given, 45.4
/ 63.7, 86.9 /
100.0, and 72.9 /
85.7.

\needspace{12\baselineskip}

\emph{Table G6.
Supervised-retriever
baseline (Section
6.7): hybrid search
with BGE fine-tuned
per customer on the
scorer's training
rows, first training
run of the retriever
(Table 2 averages
three), test, seeds
1 to 4, complete@8 /
complete@15, \%.}

\begin{longtable}[]{@{}
  >{\raggedright\arraybackslash}p{(\columnwidth - 6\tabcolsep) * \real{0.2833}}
  >{\raggedright\arraybackslash}p{(\columnwidth - 6\tabcolsep) * \real{0.1333}}
  >{\raggedright\arraybackslash}p{(\columnwidth - 6\tabcolsep) * \real{0.2333}}
  >{\raggedright\arraybackslash}p{(\columnwidth - 6\tabcolsep) * \real{0.3500}}@{}}
\toprule\noalign{}
\begin{minipage}[b]{\linewidth}\raggedright
Method
\end{minipage} &
\begin{minipage}[b]{\linewidth}\raggedright
BEAVER
\end{minipage} &
\begin{minipage}[b]{\linewidth}\raggedright
LiveSQLBench
\end{minipage} &
\begin{minipage}[b]{\linewidth}\raggedright
LiveSQLBench-Large
\end{minipage} \\
\midrule\noalign{}
\endhead
\bottomrule\noalign{}
\endlastfoot
Search alone (base
BGE) & 20.2 / 30.4 &
28.3 / 41.0 & 19.9 /
21.8 \\
Fine-tuned search
alone & 28.0 / 57.1
& 51.6 / 73.3 & 39.6
/ 58.3 \\
Fine-tuned search
alone, grouped &
27.4 / 53.6 & 60.3 /
79.8 & 41.1 /
54.6 \\
Fine-tuned search
alone, database
given & 33.3 / 59.5
& 84.4 / 100.0 &
56.5 / 75.1 \\
Cross-encoder alone
(main run) & 21.4 /
45.2 & 58.2 / 69.1 &
58.7 / 71.7 \\
NIFT-SQL-base with
cross-encoder (main
run) & 48.8 / 67.3 &
71.5 / 86.2 & 68.0 /
78.6 \\
NIFT-SQL-base with
cross-encoder (mean
of three runs) &
44.4 / 63.7 & 71.0 /
86.2 & 67.5 /
80.0 \\
\end{longtable}

Fine-tuned minus
base search: +7.7
{[}0.7, 15.0{]} /
+26.8 {[}12.7,
41.5{]}, +23.2
{[}17.1, 30.0{]} /
+32.4 {[}24.6,
40.6{]}, and +19.8
{[}14.1, 26.3{]} /
+36.6 {[}28.3,
44.9{]}.

\needspace{12\baselineskip}

\emph{Table G7.
Fusion extensions
(Section 6.8), test,
seeds 1 to 4, three
training runs,
complete@8 /
complete@15, \%.
Means over seeds of
each seed's test
mean, then over
training runs. A:
NIFT-SQL-base with
the fine-tuned
retriever in place
of base search. B:
NIFT-SQL-base plus
the reranker's logit
and rank as fusion
features. C: both.}

\begin{longtable}[]{@{}
  >{\raggedright\arraybackslash}p{(\columnwidth - 6\tabcolsep) * \real{0.3115}}
  >{\raggedright\arraybackslash}p{(\columnwidth - 6\tabcolsep) * \real{0.1311}}
  >{\raggedright\arraybackslash}p{(\columnwidth - 6\tabcolsep) * \real{0.2295}}
  >{\raggedright\arraybackslash}p{(\columnwidth - 6\tabcolsep) * \real{0.3279}}@{}}
\toprule\noalign{}
\begin{minipage}[b]{\linewidth}\raggedright
Method
\end{minipage} &
\begin{minipage}[b]{\linewidth}\raggedright
BEAVER
\end{minipage} &
\begin{minipage}[b]{\linewidth}\raggedright
LiveSQLBench
\end{minipage} &
\begin{minipage}[b]{\linewidth}\raggedright
LiveSQLBench-Large
\end{minipage} \\
\midrule\noalign{}
\endhead
\bottomrule\noalign{}
\endlastfoot
Search alone & 20.2
/ 30.4 & 28.3 / 41.0
& 19.9 / 21.8 \\
Fine-tuned
retriever, train +
validation questions
& 33.5 / 61.1 & 60.5
/ 81.7 & 50.1 /
70.4 \\
Search +
Qwen3-Reranker-8B
(search top 100) &
36.9 / 53.6 & 64.8 /
77.1 & 46.2 /
50.3 \\
Reranker, variant
chosen on validation
& 37.5 / 53.0 & 72.0
/ 90.3 & 62.2 /
78.7 \\
NIFT-SQL-base with
cross-encoder & 44.4
/ 63.7 & 71.0 / 86.2
& 67.5 / 80.0 \\
A: NIFT-SQL-base +
fine-tuned retriever
(= NIFT-SQL) & 46.4
/ 69.2 & 74.9 / 89.5
& 74.0 / 85.6 \\
B: NIFT-SQL-base +
reranker features &
51.2 / 68.5 & 78.7 /
88.3 & 74.3 /
84.0 \\
\textbf{C: A +
reranker features
(high-accuracy
mode)} &
\textbf{54.6 / 72.4}
& \textbf{81.6 /
92.5} & \textbf{78.0
/ 88.6} \\
\end{longtable}

Differences,
complete@8 /
complete@15, mean
over training runs,
95\% intervals from
a two-level
bootstrap over
training runs and
gold-table-set
clusters (BEAVER,
LiveSQLBench,
LiveSQLBench-Large).
Controls
(experiments/fusion\_\allowbreak{}controls.py),
complete@15 means:
rank average of the
two trained parts
66.3, 76.2, and
78.3, grouped 67.5,
89.3, and 81.2;
reranker alone on
NIFT-SQL's
candidates 54.4,
84.0, and 77.7,
grouped 54.4, 93.4,
and 73.8. A minus
grouped rank
average: +1.6
{[}−3.7, 6.9{]} /
+1.8 {[}−2.3,
5.9{]}, −1.7
{[}−5.3, 2.0{]} /
+0.1 {[}−3.1,
3.0{]}, +3.7
{[}−0.8, 8.3{]} /
+4.3 {[}−0.4,
9.4{]}. C minus the
reranker alone on
its candidates:
+16.9 {[}8.5,
26.5{]} / +18.1
{[}8.5, 28.0{]},
+14.6 {[}9.4,
20.3{]} / +8.4
{[}4.4, 13.3{]},
+15.6 {[}10.0,
22.3{]} / +10.7
{[}5.7, 16.7{]}. A
minus NIFT-SQL-base:
+2.0 {[}−3.2, 7.6{]}
/ +5.6 {[}0.7,
11.1{]}, +3.8
{[}−0.9, 8.7{]} /
+3.3 {[}0.6, 6.4{]},
+6.5 {[}1.5, 11.6{]}
/ +5.6 {[}2.7,
8.8{]}. B minus
NIFT-SQL-base: +6.7
{[}0.7, 14.0{]} /
+4.8 {[}−0.2,
11.0{]}, +7.7
{[}3.8, 12.0{]} /
+2.2 {[}−0.3,
5.6{]}, +7.0 {[}1.7,
12.0{]} / +4.2
{[}1.5, 7.1{]}. C
minus NIFT-SQL-base:
+10.1 {[}3.0,
18.6{]} / +8.7
{[}2.9, 15.3{]},
+10.6 {[}5.4,
15.9{]} / +6.4
{[}2.8, 10.1{]},
+10.6 {[}4.7,
16.6{]} / +8.7
{[}4.9, 12.8{]}. C
minus A: +8.1
{[}2.5, 15.3{]} /
+3.2 {[}−1.9,
8.9{]}, +6.8 {[}2.8,
11.4{]} / +3.0
{[}0.5, 5.9{]}, +4.2
{[}0.3, 8.0{]} /
+3.1 {[}0.5, 5.9{]}.
C minus B: +3.4
{[}−1.6, 9.4{]} /
+4.0 {[}0.6, 8.1{]},
+2.8 {[}−0.2, 5.9{]}
/ +4.2 {[}1.4,
7.4{]}, +3.7 {[}0.0,
7.2{]} / +4.5
{[}1.4, 7.8{]}. A
minus the fine-tuned
retriever trained on
training and
validation
questions: +12.9
{[}2.9, 23.0{]} /
+8.1 {[}1.1,
16.3{]}, +14.4
{[}9.4, 19.6{]} /
+7.7 {[}3.5,
12.7{]}, +23.7
{[}17.6, 30.2{]} /
+15.1 {[}10.0,
21.0{]}. A
(NIFT-SQL) minus the
reranker variant
chosen on
validation: +8.9
{[}−1.2, 20.0{]} /
+16.3 {[}4.5,
28.4{]}, +2.7
{[}−4.7, 9.8{]} /
−0.9 {[}−5.6,
3.7{]}, +11.5
{[}3.5, 20.0{]} /
+6.6 {[}−0.1,
13.8{]}. C minus the
reranker variant
chosen on
validation: +17.1
{[}8.3, 27.3{]} /
+19.4 {[}8.4,
30.8{]}, +9.5
{[}4.2, 15.0{]} /
+2.2 {[}−1.4,
6.1{]}, +15.6
{[}8.9, 22.6{]} /
+9.6 {[}3.8,
16.0{]}. Database
given, C minus the
reranker over the
whole database:
+17.1 {[}8.3,
27.3{]} / +17.7
{[}6.9, 28.8{]},
+9.1 {[}4.8, 14.3{]}
/ +0.6 {[}0.0,
1.9{]}, +12.9
{[}6.9, 19.7{]} /
+6.5 {[}1.0,
12.6{]}. Validation
check (two folds
inside validation,
before the test
part), complete@15:
NIFT-SQL-base 61.6,
87.2, 82.2; A 69.1,
91.8, 87.7; B 65.5,
91.0, 85.9; C 71.8,
93.6, 90.1.
Standardized fusion
weights, mean over
seeds and runs,
BEAVER /
LiveSQLBench
customer: in B,
scorer logit 1.03 /
0.28, reranker logit
0.85 / 0.08,
reranker log rank
-0.25 / -1.11. Full
output:
results/paper/FUSION\_\allowbreak{}EXTENSIONS.txt.

\hypertarget{h.-what-the-fusions-training-data-looks-like}{%
\subsubsection{H.
What the fusion's
training data looks
like}\label{h.-what-the-fusions-training-data-looks-like}}

The fusion is
trained on rows like
those below, one per
candidate table of
each validation
question (Section
3.3). The table
shows one BEAVER
validation question
from seed 1, chosen
to illustrate a
table that search
misses and the
scorer finds: the
first validation
question, by index,
with three or more
gold tables, all
inside the
candidates, and one
of them ranked 21st
or lower by search
but in the scorer's
top 10. Shown are
all gold rows, the
six non-gold rows
with the highest
fitted fusion
scores, and the
three with the
lowest scores. Ranks
are shown as ranks
(0 = first); the
fusion uses log(1 +
rank). The fused
probability is from
the fusion fitted on
this customer's
validation
questions; this one
is included.
Produced by
experiments/fusion\_\allowbreak{}example.py
(results/paper/FUSION\_\allowbreak{}EXAMPLE.txt).

Question (BEAVER):
``What is the name
of the building that
accomodates the most
students, and how
many students does
it accomodate?''
(spelling as in the
benchmark). Gold
tables: fclt\_\allowbreak{}rooms,
fclt\_\allowbreak{}building,
mit\_\allowbreak{}student\_\allowbreak{}directory.
The question has 77
candidates, 3 of
them gold. Training
data on seed 1: 63
validation questions
and 5,092 rows for
BEAVER, 6.2\%
positive.

\needspace{12\baselineskip}

\emph{Table H1.
Fusion training rows
for one BEAVER
validation question
(seed 1,
cross-encoder main
run; NIFT-SQL-base,
so the search
features come from
off-the-shelf
search).}

\begin{longtable}[]{@{}
  >{\raggedright\arraybackslash}p{(\columnwidth - 24\tabcolsep) * \real{0.1774}}
  >{\raggedright\arraybackslash}p{(\columnwidth - 24\tabcolsep) * \real{0.0484}}
  >{\raggedright\arraybackslash}p{(\columnwidth - 24\tabcolsep) * \real{0.0645}}
  >{\raggedright\arraybackslash}p{(\columnwidth - 24\tabcolsep) * \real{0.0645}}
  >{\raggedright\arraybackslash}p{(\columnwidth - 24\tabcolsep) * \real{0.0645}}
  >{\raggedright\arraybackslash}p{(\columnwidth - 24\tabcolsep) * \real{0.0645}}
  >{\raggedright\arraybackslash}p{(\columnwidth - 24\tabcolsep) * \real{0.0484}}
  >{\raggedright\arraybackslash}p{(\columnwidth - 24\tabcolsep) * \real{0.0484}}
  >{\raggedright\arraybackslash}p{(\columnwidth - 24\tabcolsep) * \real{0.0806}}
  >{\raggedright\arraybackslash}p{(\columnwidth - 24\tabcolsep) * \real{0.0806}}
  >{\raggedright\arraybackslash}p{(\columnwidth - 24\tabcolsep) * \real{0.0806}}
  >{\raggedright\arraybackslash}p{(\columnwidth - 24\tabcolsep) * \real{0.0806}}
  >{\raggedright\arraybackslash}p{(\columnwidth - 24\tabcolsep) * \real{0.0968}}@{}}
\toprule\noalign{}
\begin{minipage}[b]{\linewidth}\raggedright
Table
\end{minipage} &
\begin{minipage}[b]{\linewidth}\raggedright
Label
\end{minipage} &
\begin{minipage}[b]{\linewidth}\raggedright
Scorer logit
\end{minipage} &
\begin{minipage}[b]{\linewidth}\raggedright
Scorer rank
\end{minipage} &
\begin{minipage}[b]{\linewidth}\raggedright
Search rank
\end{minipage} &
\begin{minipage}[b]{\linewidth}\raggedright
Dense cosine
\end{minipage} &
\begin{minipage}[b]{\linewidth}\raggedright
BM25
\end{minipage} &
\begin{minipage}[b]{\linewidth}\raggedright
Rank gap
\end{minipage} &
\begin{minipage}[b]{\linewidth}\raggedright
Database best search
rank
\end{minipage} &
\begin{minipage}[b]{\linewidth}\raggedright
Database best scorer
rank
\end{minipage} &
\begin{minipage}[b]{\linewidth}\raggedright
Reranker logit
\end{minipage} &
\begin{minipage}[b]{\linewidth}\raggedright
Reranker rank
\end{minipage} &
\begin{minipage}[b]{\linewidth}\raggedright
Fused probability
\end{minipage} \\
\midrule\noalign{}
\endhead
\bottomrule\noalign{}
\endlastfoot
fclt\_\allowbreak{}building & 1 &
-2.20 & 15 & 2 &
0.524 & 0.96 & 1.67
& 0 & 0 & -1.50 & 2
& 0.124 \\
mit\_\allowbreak{}student\_\allowbreak{}directory
& 1 & 1.80 & 1 & 24
& 0.468 & 0.00 &
2.53 & 0 & 0 & -3.88
& 13 & 0.053 \\
fclt\_\allowbreak{}rooms & 1 &
-2.52 & 19 & 17 &
0.464 & 0.43 & 0.11
& 0 & 0 & -3.06 & 7
& 0.019 \\
fac\_\allowbreak{}building & 0 &
-1.38 & 8 & 0 &
0.536 & 0.96 & 2.20
& 0 & 0 & -1.25 & 1
& 0.218 \\
fclt\_\allowbreak{}building\_\allowbreak{}hist
& 0 & -5.66 & 65 & 1
& 0.527 & 0.96 &
3.50 & 0 & 0 & -2.94
& 6 & 0.110 \\
buildings & 0 &
-2.25 & 16 & 4 &
0.500 & 1.00 & 1.22
& 0 & 0 & -0.63 & 0
& 0.085 \\
fclt\_\allowbreak{}building\_\allowbreak{}hist\_\allowbreak{}1
& 0 & -5.72 & 67 & 3
& 0.523 & 0.96 &
2.83 & 0 & 0 & -2.37
& 4 & 0.078 \\
iap\_\allowbreak{}subject\_\allowbreak{}detail
& 0 & 2.20 & 0 & 42
& 0.432 & 0.00 &
3.76 & 0 & 0 & -7.06
& 56 & 0.052 \\
fac\_\allowbreak{}rooms & 0 &
0.88 & 2 & 20 &
0.441 & 0.29 & 1.95
& 0 & 0 & -3.37 & 8
& 0.042 \\
\ldots{} & & & & & &
& & & & & & \\
reservations & 0 &
-9.44 & 399 & 46 &
0.427 & 0.00 & 2.14
& 23 & 57 & -8.22 &
65 & 0.000 \\
lbaas\_\allowbreak{}l7policies &
0 & -9.06 & 144 & 43
& 0.431 & 0.00 &
1.19 & 23 & 57 &
-10.70 & 76 &
0.000 \\
backups & 0 & -4.31
& 38 & 224 & 0.340 &
0.00 & 1.75 & 38 &
38 & -8.37 & 69 &
0.000 \\
\end{longtable}

The reranker columns
are the two extra
features of Section
6.8; the fused
probability is from
the eight-feature
fusion. The rank gap
is the absolute
difference between
log(1 + scorer rank)
and log(1 + search
rank).

The question shows
both what the scorer
adds and where the
pipeline still
fails. Search ranks
mit\_\allowbreak{}student\_\allowbreak{}directory
25th (rank 24),
because the
question's words do
not appear in its
name; the trained
scorer ranks it 2nd
(rank 1). But BEAVER
holds several
near-duplicates of
the building table
(fac\_\allowbreak{}building,
fclt\_\allowbreak{}building\_\allowbreak{}hist,
fclt\_\allowbreak{}building\_\allowbreak{}hist\_\allowbreak{}1,
buildings), and the
eight-feature fusion
gives the non-gold
fac\_\allowbreak{}building the
highest probability.
fclt\_\allowbreak{}rooms, which
the scorer ranks
20th and search
18th, ends up ninth
or lower among the
candidates, so this
question misses
complete@8. The
reranker ranks
fclt\_\allowbreak{}building and
fclt\_\allowbreak{}rooms higher
than the scorer does
(3rd and 8th,
against 16th and
20th), which is the
kind of
complementary signal
Section 6.8
measures.

The fitted
standardized weights
for BEAVER on seed
1, in the order of
Section 3.3: scorer
logit 1.08, scorer
rank -0.00, search
rank -0.30, dense
cosine 0.58, BM25
0.26, rank gap 0.29,
database best search
rank -1.07, database
best scorer rank
-0.80. With the two
reranker features,
the reranker logit
(0.95) gets about
the same weight as
the scorer logit
(0.87).

\end{document}